\documentclass[11pt]{article}
\usepackage{natbib,graphicx,setspace,lscape,longtable}
\usepackage{natbib,epsfig,graphicx}
\usepackage{mathrsfs,amsmath,amsthm,amssymb,color, verbatim,xcolor}

\numberwithin{equation}{section}

\newtheorem{theorem}{Theorem}
\newtheorem{lemma}{Lemma}

\newcommand{\csection}[1]
    {\begin{center}
        \stepcounter{section}
        {\bf\large\arabic{section}. #1}
    \end{center}  }
\newcommand{\scsection}[1]
    {\begin{center}
        {\bf\large #1}
    \end{center}
}
\newcommand{\csubsection}[1]{
\begin{center}
\stepcounter{subsection}
{\it\arabic{section}.\arabic{subsection}. #1}
\end{center}
}

\newcommand{\scsubsection}[1]{
\begin{center}
\stepcounter{subsection}
{\it #1}
\end{center}
}

\def\1{\mathbf{1}}
\def\mA{\mathbb A}

\def\mR{\mathbb{R}}
\def\mS{\mathcal S}

\def\hat{\widehat}
\def\var{\mbox{var}}
\def\cov{\mbox{cov}}

\def\argmax{\mbox{argmax}}

\def\lsk{\left(}
\def\rsk{\right)}
\def\lbk{\left \{ }
\def\rbk{\right \} }
\def\lmk{\left [ }
\def\rmk{\right ] }

\def\laak{\left \| }
\def\raak{\right \| }

\def\nn{\nonumber}

\def\rmE{{\rm E}}
\def\rmP{{\rm P}}

\def\beq{\begin{equation}}
\def\eeq{\end{equation}}
\def\ben{\begin{equation*}}
\def\een{\end{equation*}}
\def\bea{\begin{eqnarray}}
\def\eea{\end{eqnarray}}

\def\bda{\begin{eqnarray*}}
\def\eda{\end{eqnarray*}}

\begin{document}
\begin{center}
{\bf\Large Mutual Influence Regression Model} \\
\bigskip
Xinyan Fan, Wei Lan, Tao Zou and Chih-Ling Tsai  \\

{\it Renmin University of China, Southwestern University of Finance and Economics, Australian National University and University of California, Davis} \


\end{center}

\begin{abstract}

In this article, we propose the mutual influence regression model (MIR) to establish the relationship between
the mutual influence matrix of actors and a set of similarity matrices induced by their
associated attributes.
This model is able to explain the heterogeneous structure of the mutual influence matrix by extending the
commonly used spatial autoregressive model
while allowing it
to change with time.
To facilitate making inferences with MIR, we establish parameter estimation, weight matrices selection and model testing.
Specifically,
we employ the quasi-maximum likelihood estimation method to estimate unknown regression coefficients,
and demonstrate that
the resulting estimator is asymptotically normal without imposing the normality assumption and while allowing the number of similarity matrices
to diverge. In addition, an extended
BIC-type criterion is introduced for selecting relevant matrices from the divergent  number of similarity matrices.
To assess the adequacy of the proposed model, we further propose an influence matrix test
and develop a novel approach in order to obtain the limiting distribution of the test.
Finally, we  extend the model to accommodate endogenous weight matrices, exogenous covariates, and both individual and time fixed effects, to broaden the usefulness of MIR.
The simulation studies support our theoretical findings, and a
real example is presented to illustrate the usefulness of the proposed MIR
model.\\


\end{abstract}

\newpage

\csection{Introduction}

Due to the possibility of relationships between subjects (such as network connections or spatial interactions),
the traditional data assumption of independent and identically distributed obervations is no longer valid, and there can be
a complex structure of mutual influence  between the subjects.
Accordingly, understanding mutual influence  has become an important topic
across various fields and applications such as business, biology, economics, medicine, sociology, political science, psychology, engineering, and science.
For example, the study of the mutual influence between actors can help to identify influential users within a network (see Trusov et al., 2010).
In addition, investigating
the mutual influence between geographic regions is essential for
exploring spillover effects in spatial data (see Golgher and Voss, 2016; Zhang and Yu, 2018),
and this type of analysis is important for understanding the spread of COVID-19  between different countries and cities (see Han et al., 2021).
Moreover, quantifying mutual influence in mobile social networks is helpful to provide important insights into the design
of social platforms and applications (see Peng et al., 2017).
These examples motivate us to introduce the mutual influence regression model so that we are able to effectively and systematically study mutual influence.

Let $Y_{1t}, \cdots, Y_{nt}$ be the responses of $n$ actors observed at time $t$ for $t=1, \cdots, T$.
To characterize the mutual influence among the $n$ actors, the following regression model
can be considered for each actor $i=1,\cdots,n$
at $t=1, \cdots, T$,
\begin{equation}Y_{it}=b_{i1t}Y_{1t}+\cdots+b_{i(i-1)t}Y_{(i-1)t}+b_{i(i+1)}Y_{(i+1)t}+\cdots+b_{int}Y_{nt}+\epsilon_{it},\end{equation}
where $b_{ijt}$ presents the influence effect of $Y_{jt}$ on $Y_{it}$ and $\epsilon_{it}$ is the random noise.
Define $Y_t=(Y_{1t}, \cdots, Y_{nt})^\top\in\mR^n$, $\epsilon_t=(\epsilon_{1t}, \cdots, \epsilon_{nt})^\top\in\mR^n$ and $B_t=(b_{ijt})\in\mR^{n\times n}$ with $b_{iit}=0$.
Then
we have the matrix form of (1.1),
\begin{equation} Y_t=B_t Y_t+\epsilon_t,\end{equation}
where $B_t$ is called the mutual influence matrix and it characterizes the degree of mutual influence among the $n$ actors at time $t$.

Estimating model (1.2) is a challenging task
since it involves a large number of  parameters, specifically $n(n-1)$ for each $t$.
To avoid the issue of high dimensionality,
one commonly used approach
is to employ the spatial autoregressive  (SAR) model, which parameterizes the mutual influence matrix $B_t$ by $B_t=\rho W^{(t)}$, where $W^{(t)}$ is the adjacency matrix of a known network
or a spatial weight matrix whose elements
are a function of geographic or economic distances.
In addition, $\rho$ is the single influence parameter that characterizes the influence
power among the $n$ actors; see,
for example, Lee (2004), Zhou et al. (2017) and Huang et al. (2019)
for detailed discussions and the references therein. Accordingly, model (1.2) becomes estimable since the number of parameters is greatly reduced from $n(n-1)$ to 1.

Because the SAR model only involves a single influence parameter $\rho$, it may not fully
capture the influential information of $B_t$. Hence, Lee and Liu (2010),
Elhorst et al. (2012), Liu et al. (2014), Badinger and Egger (2011, 2015) and Lam and Souza (2020) considered a higher-order SAR model that includes
multiple weight matrices (i.e., $W^{(t)}$s) along with their associated parameters. Gupta and Robinson (2015, 2018) further extended it
by allowing the number of weight matrices
to diverge.
In general, the elements of weight matrix $W^{(t)}$ are functions of the geographic or economic distances among the $n$ actors.
For example, a typical choice of distance measure for spatial data
is geographic distance (Dou et al., 2016; Zhang and Yu, 2018; Gao et al., 2019).
In addition, one natural choice of distance measure for network data is whether there exists a link between the actors through the adjacency matrix (Zhou et al., 2017; Zhu et al., 2017; Huang et al., 2019).
However, the above weight settings cannot be directly applied to the higher-order SAR model
for non-geographic or non-network data since these distance measure are not well defined for other types of data. Accordingly, how to parameterize the
mutual influence matrix for non-geographic and non-network data is an unsolved problem that needs further investigation. This motivates us to study the following two important and challenging subjects:
(i) How to define weight matrices for general non-geographic and non-network data? (ii) How to assess the adequacy of the selected weight matrices?

To resolve challenge (i),
we propose using similarity matrices induced from attributes
(e.g., gender or income) to be our weight matrices to accommodate non-geographic and non-network data. Specifically, let
$\mathbf{Z}^{(t)}=(z^{(t)}_{1}, \cdots, z^{(t)}_{n})^\top\in\mR^n$ denote the vector of
values obtained from the $n$ actors for a given attribute.
Then, for any two actors $j_1$ and $j_2$, the squared distance between $j_1$ and $j_2$ can be defined as the distance between $z^{(t)}_{j_1}$ and
$z^{(t)}_{j_2}$, e.g., $(z^{(t)}_{j_1}-z^{(t)}_{j_2})^2$. Following the suggestion of Jenish and Prucha (2012) and Qu and Lee (2015),
we consider the similarity matrix as a non-increasing function of the squared distance between actors
$j_1$ and $j_2$, i.e., $A^{(t)}=(a\{-(z^{(t)}_{j_1}-z^{(t)}_{j_2})^2\})_{n\times n}$ for some bounded and non-decreasing function $a(\cdot)$.
Furthermore, we can employ the same procedure to create a set of similarity
matrices $A^{(t)}$s deriving from the actors' attributes. In practice, those similarity matrices change along with time $t$. To this end, we introduce the time heterogeneous
matrices, $A^{(t)}$s, which naturally link to the mutual influence matrix $B_t$.
To overcome the aforementioned challenge (ii), we introduce an influence matrix test
to examine the adequacy of the selected similarity matrices (i.e., weight matrices) for the high dimensional and time varying mutual influence matrix.

The main contribution of this paper is two-fold.
The first is to propose a mutual influence regression (MIR) model that establishes a relationship between
the mutual influence matrix and a set of similarity matrices induced by associated attributes of the actors.
The emerging model not only broadens the usefulness of the traditional spatial autoregressive model, but also
captures the
heterogeneous structure of the mutual influence matrix by allowing it
to change with time.  Accordingly, we  study the parameter space of the model and then employ the quasi-maximum likelihood estimation method (see, e.g., Wooldridge, 2002) to estimate unknown regression coefficients.
By thoroughly studying the convergence of the Hessian matrix in Frobenius norm, we are able to show that
the resulting estimator is asymptotically normal under some
mild conditions without imposing the normality assumption while allowing the number of similarity matrices
to diverge.
Since the number of similarity matrices is diverging,
an extended BIC-type criterion motivated from Chen and Chen (2008) is introduced to select relevant matrices.
We show that this extended BIC-type criterion is consistent based on a novel result of the exponential tail probability for the general form of
quadratic functions. To expand the usefulness of MIR, we further extend the model to accommodate endogenous weight matrices, exogenous covariates, and both individual and time
fixed effects.

The second is to introduce an influence matrix test for assessing whether the mutual influence matrix $B_t$ satisfies  a linear structure of the  time-varying weight matrices. Based on this setting,
$\cov(Y_t)$ is a nonlinear function of the time-varying weight matrices. Thus, our test
is different from the common hypothesis test for testing whether $\cov(Y_t)$
is a linear structure of the weight matrices (e.g.,  see Zheng et al., 2019).
Under a nonlinear structure for the mutual influence matrix $B_t$, however, the quasi-maximum likelihood estimators of regression coefficients can result in
 a larger variance in the test statistic. As a result, obtaining the asymptotic distribution of the test statistic becomes a challenging task,
especially when the number of similarity matrices is diverging.
To overcome such difficulties, we develop a novel approach in order to show
the asymptotic normality of a summation of the product of quadratic forms with
a diverging number of similarity matrices.

The remainder of this article is organized as follows. Section 2 introduces the
mutual influence regression model, studies the parameter space, and obtains quasi-maximum likelihood estimators of regression coefficients,
which are asymptotically normal. Section 3 presents the extended BIC-type selection criterion as well as its consistency property.
In addition, a high dimensional covariance test is given to examine the model adequacy. The theoretical property of this test is provided.
Section 4 studies an endogeneity-adjusted quasi-maximum
likelihood estimation  method that is modified from Qu and Lee (2015) to accommodate endogenous weight matrices.
Section 5 further extends the model to accommodate  exogenous covariates and both individual and time fixed effects.
Simulation studies
and an empirical example are presented in Sections 6 and 7, respectively, while
Section 8 concludes the article with a discussion. All theoretical
proofs are relegated to the Appendices and supplementary material.


\csection{Mutual Influence Regression Model and Estimation}

\csubsection{Model and Notation}

We first construct similarity matrices before modeling the mutual influence matrix $B_t$ as a regression function of them.
Let $Z_{k}^{(t)}$ be the $k$-th $n\times 1$ continuous attribute vector collected at the $t$-th time for $k=1,\cdots,d$. Adapting
Qu and Lee's (2015) approach in order to incorporate the time effect $t$, we then obtain heterogeneous
similarity matrices:
$A_k^{(t)}=A_k^{(t)}(Z_{k}^{(t)})
=(a\{-(Z_{k{j_1}}^{(t)}-Z_{k{j_2}}^{(t)})^2\})_{n\times n}$
 for $j_1=1,\cdots, n$ and $j_2=1,\cdots, n$, where $a(\cdot)$ is a bounded and non-decreasing function and
$Z_{k{j_1}}^{(t)}$ and $Z_{k{j_2}}^{(t)}$ are the $j_1$-th and $j_2$-th elements of  $Z_{k}^{(t)}$, respectively.
For continuous attributes,
we consider $a(\cdot)$ equal to  the exponential function with
$a\{-(Z_{k{j_1}}^{(t)}-Z_{k{j_2}}^{(t)})^2\}=\exp\{-(Z_{{kj_1}}^{(t)}-Z_{{kj_2}}^{(t)})^2\}$ when $|Z_{{kj_1}}^{(t)}-Z_{{kj_2}}^{(t)}|<\phi^{(t)}_k$ for some pre-specified
positive constant $\phi^{(t)}_k$, and $a\{-(Z_{k{j_1}}^{(t)}-Z_{k{j_2}}^{(t)})^2\}=0$ otherwise.
That is, once the distance
between any two actors measured by their associated attributes in $Z_{k}^{(t)}$
exceeds a threshold, the two actors are not mutually influenced.
For discrete attributes $Z_{k}^{(t)}$, we define
$a(Z_{kj_1}^{(t)},Z_{{kj_2}}^{(t)})=1$ if $Z_{kj_1}^{(t)}$ and $Z_{kj_2}^{(t)}$ belong to the same
class, and  $a(Z_{kj_1}^{(t)},Z_{{kj_2}}^{(t)})=0$ otherwise. In this case,
$A_k^{(t)}$ can be regarded as the adjacency matrix of the network induced by attributes $Z_{k}^{(t)}$.
Note that in this paper we consider the attributes $Z_{k}^{(t)}$ to be exogenous, except in Section 4.

To establish the relationship between the mutual influence matrix and a set of similarity matrices, motivated from Anderson (1973), Qu et al. (2000) and Zheng et al. (2019),
we parameterize
the mutual influence matrix $B_t$ as a function of attributes $Z_{k}^{(t)}$s ($k=1, \cdots, d$) given below.
\begin{equation}B_t(\lambda) \triangleq B_t(Z_{1}^{(t)}, \cdots, Z_{d}^{(t)}, \lambda)=\lambda_1 W^{(t)}_1+\cdots+\lambda_d W^{(t)}_d,\end{equation}
where $W_k^{(t)}=(w(Z_{k{j_1}}^{(t)},Z_{k{j_2}}^{(t)}))_{n\times n}$
and $w(Z_{k{j_1}}^{(t)},Z_{k{j_2}}^{(t)})=a(Z_{k{j_1}}^{(t)},Z_{k{j_2}}^{(t)})/\sum_{j_2} a(Z_{k{j_1}}^{(t)},Z_{k{j_2}}^{(t)})$
is the row-normalized version of $A_k^{(t)}$.
We name $W_k^{(t)}$ as the weight matrix for $k=1,\cdots,d$, which is also called the similarity matrix in the rest of the article.
The reason for adopting the row-normalization method
is primarily its wide applicability (see, e.g., Lee, 2004 and Lee and Yu, 2010a, 2010b).
In practice, there are several alternative normalization methods that can be considered, such as the column normalization and the normalization based on the maximum absolute row (or column) sum norm; see Leenders (2002) and Kelejian and Prucha (2010) for detailed discussions.

Substituting (2.1) into (1.2), we
introduce the following mutual influence regression (MIR) model,
\begin{equation}Y_t=B_t(Z_{1}^{(t)}, \cdots, Z_{d}^{(t)}, \lambda)Y_t+\epsilon_t=\big(\lambda_1 W^{(t)}_1+\cdots+\lambda_d W^{(t)}_d\big)Y_t+\epsilon_t,\end{equation}
where $\lambda_1, \cdots, \lambda_d$ are unknown regression coefficients.
This model is able to explain the structure of the mutual influence matrix $B_t$ at each time $t$ via a set of similarity matrices $W^{(t)}_k$,
induced by the covariates $Z_k^{(t)}$ and their associated
influence parameter $\lambda_k$. For the sake of simplicity, we refer to the above model as MIR in the rest of the paper.
To ease notation, we use $B_t$ rather than $B_t(\lambda)$ in the rest of article.
Define $\Delta_t(\lambda)=I_n-B_t=I_n-\big(\lambda_1 W^{(t)}_1+\cdots+\lambda_d W^{(t)}_d\big)$, where $I_n$ is the identity matrix of dimension $n$. Then,  model (2.2) leads to
$\Delta_t (\lambda)Y_t=\epsilon_t$. To assure (2.2) identifiable, we require that $\Delta_t (\lambda)$ is invertible.

It is worth noting that,  for $d=1$ and $W^{(t)}_1=W$ constructed by network or spatial data, MIR is the
classical spatial autoregressive model of LeSage and Pace (2009).
Furthermore,  by model (2.1), we have
$b_{j_1j_2t}=\lambda_1w(Z_{1j_1}^{(t)},Z_{{1j_2}}^{(t)})+\cdots+\lambda_d w(Z_{dj_1}^{(t)},Z_{{dj_2}}^{(t)})$. Accordingly,
the influence effect of node $j_2$ on $j_1$, $b_{j_1j_2t}$, is the linear combination of
similarity matrices at time $t$.
Specifically, for $k=1,\cdots, d$, the similarity matrix
$w(Z_{kj_1}^{(t)},Z_{{kj_2}}^{(t)})$ measures the distance between
nodes $j_1$ and $j_2$, and its effect is determined by the influence parameter  $\lambda_k$.
Suppose $\lambda_k>0$.
Based on the MIR model (2.2), for any two actors $j_1$ and $j_2$,  the smaller
the distance between $Z_{kj_1}^{(t)}$ and $Z_{{kj_2}}^{(t)}$, the larger the influence effect between
$Y_{j_1t}$ and $Y_{j_2t}$. Therefore,
the covariate $Z_k^{(t)}$ yields a positive effect on the mutual influence between responses of the $n$ actors.
In sum,  models (2.1) and (2.2) link the mutual influence matrix with a large number of exogenous attributes to responses, which can lead to insightful findings and provide practical interpretations.

\csubsection{Parameter Estimation}

In this paper,
we assume that
$\epsilon_t$s are iid random variables with mean 0 and covariance matrix $\sigma^2 I_n$
for $t=1,\cdots, T$,
where $\sigma^2$ is a scaled parameter.
By (2.2), we have
$Y_t=\Delta_t^{-1}(\lambda)\epsilon_t$.
Then $E(Y_t)=0$ and $\var(Y_t)\triangleq \Sigma_t=\sigma^2 \Delta_t^{-1}(\lambda)\{\Delta_t^\top(\lambda)\}^{-1}$,
and we obtain the
quasi-loglikelihood function,
\begin{equation}\ell(\theta)=-\frac{nT}{2}\log (2\pi)-\frac{nT}{2}\log(\sigma^2)+\sum_{t=1}^T\log|\mbox{det}(\Delta_t(\lambda))|-\frac{1}{2\sigma^2}\sum_{t=1}^T Y_t^\top \Delta_t^\top(\lambda)\Delta_t(\lambda)Y_t,\end{equation}
where $\theta=(\lambda^\top,\sigma^2)^\top$.

We next employ  the concentrated quasi-likelihood approach to estimate
$\theta$.
Specifically, given $\lambda$, one can estimate $\sigma^2$ by
\[\hat\sigma^2(\lambda)=(nT)^{-1}\sum_t Y_t^\top \Delta_t^\top(\lambda)\Delta_t(\lambda)Y_t.\]
Plugging this into (2.3), the resulting quasi-concentrated log-likelihood function is
\begin{equation}\label{lc}
\ell_c(\lambda)=-\frac{nT}{2}\log (2\pi)-\frac{nT}{2}-\frac{nT}{2}\log\big\{\hat\sigma^2(\lambda)\big\}+\sum_{t=1}^T\log|\mbox{det}(\Delta_t(\lambda))|.
\end{equation}
Accordingly, we obtain the quasi-maximum likelihood estimator of $\lambda$, which is
$\hat\lambda=\mbox{argmax}_{\lambda\in\Lambda} \ell_c(\lambda)$ and  $\Lambda$ is the parameter space.
To make $\hat\lambda$ estimable, it is necessary to specify the parameter space $\Lambda$.
Based on  model (2.2) and the definition of $\Delta_t(\lambda)$, one should naturally require that,
for any $\lambda\in \Lambda$, $\Delta_t(\lambda)$ is invertible. It is worth noting that a sufficient condition for the invertibility of $\Delta_t(\lambda)$ is
$\|\sum_{k=1}^{d}\lambda_kW_k^{(t)}\|<1$, where $\|\cdot\|$ denotes the $L_2$ (i.e., spectral) norm.
Using the fact that $W_k^{(t)}$
is row-normalized, we  have that
$\|\sum_{k=1}^{d}\lambda_k W_k^{(t)}\|\leq \max_{k}\|W_k^{(t)}\|\sum_{k=1}^{d}|\lambda_k|
\leq \sum_{k=1}^{d}|\lambda_k|.$
Accordingly, a sufficient condition for the invertibility of $\Delta_t(\lambda)$ is
$\sum_{k=1}^{d}|\lambda_k|<1$. This leads us to define the parameter space of $\lambda$ as follows:
\[\Lambda=\big\{\lambda: \sum_{k=1}^{d}|\lambda_k|<1-\varsigma\big\},\]
where $\varsigma$ is some sufficiently small positive number.
This type of parameter space is also adapted by Gupta and Robinson (2018).

Using the assumption of $\sigma^2>0$,  the parameter space of $\theta$
is
\[\Theta=\big\{\theta=(\lambda^\top, \sigma^2)^\top: \lambda\in\Lambda \mbox{~and~} \sigma^2>0\big\}.\]
In addition, $\sigma^2$ can be estimated by
$\hat\sigma^2=\hat\sigma^2(\hat\lambda$),
which leads to the quasi-maximum likelihood estimator (QMLE), i.e.,  $\hat\theta=(\hat\lambda^\top,\hat\sigma^2)^\top$.
Denote by $\theta_0=(\lambda_0^\top,\sigma_0^2)^{\top}$
the unknown true parameter vector, where  $\lambda_0=(\lambda_{01}, \cdots, \lambda_{0d})^\top\in \Lambda$ and $\sigma_0^2>0$.
By Lemma 3 and Condition (C4),
the second order derivative matrix of $\ell(\theta)$ is negative definite
for sufficiently large $nT$  in a small neighborhood of $\theta_0$.
Accordingly, the  parameter estimator $\hat\theta$ exists and lies in $\Theta$.
The asymptotic property of $\hat\theta$ is given in the following theorem.

\begin{theorem}
Under Conditions (C1)--(C5) in Appendix A, as $nT\rightarrow\infty$, $\sqrt{nT/d} D \mathcal{I}(\theta_0)(\hat\theta-\theta_0)$ is asymptotically normal with mean 0 and covariance
matrix $G(\theta_0)$, where $D$ is an arbitrary $M\times(d+1)$ matrix with $M<\infty$ satisfying  $\|D\|<\infty$
and  $d^{-1}D\mathcal{J}(\theta_0)D^{\top}\to G(\theta_0)$, and $\mathcal{I}(\theta_0)$ and $\mathcal{J}(\theta_0)$ are defined in Condition (C4).
\end{theorem}

\noindent
Note that $nT\rightarrow\infty$ in the above theorem  means that either $n$ or $T$ go to infinity.
To make this theorem practically useful, one needs to estimate
$\mathcal{I}(\theta_0)$ and $\mathcal{J}(\theta_0)$ consistently.
For $k=1,\cdots,d+1$ and $l=1,\cdots,d+1$, define
$\mathcal{I}_{nT}(\theta_0)=-(nT)^{-1}E\{\frac{\partial^2 \ell({\theta_0})}{\partial \theta\partial\theta^{\top}}\}\triangleq (\mathcal{I}_{nT, kl})\in\mR^{(d+1)\times (d+1)}$
and $\mathcal{J}_{nT}(\theta_0)=(nT)^{-1}\var(\frac{\partial \ell({\theta_0})}{\partial \theta})
\triangleq(\mathcal{J}_{nT, kl})\in\mR^{(d+1)\times (d+1)}$.
By Condition (C4) in Appendix A, it suffices to show that the plug-in estimators
$\mathcal{I}_{nT}(\hat\theta)$
and $\mathcal{J}_{nT}(\hat\theta)$ are
consistent  of $\mathcal{I}({\theta_0})$ and $\mathcal{J}({\theta_0})$, respectively.

After simple calculation, we have
that, for any $k=1, \cdots, d$ and $l=1 ,\cdots, d$,
\[\mathcal{I}_{nT, k(d+1)}\triangleq-(nT)^{-1}E\Big\{\frac{\partial^2 \ell({\theta_0})}{\partial \lambda_k\partial\sigma^2}\Big\}=\frac{1}{nT\sigma^2}\sum_{t=1}^T tr(W_k^{(t)}\Delta^{-1}_t({\lambda_0}))=\frac{1}{nT\sigma^2} tr(U_k) \quad\mbox{and}\]
\[\mathcal{I}_{nT, kl}\triangleq-(nT)^{-1}E\Big\{\frac{\partial^2 \ell({\theta_0})}{\partial \lambda_k\partial\lambda_l}\Big\}=(nT)^{-1}\sum_{t=1}^T tr\{\Delta^{-1\top}_t({\lambda_0})W_k^{(t)\top}W_l^{(t)}\Delta^{-1}_t({\lambda_0})\}\]
\[+(nT)^{-1}\sum_{t=1}^T tr\{W_k^{(t)}\Delta^{-1}_t(\lambda_0)W_l^{(t)}\Delta^{-1}_t(\lambda_0)\}=\frac{2}{nT}tr(U_k U_l),\]
where $U_k=\mbox{diag}\big\{s(W_k^{(1)}\Delta^{-1}_t({\lambda_0})), \cdots, s(W_k^{(T)}\Delta^{-1}_T({\lambda_0}))\big\}\in\mR^{(nT)\times (nT)}$ and
$s(A)=(A+A^\top)/2$ for any arbitrary matrix $A$.
In addition,
\[\mathcal{I}_{nT, (d+1)(d+1)}\triangleq-(nT)^{-1}E\Big\{\frac{\partial^2 \ell({\theta_0})}{\partial^2 \sigma^2}\Big\}=\frac{1}{2{\sigma_0^4}}.\]
Using the result $\hat\theta\rightarrow_p {\theta_0}$ in Theorem 1, we have $\mathcal{I}_{nT}(\hat\theta)\rightarrow_p \mathcal{I}_{nT}({\theta_0})$. This, together with
Condition (C4), implies $\mathcal{I}_{nT}(\hat\theta)\rightarrow_p \mathcal{I}({\theta_{0}})$.

After algebraic calculation, we next obtain that,
for any $k=1, \cdots, d$ and $l=1 ,\cdots, d$,
\[\mathcal{J}_{nT, k(d+1)}\triangleq(nT)^{-1}\cov\Big\{\frac{\partial \ell({\theta_0})}{\partial\lambda_k},\frac{\partial \ell({\theta_0})}{\partial\sigma^2}\Big\}=\frac{1}{2nT{\sigma_0^2}}\big\{(\mu^{(4)}-1)tr(U_k)\big\} \quad \mbox{and}\]
\[\mathcal{J}_{nT, kl}\triangleq(nT)^{-1}\cov\Big\{\frac{\partial \ell({\theta_0})}{\partial\lambda_k},\frac{\partial \ell({\theta_0})}{\partial\lambda_l}\Big\}=\frac{2}{nT}tr(U_kU_l)+\frac{\mu^{(4)}-3}{nT}tr(U_k\otimes U_l),\]
where $\mu^{(4)}=E(\epsilon_{it}^4)/{\sigma_0^4}$ can be consistently estimated by $\hat\mu^{(4)}=(nT)^{-1}\sum_{i=1}^n\sum_{t=1}^T \hat\epsilon_{it}^4/{\hat{\sigma}^4}$
with $\hat\epsilon_t=\Delta^{-1}_t({\hat{\lambda}})Y_t$ and $\hat\epsilon_t=(\hat\epsilon_{1t}, \cdots, \hat\epsilon_{nt})^\top$.
Furthermore,
\[\mathcal{J}_{nT, (d+1)(d+1)}\triangleq(nT)^{-1}\var\Big\{\frac{\partial \ell({\theta_0})}{\partial\sigma^2}\Big\}=\frac{1}{4{\sigma^4_0}}\big\{2+(\mu^{(4)}-3)\big\}.\]
 As a result, $\mathcal{J}({\theta_0})$ can be consistently estimated by
$\mathcal{J}_{nT}(\hat\theta)$.
In sum, one can practically apply Theorem 1 by replacing
$\mathcal{I}(\theta_0)$ and $\mathcal{J}(\theta_0)$ with their corresponding estimators $\mathcal{I}_{nT}(\hat\theta)$ and $\mathcal{J}_{nT}(\hat\theta)$, respectively.

According to Theorem 1, we are able to assess the significance of ${\lambda_{0k}}$, which allows us to determine  the influential similarity matrices, $W^{(t)}_k$, induced by their associated covariates
$Z_k^{(t)}$ for $k=1, \cdots, d$. In addition,
based on the estimated $\hat\lambda$, the mutual influence matrix $B_t$ can be estimated by
$\hat B_t=\hat\lambda_1 W^{(t)}_1+\cdots+\hat\lambda_d W^{(t)}_d$, whose asymptotic property is given below.

\begin{theorem}
Under Conditions (C1)--(C5) in Appendix A, as $nT\rightarrow\infty$, $\sup_{t\leq T}\|\hat B_t-B_t\|=O_p\{d/\sqrt{nT}\}$.
\end{theorem}

\noindent The above theorem indicates that the estimated mutual influence matrix $\hat B_t$ is consistent uniformly for any $t$ under the $L_2$ norm,
as either $n$ or $T$ goes to infinity and  $d=o\{(nT)^{1/4}\}$ is from Condition (C5).
 Hence, $\hat B_t$ can be a consistent estimator of $B_t$
even for finite $T$. After estimating the mutual influence matrix, we next study the selection of similarity matrices and test the fitness of $B_t$.

\csection{Similarity Matrix Selection and Influence Matrix Test}

\csubsection{Selection Consistency}

In MIR, the number of similarity matrices is diverging, which motivates us to consider the similarity matrix selection; see also Han et al. (2017). Note that
the traditional Bayesian information criterion (BIC) becomes overly liberal when $d$ is diverging as demonstrated by Chen and Chen (2008). Hence, we modify
the extended Bayesian information criterion (EBIC) to select  similarity matrices.
To this end,
we define the true model
$\mS_T=\{k: \lambda_{0k}\not=0\}$, which consists of all relevant $W_k^{(t)}$s.
In addition, let
$\mS_F=\{1, \cdots, d\}$ denote the full model and  $\mS$ represent an arbitrary
candidate model such that $\mS\subset\mS_F$. Moreover,
let $\hat\theta_{\mS}=(\hat\theta_{k,\mS}: k\in\mS)$ be the
maximum likelihood estimator of $\theta_{0\mS}=(\theta_{0k}: k\in\mS)\in\mR^{|\mS|}$.
In practice, the true model $\mS_T$ is unknown.
Motivated from Chen and Chen (2008), we propose the information criterion given below
to select similarity matrices,
\[
\mbox{EBIC}_{\gamma}(\mS)=-2\ell(\hat\theta_{\mS})+|\mS|\log(nT)+\gamma|\mS|\log(d)
\]
for some $\gamma>0$.
Based on this criterion, one can select the optimal model, which is
$\hat\mS=\mbox{argmin}_{\mS} \mbox{EBIC}_{\gamma}(\mS)$.

 Define as $\mA_0=\{\mS:\mS_T\subset \mS, |\mS|\leq q\}$ and $\mA_1=\{\mS:\mS_{T}\not \subset \mS, |\mS|\leq q\}$ the sets of
the overfitted and underfitted models, respectively, where the size of any candidate model is no larger than the positive constant $q$ defined in Condition (C7).
Then, we obtain the theoretical properties of $\mbox{EBIC}_{\gamma}$ given below.

\begin{theorem}
Under Conditions (C1)--(C7) in Appendix A, as $nT\rightarrow \infty$,
we have
$$
P\Big\{\min_{\mS\in \mA_1}{\rm EBIC}_{\gamma}(\mS)\leq {\rm EBIC}_{\gamma}(\mS_T)\Big\}\to 0
$$ for any $\gamma>0$ and
$$
P\Big\{\min_{\mS\in \mA_0,\mS\neq\mS_T}{\rm EBIC}_{\gamma}(\mS)\leq {\rm EBIC}_{\gamma}(\mS_T)\Big\}\to 0
$$
for $\gamma>q^2C_w^2/\tau_2c_{\min,3}\sigma_0^4 -4$, where $C_w$, $c_{\min,3}$ and $\tau_2$ are
finite positive constants which are defined in Conditions (C3), (C7) and Lemma 3 (ii) in Appendix A, respectively.
\end{theorem}
\noindent The above theorem holds as long as either $n$ or $T$ go to infinity.
Note that the assumption
$\min_{k\in\mS_T}|\lambda_{0k}|\sqrt{nT/\log(nT)}\to\infty$ given in Condition (C6) is modified from Chen and Chen (2008).
According to Theorem 3, if the minimum value of $\lambda_{0k}$ for $k\in\mS_T$ is relatively large, EBIC can identify the true model consistently by choosing $\gamma$ appropriately
even with the diverging number of
similarity matrices. Our simulation results indicate that $\gamma=2$ performs satisfactorily under various settings.

\csubsection{Influence Matrix Test}

To examine the adequacy of model (2.1) for modeling the mutual influence matrix $B_t$ as a
linear combination of weight matrices $W^{(t)}_k$ ($k=1, \cdots, d$),
we consider the following hypotheses,
\[H_{0}: B_t=\lambda_{01} W^{(t)}_1+\cdots+\lambda_{0d} W^{(t)}_d \mbox{~for all~} t=1, \cdots, T, \mbox{~vs~}\]
\beq H_1: B_t\not=\lambda_{01} W^{(t)}_1+\cdots+\lambda_{0d} W^{(t)}_d \mbox{~for some~} t=1, \cdots, T.\eeq
Note that, under  $H_0$, we have $\Sigma_t=\sigma^2_0 (I_n-B_t)^{-1}(I_n-B_t^\top)^{-1}$, which is a nonlinear function of
the weight matrices $W^{(t)}_k$s. This is different from the covariance structure considered in Qu et al. (2000) and Zheng et al. (2019) which assumes that $\Sigma_t$
is a linear function of the weight matrices.

To test (3.1), it is natural to compare the estimates of $B_t$ calculated under the null and alternative hypotheses, respectively.
 Then reject the null hypothesis of (3.1) if their difference is relatively large.
However, the computation of $B_t$ under the alternative hypothesis is infeasible since it involves  $n(n-1)T$ unknown parameters.
Hence, we propose to test (3.1) by comparing the covariance matrix of $Y_t$ under the null and alternative hypotheses, respectively.
Under  $H_0$,
we have $\cov(Y_t)=\Sigma_t=\sigma^2_0 (I_n-B_t)^{-1}(I_n-B_t^\top)^{-1}$. Based on Theorem 2,  $B_t$ can be consistently estimated by $\hat B_t=B_t(\hat\lambda)$. Accordingly, one can approximate $\cov(Y_t)$ by $\hat\Sigma_t=\hat\sigma^2 (I_n-\hat B_t)^{-1}(I_n-\hat B_t^\top)^{-1}$,
where $\hat\sigma^2=(nT)^{-1}\sum_t Y_t^\top \Delta_t^\top(\hat\lambda)\Delta_t(\hat\lambda)Y_t$.
On the other hand,  $\cov(Y_t)$ can be approximated by its sample version under the alternative, and
we expect that
$E(Y_t Y_t^\top)\approx \hat\Sigma_t$ under the null hypothesis, which motivates us to
employ the quadratic loss function $tr(Y_tY_t^\top\hat\Sigma_t^{-1}-I_n)^2$ to measure the difference between $Y_tY_t^\top$ and $\hat\Sigma_t$.
It is expected that, under $H_0$,
the difference should be small across $t=1, \cdots, T$.
Hence, we propose the following test statistic,
\[T_{ql}=(nT)^{-1}\sum_{t=1}^T tr(Y_tY_t^\top\hat\Sigma_t^{-1}-I_n)^2,\]
to assess the adequacy of (2.1).

To show the asymptotic distribution
 of $T_{ql}$, let
$\mu_{ql}=n+\mu^{(4)}-2$ and
\beq \sigma_{ql}^2=(4\mu^{(4)}-4)n/T+ 4n^{-2}T^{-4}{\sigma^4_0}\sum_{t_1\not=t_2\not=t_3}\sum_{k_1,l_1}\sum_{k_2,l_2}\mathcal{I}^{-1}_{k_1l_1}({\theta_0})\mathcal{I}_{k_2l_2}^{-1}({\theta_0})\{tr(U_{t_1k_1}U_{t_1k_2})
+(\mu^{(4)}-3)\nonumber \eeq \beq \times tr(U_{t_1k_1}{\otimes}U_{t_1k_2})\} tr(V_{t_2l_1})tr(V_{t_3l_2})+
(8\mu^{(4)}-8)n^{-1}T^{-3}{\sigma^4_0}\sum_{t_1\not=t_2} \sum_{k,l}\mathcal{I}^{-1}_{kl}({\theta_0})tr(U_{t_1k})tr(V_{t_2l}),\eeq where
$\mathcal{I}_{kl}^{-1}({\theta_0})$ is the $kl$-th element of $\mathcal{I}^{-1}({\theta_0})$,
$U_{tk}={s\{W_k^{(t)}\Delta_t^{-1}(\lambda_0)\}}$, $V_{tk}=\{\Delta_t^{-1}({\lambda_0})\}^\top\widetilde\Lambda_{tk} \Delta_t^{-1}({\lambda_0})$ and
$\widetilde\Lambda_{tk}$ is the matrix form of $\partial \mbox{vec}\{\Sigma_t^{-1}({\theta_0})\}/\partial \theta_k$ for $t_1, t_2, t_3, t=1,\cdots,T$, $k_1, k_2, k=1,\cdots,d$, and $l_1, l_2, l=1,\cdots,d$.
Then, the next theorem presents the asymptotic property of $T_{ql}$.

\begin{theorem}
Under the null hypothesis of $H_0$, Conditions (C1)--(C5) in Appendix A and assuming that $n/T\rightarrow c$ for some finite positive constant $c$,
we have \[(T_{ql}-\mu_{ql})/\sigma_{ql}\rightarrow_d N(0,1)\] as $nT\rightarrow\infty$.
\end{theorem}

\noindent Unlike Theorems 1--3, the above result requires that both $n$ and $T$ tend to infinity with $n/T\rightarrow c$ for some finite positive constant $c$.
This condition is reasonable
 since we need the replications of similarity matrices to
test the adequacy of MIR. Note that this condition is
commonly used for testing high dimensional covariance structures (see, e.g., Ledoit and Wolf, 2002; Zheng et al., 2019).
The above theorem indicates that  the asymptotic variance of $T_{ql}$ is  $\sigma^2_{ql}$, which is given in (3.2) and it includes three components.
The first component $(4\mu^{(4)}-4)c$
is the leading term of variance of $(nT)^{-1}\sum_{t=1}^T tr(Y_tY_t^\top\Sigma_t^{-1}-I_n)^2$ obtained by assuming that ${\lambda_0}$ is known, while
the last two components are of orders $O(d^2)$ and $O(d)$, respectively, and cannot be ignored. These two
non-negligible components are
mainly induced by the estimator $\hat\lambda$, which makes the proof of Theorem 4 more complicated. Thus, we develop Lemma 4 in Appendix A to resolve this challenging task.

To make the above theorem practically useful, one needs to estimate the two unknown terms $\mu_{ql}$ and $\sigma_{ql}$.
Note that $\mu^{(4)}$ in $\mu_{ql}$ can be consistently estimated by $\hat\mu^{(4)}$, which is  defined in the explanation of Theorem 1. As a result,
$\hat\mu_{ql}=n+\hat\mu^{(4)}-2$ is a consistent estimator of $\mu_{ql}$.
It is also worth noting that
 $U_{tk}$, $V_{tk}$ and $\mathcal{I}^{-1}_{kl}({\theta_0})$ can be consistently estimated by
$\hat U_{tk}={s(}W_k^{(t)}\Delta_t^{-1}(\hat\lambda))$,
$\hat V_{tk}=\{\Delta_t^{-1}(\hat\lambda)\}^\top\hat\Lambda_{tk} \Delta_t^{-1}(\hat\lambda)$ and
$\mathcal{I}^{-1}_{kl}(\hat\theta)$, respectively,
for $t=1,\cdots,T$ and $k, l=1,\cdots,d$, where
$\hat\Lambda_{tk}$
is the matrix form of $\partial \mbox{vec}\{\Sigma_t^{-1}(\hat\theta)\}/\partial \theta_k$
and $s(A)=(A+A^\top)/2$ for any arbitrary matrix $A$ defined in Section 2.2.
Accordingly, $\hat\sigma_{ql}$, obtained by replacing unknown parameters with their corresponding estimators, is a consistent estimator of  $\sigma_{ql}$. Consequently, for a
given significance level $\alpha$, we are able to reject the null hypothesis of $H_0$
if $|T_{ql}-\hat\mu_{ql}|>\hat\sigma_{ql}z_{1-\alpha/2}$, where $z_{\alpha}$ stands for the $\alpha$-th quantile of the standard normal distribution.

\csection{Mutual Influence Regression with Endogenous Weight Matrices}

In the previous two sections, we assumed that the
weight matrices $W_k^{(t)}$s were exogenous. However, as noted by Qu and Lee (2015) and
Qu et al. (2017), the weight matrices constructed from actors' attributes
can be endogenous.
As defined before, let $\mathbf{Z}^{(t)}=(Z_1^{(t)}, \cdots, Z_d^{(t)})\in\mR^{n\times d}$ be the collection of attributes, and
$z_i^{(t)\top}$ be the $i$-th row of $\mathbf{Z}^{(t)}$. Then, assume that the endogeneity of weight matrices $W_k^{(t)}$s is due to the correlation between $z_i^{(t)\top}$ and $\epsilon_{it}$, where
$\epsilon_{it}$ is the $i$-th element of the error term $\epsilon_i$. When the weight matrices are endogenous, the
QMLE calculated from exogenous weight matrices could be biased, as noticed by Qu and Lee (2015).
Hence, we modify the approaches used in Qu and Lee (2015) and
Qu et al. (2017) to accommodate endogenous weight matrices. To model the correlation structures between $z_i^{(t)}$ and $\epsilon_{it}$, we assume the following condition,
which is directly borrowed from Qu and Lee (2015) and
Qu et al. (2017).
\begin{itemize}
\item [(E1)] Assume that $(z_i^{(t)\top},\epsilon_{it})^\top$ is independent and identically distributed with mean 0 and covariance matrix $\Sigma_{z\epsilon}$ for any $i$ and
$t$, where $\Sigma_{z\epsilon}=(\Sigma_z, \sigma_{z\epsilon}^\top;\sigma_{z\epsilon}, \sigma^2)$.
\end{itemize}

\noindent
We further assume that there exist a finite constant vector $\delta\in\mR^{d}$ and a positive constant
$\sigma_z$ such that $E(\epsilon_{it}|z_i^{(t)})=z_i^{(t)\top}\delta$ and $\var(\epsilon_{it}|z_i^{(t)})=\sigma_z^2$.

Denote $\upsilon_{it}=\epsilon_{it}-z_i^{(t)\top}\delta$ and $\upsilon_t=(\upsilon_{1t}, \cdots, \upsilon_{nt})^\top$. Then, by Condition (E1), we have
$\delta=\Sigma_z^{-1}\sigma_{z\epsilon}$ and $\sigma_z^2=\sigma^2-\sigma_{z\epsilon}^\top \Sigma_z^{-1}\sigma_{z\epsilon}$.
Accordingly, $E(\upsilon_{t}|\mathbf{Z}^{(t)})=0$ and
$\mbox{cov}(\upsilon_{t}|\mathbf{Z}^{(t)})=\sigma_z^2 I_n$.
In addition,
$\upsilon_{t}$ is uncorrelated with the attributes $\mathbf{Z}^{(t)}$. Plugging
$\upsilon_{it}$ into model (2.2), we then have
\beq Y_t=(\lambda_1 W_1^{(t)}+\cdots+\lambda_d W_d^{(t)})Y_t+\mathbf{Z}^{(t)}\delta+\upsilon_{t}.\eeq
Note that the noise term $\upsilon_{t}$ in (4.1) is uncorrelated with the weight matrices $W_k^{(t)}$s for $k=1, \cdots, d$. This is due to
incorporating the variables $\mathbf{Z}^{(t)}$ in order to control for possible endogeneity.

We next employ the quasi-maximum likelihood estimation method to estimate model (4.1). Define
$\theta_z=(\lambda^\top, \sigma_z^2, \delta^\top, \mbox{vec}^*(\Sigma_z)^\top)^\top$, where $\mbox{vec}^*(\Sigma_z)$
means the unique values of the vectorization of $\Sigma_z$.
Then the quasi-loglikelihood function of model (4.1) is
\begin{align}
{\ell_z(\theta_z)}=& -\frac{nT}{2}\log (2\pi)-\frac{nT}{2}\log(\sigma_z^2)+\sum_{t=1}^T\log|\mbox{det}(\Delta_t(\lambda))|\notag-\frac{nT}{2}\log|\Sigma_z|\\
& -\frac{1}{2}\sum_{t=1}^T tr\big\{\mathbf{Z}^{(t)\top}\Sigma_z^{-1}\mathbf{Z}^{(t)}\big\}-\frac{1}{2\sigma_z^2}\sum_{t=1}^T \big\{\Delta_t(\lambda)Y_t-\mathbf{Z}^{(t)}\delta\big\}^\top\big\{\Delta_t(\lambda)Y_t-\mathbf{Z}^{(t)}\delta\big\}.
\end{align}
Given $\lambda^\top$, $\sigma_z^2$ and $\delta^\top$,
the estimator of $\Sigma_z$ can be obtained by optimizing (4.2), and denote it $\hat\Sigma_z=(nT)^{-1}\sum_{t=1}^T \mathbf{Z}^{(t)}\mathbf{Z}^{(t)\top}$.
Plugging this estimator into  (4.2) and ignoring some irrelevant constants, the resulting quasi-concentrated log-likelihood
function is
\[\ell_{zc}(\theta_{zc})=-\frac{nT}{2}\log(\sigma_z^2)+\sum_{t=1}^T\log|\mbox{det}(\Delta_t(\lambda))|\notag-\frac{1}{2\sigma_z^2}\sum_{t=1}^T \big\{\Delta_t(\lambda)Y_t-\mathbf{Z}^{(t)}\delta\big\}^\top\big\{\Delta_t(\lambda)Y_t-\mathbf{Z}^{(t)}\delta\big\},\]
where $\theta_{zc}=(\lambda^\top, \sigma_z^2, \delta^\top)^\top\in\mR^{2d+1}$.
Accordingly, we obtain the quasi-maximum likelihood estimator of
$\theta_{zc}$, which  is $\hat\theta_{zc}=\argmax_{\theta_{zc}\in \Theta_{zc}} \ell_{zc}(\theta_{zc})$,
where $\Theta_{zc}=\big\{\theta_{zc}=(\lambda^\top, \sigma_z^2, \delta^\top)^\top: \lambda\in\Lambda, \sigma_z^2>0 \mbox{~and~} \delta\in\mR^d\big\}$ is the parameter space.
Based on
$\delta=\Sigma_z^{-1}\sigma_{z\epsilon}$ and $\sigma_z^2=\sigma^2-\sigma_{z\epsilon}^\top \Sigma_z^{-1}\sigma_{z\epsilon}$, we subsequently obtain the estimators  $\hat\sigma_{z\epsilon}$ and $\hat \sigma^2$.
Hereafter, we name the resulting estimator as the endogeneity-adjusted QMLE (EA-QMLE).

To establish the asymptotic properties of $\hat\theta_{zc}$, we evaluate the first derivatives of $\ell_{zc}(\theta_{zc})$ as follows,
\[\frac{\partial \ell_{zc}(\theta_{zc})}{\partial \lambda_k}=\sum_{t=1}^T \Big[\frac{1}{\sigma_z^2}Y_t^\top W_k^{(t)\top} \big\{\Delta_t(\lambda)Y_t-\mathbf{Z}^{(t)}\delta\big\}-tr\big\{\Delta^{-1}_t(\lambda)W_k^{(t)}\big\}\Big],\]
\[\frac{\partial \ell_{zc}(\theta_{zc})}{\partial \delta}=\frac{1}{\sigma_z^2}\sum_{t=1}^T \mathbf{Z}^{(t)\top}\big\{\Delta_t(\lambda)Y_t-\mathbf{Z}^{(t)}\delta\big\}, \mbox{~and~}\]
\[\frac{\partial \ell_{zc}(\theta_{zc})}{\partial \sigma_z^2}=-\frac{nT}{2\sigma_z^2}+\frac{1}{2\sigma_z^4} \sum_{t=1}^T \big\{\Delta_t(\lambda)Y_t-\mathbf{Z}^{(t)}\delta\big\}^\top\big\{\Delta_t(\lambda)Y_t-\mathbf{Z}^{(t)}\delta\big\}.\]
Under Condition (E1), one can verify that $E\big\{\frac{\partial \ell_{zc}(\theta_{0zc})}{\partial \theta_{zc}}\big\}=0$. This indicates that
the first-order condition is satisfied.

Denote $\theta_{0zc}=(\lambda_0^{\top},\sigma_{0z}^2,\delta_0^{\top})^{\top}$ the true parameter vector.
In addition, for $k=1,\cdots,2d+1$ and $l=1,\cdots,2d+1$, define
$\mathcal{I}_{znT}(\theta_{0zc})=-(nT)^{-1}E\{\frac{\partial^2 \ell_{zc}(\theta_{0zc})}{\partial \theta_{zc}\partial\theta_{zc}^{\top}}\}\triangleq (\mathcal{I}_{znT, kl})\in\mR^{(2d+1)\times (2d+1)}$
and $\mathcal{J}_{znT}(\theta_{0zc})=(nT)^{-1}\var(\frac{\partial \ell_{zc}(\theta_{0zc})}{\partial \theta_{zc}})
\triangleq(\mathcal{J}_{znT, kl})\in\mR^{(2d+1)\times (2d+1)}$.
Then, the asymptotic properties of $\hat\theta_{zc}$ and $\hat{B}_t$ are given in the following two theorems.
\begin{theorem}
Under Conditions (E1)  and Conditions (C3) and (E2)--(E5) in Appendix A, as $n\rightarrow\infty$ and $d^2/n\rightarrow 0$, $\sqrt{nT/d} D_z \mathcal{I}_{zc}(\theta_{0zc})(\hat\theta_{zc}-\theta_{0zc})$ is asymptotically normal with mean 0 and covariance
matrix $G_z(\theta_{0zc})$, where $D_z$ is an arbitrary $M\times(2d+1)$ matrix with $M<\infty$ satisfying  $\|D_z\|<\infty$
and  $d^{-1}D_z\mathcal{J}_{zc}(\theta_{0zc})D_z^{\top}\to G_z(\theta_{0zc})$, and $\mathcal{I}_{zc}(\theta_{0zc})$ and $\mathcal{J}_{zc}(\theta_{0zc})$ are defined in Condition (E5).
\end{theorem}
\begin{theorem}
Under the same conditions as in Theorem 5, we have that~ $\sup_{t\leq T}\|\hat B_t-B_t\|=O_p\{d(nT)^{-1/2}\}$.
\end{theorem}

Note that Theorem 5 holds as $n$ tends to infinity, whether $T$ is fixed or divergent,
and this is also applicable when $d$ is divergent.
In practice, $\mathcal{I}_{zc}(\theta_{0zc})$ and $\mathcal{J}_{zc}(\theta_{0zc})$ can be consistently
estimated by $\mathcal{I}_{znT}(\hat\theta_{zc}) $ and $\mathcal{J}_{znT}(\hat\theta_{zc})$, respectively.

\noindent\textbf{Remark 1:}
It is worth noting that our model has multiple weight matrices, and hence,
we cannot apply the approach of Qu and Lee (2015) and
Qu et al. (2017) to show the theoretical property of parameter estimators.
This is because they adopted the near-epoch dependence assumption
from Jenish and Prucha (2012) and assumed the number of weight matrices, induced by the total
``distance" of attributes $\mathbf{Z}^{(t)}$, to be $d=1$. However, when $d>1$,
the near-epoch dependence assumption cannot be applied directly.
The main reason is that, for any two nodes
$j_1,j_2=1, \cdots, n$,  the magnitude of total ``distance" between their attributes
 $\sum_{k=1}^d{(Z_{k{j_1}}^{(t)}-Z_{k{j_2}}^{(t)})^2}$ does not imply the magnitude of the ``distance" between  each of their individual attributes $(Z_{k{j_1}}^{(t)}-Z_{k{j_2}}^{(t)})^2$,
where $k=1,\cdots, d$.
Therefore, we propose a novel approach by introducing a general linear-quadratic form with random weight matrices and then employing the
martingale central limit theorem to demonstrate the theoretical property of parameter estimators in Theorem 5.

\csection{Mutual Influence Regression with Exogenous Covariates, Individual, and Time Fixed Effects}

In this section, we extend MIR to accommodate the external determinants of the responses by adding exogenous covariates and  individual and time fixed effects.
As a result, we have considered four
models in Sections 5.1--5.4 and eight theorems, Theorems 7--14. To save space,
we only present the proof of Theorem 11, and it is in the supplementary material.
This is because the proofs of Theorems 7, 9 and 13 are similar to that of Theorem 11 by applying Lemmas S.3, S.4 and S.7, respectively, in the supplementary material.
In addition, the proofs of Theorems 8, 10, 12 and 14 are similar to that of Theorem 2.

\csubsection{Mutual Influence Regression with Exogenous Covariates}

To extend MIR augmented with exogenous covariates, we introduce the following model,
\beq
Y_t=(\lambda_1W_1^{(t)}+\cdots+\lambda_dW_d^{(t)})Y_t+X_{t1}\beta_1+\cdots+X_{tp}\beta_p+\epsilon_t,
\eeq
where $X_{tj}\in \mR^{n}$ is the $j$-th covariate of the $n$ actors observed at time $t$,
$\beta_1,\cdots,\beta_p$ are unknown regression coefficients, and $p$ is fixed.
Denote $X_t=(X_{t1},\cdots,X_{tp})\in\mR^{n\times p}$ and $\beta=(\beta_1,\cdots,\beta_p)^{\top}\in\mR^p$.
 The resulting quasi-loglikelihood function is
\begin{align}
{\ell_e(\theta_e)}=& -\frac{nT}{2}\log (2\pi)-\frac{nT}{2}\log(\sigma^2)+\sum_{t=1}^T\log|\mbox{det}(\Delta_t(\lambda))|\notag\\
& -\frac{1}{2\sigma^2}\sum_{t=1}^T \big\{\Delta_t(\lambda)Y_t-X_t\beta\big\}^\top\big\{\Delta_t(\lambda)Y_t-X_t\beta\big\},\label{eq IMRC}
\end{align}
where $\theta_e=(\lambda^\top,\sigma^2,\beta^{\top})^\top\in\mR^{p+d+1}$ and $\Delta_t(\lambda)=I_n-(\lambda_1W_1^{(t)}+\cdots+\lambda_dW_d^{(t)})$ as defined in Section 2.1.
For given $\lambda$, we can estimate $\beta$ and $\sigma^2$ by
\begin{align*}
\hat{\beta}(\lambda) &=\big(\sum_{t=1}^{T}X_t^\top X_t\big)^{-1}\sum_{t=1}^{T}X_t^{\top}\Delta_t(\lambda)Y_t \mbox{~and~}\\
\hat{\sigma}^2(\lambda)&=\frac{1}{nT}\sum_{t=1}^T \big\{\Delta_t(\lambda)Y_t-X_t\hat{\beta}(\lambda)\big\}^\top\big\{\Delta_t(\lambda)Y_t-X_t \hat{\beta}(\lambda)\big\},
\end{align*}
respectively.
Plugging them into (\ref{eq IMRC}), the resulting quasi-concentrated log-likelihood function is
$$
\ell_{ec}(\lambda)=-\frac{nT}{2}\log (2\pi)-\frac{nT}{2}-\frac{nT}{2}\log\big\{\hat\sigma^2(\lambda)\big\}+\sum_{t=1}^T\log|\mbox{det}(\Delta_t(\lambda))|.
$$
Then the  quasi-maximum likelihood estimator of $\lambda$ is $\hat{\lambda}=\mbox{argmax}_{\lambda\in \Lambda}{\ell_{ec}(\lambda)}$. Subsequently, $\beta$ and $\sigma^2$
can be correspondingly estimated by $\hat{\beta}=\hat{\beta}(\hat{\lambda})$ and $\hat{\sigma}^2=\hat{\sigma}^2(\hat\lambda)$, which leads to ${\hat{\theta}_e}=(\hat{\lambda}^{\top},\hat{\sigma}^2,\hat{\beta}^{\top})^{\top}$. Subsequently, $\hat{B}_t$ is obtained  by evaluating ${B}_t$ at ${\hat{\theta}_e}$.

Denote $\theta_{e0}=(\lambda_0^{\top},\sigma^2_0,\beta_0^{\top})^{\top}$  the true parameters.
The asymptotic properties of ${\hat{\theta}_e}$ and $\hat{B}_t$ are given in the following two theorems.
\begin{theorem}
Under Conditions (C1)--(C3) and (C5) in Appendix A and (C4-I) and (C8-I) in the supplementary material,
$\sqrt{nT/d}D_1\mathcal{I}^e(\theta_{e0})(\hat\theta_{e}-\theta_{e0})$ is asymptotically normal with mean 0 and covariance matrix ${G_1(\theta_{e0})}$,
where $D_1$ is an arbitrary $M\times(d+1+p)$ matrix   satisfying $\|D_1\|<\infty$ and $d^{-1}D_1\mathcal{J}^e({\theta_{e0}})D_1^{\top}\to G_1(\theta_{e0})$,  $M<\infty$,
and $\mathcal{I}^e(\theta_{e0})$
and $\mathcal{J}^e(\theta_{e0})$ are defined in Condition (C4-I) in the supplementary material.
\end{theorem}
\begin{theorem}
Under Conditions (C1)--(C3) and (C5) in Appendix A and (C4-I) and (C8-I) in the supplementary material, we have that $\sup_{t\leq T}\|\hat B_t-B_t\|=O_p\{d(nT)^{-1/2}\}$.
\end{theorem}
\noindent
Theorems 7 and 8 are counterparts of Theorems 1 and 2, respectively, and they indicate that $\hat{\theta}_e$ is consistent and asymptotically normal and $\hat{B}_t$ is consistent uniformly.

\csubsection{Mutual Influence Regression with Exogenous Covariates and Interaction of Covariates with Weight Matrices}

In the previous section, we allowed the response $Y_{it}$ of node $i$ to be affected by its associated covariates $X_{it}=(X_{it1},\cdots,X_{itp})^{\top}$. However, the covariate of node $j$ for $j\not=i$ may also influence $Y_{it}$, and this influence effect depends on the ``distance" between nodes $i$ and $j$.
Thus, if two nodes $i$ and $j$ are close to each other, then we expect that the influence effect of the covariate of node $j$ on $Y_{it}$ is relatively large,
otherwise it is relatively small.
In this paper, we assume that
the ``distance" between nodes $i$ and $j$ induced by covariates $Z_k^{(t)}$ can be measured  by the weight matrices $W_k^{(t)}$ for $k=1, \cdots, d$.
Accordingly,
we introduce the following model
\begin{align}
Y_t= &(\lambda_1W_1^{(t)}+\cdots+\lambda_dW_d^{(t)})Y_t+(\beta_{10}I_n+\beta_{11}W_1^{(t)}+\cdots+\beta_{1d}W_d^{(t)})X_{t1}\notag\\
        &+\cdots+(\beta_{p0}I_n+\beta_{p1}W_1^{(t)}+\cdots+\beta_{pd}W_d^{(t)})X_{tp}+\epsilon_t,\label{equ IT}
\end{align}
where $\tilde{\beta}=(\beta_{10},\cdots,\beta_{p0},\beta_{11},\cdots,\beta_{pd})^{\top}\in\mathbb{R}^{p(d+1)}$ is the unknown vector of regression coefficients that describe the interaction effects between the
covariates $X_t$ and the
weight matrices $W_k^{(t)}$ for $k=1, \cdots, d$.
Define
$\tilde{X}^{(t)}=[I_n,W_1^{(t)},\cdots,W_d^{(t)}]\times [I_{d+1}\otimes X_t].$
Then (\ref{equ IT}) can be expressed as
$$
Y_t=(\lambda_1W_1^{(t)}+\cdots+\lambda_dW_d^{(t)})Y_t+\tilde{X}^{(t)}\tilde{\beta}+\epsilon_t.
$$
The resulting quasi-loglikelihood function is
\begin{align}
\ell_s(\theta_s)=& -\frac{nT}{2}\log (2\pi)-\frac{nT}{2}\log(\sigma^2)+\sum_{t=1}^T\log|\mbox{det}(\Delta_t(\lambda))|\notag\\
& -\frac{1}{2\sigma^2}\sum_{t=1}^T \big\{\Delta_t(\lambda)Y_t-\tilde{X}^{(t)}\tilde{\beta}\big\}^\top\big\{\Delta_t(\lambda)Y_t-\tilde{X}^{(t)}\tilde{\beta}\big\},\notag
\end{align}
where
$\theta_{s}=(\lambda^{\top},\sigma^2,\tilde{\beta}^{\top})^{\top}\in\mR^{(p+1)(d+1)}$.
Following the same approach as in Section 5.1, we can obtain
 the quasi-maximum likelihood
estimator  $\hat{\theta}_s$. In addition, $\hat{B}_t$ can  obtained  by evaluating ${B}_t$ at ${\hat{\theta}_s}$.

Denote the true parameter $\theta_{s0}=(\lambda_0^{\top},\sigma_0^{2},\tilde{\beta}_0^{\top})^{\top}$.
The asymptotic properties of $\hat{\theta}_s$ and $\hat{B}_t$ are given in the following theorems.
\begin{theorem}
Under Conditions (C1)--(C3) and (C5) in Appendix A and (C4-II) and (C8-II) in the supplementary material, $\sqrt{nT/d}D_2\mathcal{I}^s(\theta_{s0})(\hat\theta_{s}-{\theta_{s0}})$ is
asymptotically normal with mean 0 and covariance matrix $G_2(\theta_{s0})$, where  $D_2$ is an arbitrary $M\times\{(d+1)(p+1)\}$ matrix
satisfying $\|D_2\|<\infty$ and $d^{-1}D_2\mathcal{J}^s(\theta_{s0})D_2^{\top}\to G_2(\theta_{s0})$, $M<\infty$, and
$\mathcal{I}^s(\theta_{s0})$ and $\mathcal{J}^s(\theta_{s0})$ are defined in Condition (C4-II) in the supplementary material.
\end{theorem}
\begin{theorem}
Under Conditions (C1)--(C3) and (C5) in Appendix A and (C4-II) and (C8-II) in the supplementary material, we have that $\sup_{t\leq T}\|\hat B_t-B_t\|=O_p\{d(nT)^{-1/2}\}$.
\end{theorem}
\noindent
Theorems 9 and 10 are counterparts of Theorems 1 and 2, respectively, and they indicate that $\hat{\theta}_s$ is consistent and asymptotically normal and $\hat{B}_t$ is consistent uniformly.

\csubsection{Mutual Influence Regression with {Individual} Effects}

In Section 5.1, we have extended MIR to include exogenous covariates. To further accommodate  individual fixed effects (see Lee and Yu, 2010a),
we propose
 the following model
\begin{equation}\label{equ imrf}
Y_t=(\lambda_1W_1^{(t)}+\cdots+\lambda_d W_d^{(t)})Y_t+X_{t1}\beta_1+\cdots+X_{tp}\beta_p+\omega+\epsilon_t,
\end{equation}
where $\omega$ is the $n\times 1$ vector of  individual fixed effects.
The resulting quasi-loglikelihood function of model (5.4) is
\begin{align}
\ell_{f,\omega}(\theta_f,\omega)=& -\frac{nT}{2}\log (2\pi)-\frac{nT}{2}\log(\sigma^2)+\sum_{t=1}^T\log|\mbox {det}(\Delta_t(\lambda))|\notag\\
& -\frac{1}{2\sigma^2}\sum_{t=1}^T \big\{\Delta_t(\lambda)Y_t-X_{t}\beta-\omega\big\}^\top\big\{\Delta_t(\lambda)Y_t-X_{t}\beta-\omega\big\},\label{equ imrf}
\end{align}
where $\theta_{f}=(\lambda^{\top},\sigma^2,\beta^{\top})^{\top}\in\mR^{d+p+1}$.
Given $\theta_{f}$, one can estimate $\omega$ by
$\hat{\omega}(\theta_{f})=T^{-1}\sum_{t=1}^{T}\big\{\Delta_{t}(\lambda)Y_t-X_{t}\beta\big\}$. Plugging this estimator into (\ref{equ imrf}), we obtain
\begin{align}
\ell_{f}(\theta_f)=& -\frac{nT}{2}\log (2\pi)-\frac{nT}{2}\log(\sigma^2)+\sum_{t=1}^T\log|\mbox{det}(\Delta_t(\lambda))|\notag\\
& -\frac{1}{2\sigma^2}\sum_{t=1}^T \big\{\Delta_t(\lambda)Y_t-X_t\beta-\hat{\omega}(\theta_f)\big\}^\top\big\{\Delta_t(\lambda)Y_t-X_t\beta-\hat{\omega}(\theta_f)\big\}.\notag
\end{align}
Following the same approach as in Section 5.1, we subsequently obtain
 the quasi-maximum likelihood
estimator  $\hat{\theta}_f$. Then $\hat{\omega}=\hat{\omega}(\hat{\theta}_f)$ and
$\hat{B}_t$ can be obtained
by evaluating ${B}_t$ at ${\hat{\theta}_f}$.

Denote the true parameters by $\theta_{f0}=(\lambda_0^{\top},\sigma^2_0,\beta_0^{\top})^{\top}$ and $\omega_0$. In addition, define $\theta_{fT}=\theta_{f0}-(0^{\top},T^{-1}\sigma^2_0,0^{\top})^{\top}$.
We then have the following results.
\begin{theorem}
Under Conditions (C1)--(C3) and (C5) in Appendix A and (C4-III) and (C8-I) in the supplementary material, we have that
(i) $\sqrt{nT/d}D_3\mathcal{I}^{f}(\theta_{f0})(\hat\theta_{f}-{\theta_{fT}})$ is asymptotically normal with mean 0 and covariance matrix $G_3(\theta_{f0})$, where  $D_3$ is an arbitrary  $M\times(d+p+1)$ matrix satisfying $\|D_3\|<\infty$ and $d^{-1}D_3\mathcal{J}^{f}(\theta_{f0})D_3^{\top}\to G_3(\theta_{f0})$,
$M<\infty$, and $\mathcal{I}^f(\theta_{f0})$ and $\mathcal{J}^f(\theta_{f0})$ are defined in Condition (C4-III) in the supplementary material; (ii) if $d=o(n^{1/2})$, then $\sqrt{T}(\hat{\omega}_{i}-\omega_{0i})$ is asymptotically normal with mean 0 and variance $\sigma^2_{0}$ for $i=1,\cdots,n$, and  it is also asymptotically independent from $\sqrt{T}(\hat\omega_{j}-\omega_{0j})$ for $j\ne i$.
\end{theorem}
\begin{theorem}
Conditions (C1)--(C3) and (C5) in Appendix A and (C4-III) and (C8-I) in the supplementary material, we have $\sup_{t\leq T}\|\hat B_t-B_t\|=O_p\{d(nT)^{-1/2}\}$.
\end{theorem}
\noindent Theorems 11 and 12 are counterparts of Theorems 1 and 2, respectively.
By Theorem 11 (i), we have that  $\hat{\theta}_f$  converges to $\theta_{f0}$  as $T\to \infty$ and $\hat{\theta}_f$ is asymptotically normal as $n/(dT)\to 0$.
When $T$ is fixed or $n/(dT)\not\rightarrow 0$, $\sqrt{nT/d}D_3\mathcal{I}^{f}(\theta_{f0})(\hat\theta_{f}-\theta_{f0})$ is not centered at 0 and therefore  bias-correction is needed.
We then construct the bias-corrected estimator as $\hat{\theta}_{f,ba}=(\hat\lambda_f^\top, \frac{T}{T-1}\hat\sigma_f^2, \hat\beta_f^\top)^\top$,
where $\hat{\theta}_f=(\hat\lambda_f^\top, \hat\sigma_f^2, \hat\beta_f^\top)^\top$. By Theorem 11, we immediately obtain that
$\sqrt{nT/d}D_3\mathcal{I}^{f}(\theta_{f0})(\hat\theta_{f,ba}-\theta_{f0})$ is asymptotically normal with mean 0 and covariance matrix $G_3(\theta_{f0})$.
The above results are consistent with the findings in Lee and Yu (2010a) for $d=1$.
Theorem 12 indicates that $\hat{B}_t$ is consistent uniformly.

\csubsection{Mutual Influence Regression with Both Individual and Time Effects}

To further accommodate both individual and time effects, we consider
 the following model
\begin{equation}\label{equ ite}
Y_t=(\lambda_1W_1^{(t)}+\cdots+\lambda_d W_d^{(t)})Y_t+X_{t1}\beta_1+\cdots+X_{tp}\beta_p+\omega+g_t
\mathbf{1}_n+\epsilon_t,
\end{equation}
where $\omega$ is the $n\times 1$ vector of individual fixed effect,
$g_t$ is the fixed effect for time $t$, and $\mathbf{1}_n=(1,\cdots,1)^{\top}$.

To estimate the unknown parameters, we follow the idea of Lee and Yu (2010b)
and eliminate the time effects  in model (5.6).
 Let $(F_{n,n-1},\mathbf{1}_n/\sqrt{n})$ be the eigenvector matrix of
 $J_n=I_n-n^{-1} \mathbf{1}_n\mathbf{1}_n^{\top}$, where $F_{n,n-1}$ is the $n\times (n-1)$ sub-matrix corresponding to the eigenvalues of 1s.
 Since $W_k^{(t)}$ has been row normalized for any $k=1, \cdots, d$, (\ref{equ ite}) can be transformed into
\begin{equation}\label{equ ite2}
Y_t^{*}=(\lambda_1W_1^{*(t)}+\cdots+\lambda_d W_d^{*(t)})Y_t^*+X_t^{*}\beta+\omega^*+\epsilon_t^{*},
\end{equation}
where $Y_t^{*}=F_{n,n-1}^{\top}Y_t$, $W_k^{*(t)}=F_{n,n-1}^{\top}W_k^{(t)}F_{n,n-1}$, $X_t^{*}=F_{n,n-1}^{\top}X_t$, $\omega^*=F_{n,n-1}^{\top}\omega$ and $\epsilon_t^{*}=F_{n,n-1}^{\top}\epsilon_t$. Denote $\theta_g=(\lambda^{\top},\sigma^2,\beta^{\top})^{\top}$. Following the same approach as in Section 5.3, we obtain
the concentrated quasi-loglikelihood function of model (\ref{equ ite2}), which is
\begin{align}
\ell_{g}(\theta_g)=& -\frac{(n-1)T}{2}\log (2\pi)-\frac{(n-1)T}{2}\log(\sigma^2)+\sum_{t=1}^T\log|\mbox{det}(\Delta_t^{*}(\lambda))|\notag\\
& -\frac{1}{2\sigma^2}\sum_{t=1}^T \big\{\Delta_t^{*}(\lambda)Y_t^{*}-X_t^{*}\beta-\hat{\omega}^{*}(\theta_g)\big\}^\top\big\{\Delta_t^{*}(\lambda)Y_t^{*}-X_t^{*}\beta-\hat{\omega}^{*}(\theta_g)\big\},\notag
\end{align}
where $\Delta_t^{*}(\lambda)=I_{n-1}-\sum_{k=1}^d \lambda_kW_k^{*(t)}$ and $\hat{\omega}^{*}(\theta_g)=T^{-1}\sum_{t=1}^{T}\big\{\Delta^{*}_{t}(\lambda)Y_t^{*}-X^{*}_{t}\beta\big\}$.
By maximizing $\ell_{g}(\theta_g)$, we obtain the quasi-maximum likelihood
estimator of $\theta_g$ and denote it as $\hat{\theta}_g$. Then, $\hat{B}_t$ can be obtained
by evaluating ${B}_t$ at ${\hat{\theta}_g}$.

Denote the true parameters by $\theta_{g0}=(\lambda_0^{\top},\sigma^2_0,\beta_0^{\top})^{\top}$, and $\theta_{gT}=\theta_{g0}-(0^{\top},T^{-1}\sigma^2_0,0^{\top})^{\top}$.
We then have the following results.
\begin{theorem}
Under Conditions (C1)--(C3) and (C5) in Appendix A and (C4-IV) and (C8-I) in the supplementary material, we have that
$\sqrt{nT/d}D_4\mathcal{I}^{g}(\theta_{g0})(\hat\theta_{g}-\theta_{gT})$ is asymptotically normal with mean 0 and covariance matrix $G_4(\theta_{g0})$, where  $D_4$ is an arbitrary  $M\times(d+p+1)$ matrix satisfying $\|D_4\|<\infty$ and $d^{-1}D_4\mathcal{J}^{g}(\theta_{g0})D_4^{\top}\to G_4(\theta_{f0})$,
$M<\infty$, and $\mathcal{I}^g(\theta_{g0})$ and $\mathcal{J}^g(\theta_{g0})$ are defined in Condition (C4-IV) of the supplementary material.
\end{theorem}
\begin{theorem}
Under Conditions (C1)--(C3) and (C5) in Appendix A and (C4-IV) and (C8-I) in the supplementary material, we have $\sup_{t\leq T}\|\hat B_t-B_t\|=O_p\{d(nT)^{-1/2}\}$.
\end{theorem}
\noindent According to Theorem 13, we have that  $\hat{\theta}_g$  converges to $\theta_{g0}$  as $T\to \infty$ and $\hat{\theta}_g$ is asymptotically normal as long as
$n/(dT)\to 0$.
When $T$ is fixed or $n/(dT)\not\rightarrow 0$, $\sqrt{nT/d}D_4\mathcal{I}^{g}(\theta_{g0})(\hat\theta_{g}-\theta_{g0})$ is not centered at 0 and therefore bias-correction is needed.
We construct the bias-corrected estimator as $\hat{\theta}_{g,ba}=(\hat\lambda_g^\top, \frac{T}{T-1}\hat\sigma_g^2, \hat\beta_g^\top)^\top$,
where $\hat{\theta}_g=(\hat\lambda_g^\top, \hat\sigma_g^2, \hat\beta_g^\top)^\top$. By Theorem 13, we immediately obtain that
$\sqrt{nT/d}D_4\mathcal{I}^{g}(\theta_{g0})(\hat\theta_{g,ba}-\theta_{g0})$ is asymptotically normal with mean 0 and covariance matrix $G_4(\theta_{g0})$.
Theorem 14 indicates that $\hat{B}_t$ is consistent uniformly.

\csection{Simulation Studies}

To demonstrate the finite sample performance of our proposed MIR model,
we conduct simulation studies with the following two settings for
exogenous and endogenous weight matrices, respectively.

\begin{center}
Setting I: Exogenous Weight Matrices
\end{center}

The similarity matrices
$A_k^{(t)}=(a(Z^{(t)}_{kj_1},Z^{(t)}_{kj_2}))\in\mR^{n\times n}$ with zero diagonal elements and $a(Z^{(t)}_{kj_1},Z^{(t)}_{kj_2})=\exp\{-(Z^{(t)}_{kj_1}-Z^{(t)}_{kj_2})^2\}$ if
$|Z^{(t)}_{kj_1}-Z^{(t)}_{kj_2}|<\phi^{(t)}_k$, $a(Z^{(t)}_{kj_1},Z^{(t)}_{kj_2})=0$ otherwise,
where $j_1$ and $j_2$ range from 1 to $n$ and $Z^{(t)}_k=(Z^{(t)}_{k1}, \cdots, Z^{(t)}_{kn})^\top$
are iid according
to a multivariate normal
distribution with mean 0 and covariance matrix $I_n$
for $k=1, \cdots, d$ and $t=1,\cdots,T$, and $\phi^{(t)}_k$ is selected to control the density of $A_k^{(t)}$ (i.e., the proportion of nonzero elements)
defined as $10/n$ for any $k$ and $t$ (see, e.g.,
Zou et al. 2017).
Accordingly, we obtain $W_k^{(t)}=(w(Z_{k{j_1}}^{(t)},Z_{k{j_2}}^{(t)}))_{n\times n}$
with $w(Z_{k{j_1}}^{(t)},Z_{k{j_2}}^{(t)})=a(Z_{k{j_1}}^{(t)},Z_{k{j_2}}^{(t)})/\sum_{j_2} a(Z_{k{j_1}}^{(t)},Z_{k{j_2}}^{(t)})$.
The random errors $\epsilon_{it}$ are iid and simulated from three distributions:
(i) the standard normal distribution $N(0,1)$; (ii) the mixture distribution
$0.9N(0,5/9)+0.1N(0,5)$; (iii) the standardized exponential distribution. The last two distributions allow us to examine the
robustness of parameter estimates to other distributions.
Finally, the response vectors $Y_t$ are generated by
$Y_t=(I_n-\lambda_1W_1^{(t)}-\cdots-\lambda_d W_d^{(t)})^{-1}\epsilon_t$ for $t=1,\cdots,T$.
Since the random error $\epsilon_t$ is independent of $Z^{(t)}_k$ for any $k=1,\cdots,d$
and $t=1, \cdots, T$, the weight matrices
$W_k^{(t)}$ are exogenous in this setting.

For each setting, we consider three different numbers of observations $T=25, 50$ and 100, three different
numbers of actors $n=25, 50$ and 100, and
all of the results are generated with 500 realizations.
Since the results for all three error distributions  are qualitatively similar, we only present the results for the standard normal distribution, and the results for the mixture normal and the standardized exponential
distributions are relegated to the supplementary material.

To assess the performance of parameter estimators, we consider three different numbers of covariates $d=2, 6$ and 12,
where $d=2$ is borrowed from Zou et al. (2017), $d=6$ is used in our real data analysis, and $d=12$ is an exploration of larger similarity matrices. Since the simulation results for $d=12$ are qualitatively similar to those for $d=2$ and 6, we report them in the
supplementary material. The regression coefficients are $\lambda_k=0.2$ for
 $k=1,\cdots, d$.
In addition,
let $\hat\lambda^{(m)}=(\hat\lambda_1^{(m)},\cdots,\hat\lambda_d^{(m)})^\top\in\mR^d$
be the parameter estimate in the $m$-th realization obtained via the proposed QMLE. For each
$k=1, \cdots, d$, we evaluate the average bias of $\hat\lambda_k^{(m)}$ by BIAS=$500^{-1}\sum_m (\hat\lambda_k^{(m)}-\lambda_k)$.
Using the results of Theorem 1, we compute the standard error of $\hat\lambda_k^{(m)}$ via its asymptotic distribution,  and denote it
$\mbox{SE}^{(m)}$. Then, the average of the estimated standard errors is SE=$500^{-1}\sum_m \mbox{SE}^{(m)}$.
To assess the validity of the estimated standard errors, we also calculate the true standard error via the 500 realizations and denote it
$\mbox{SE}^*=500^{-1}\sum_m (\hat\lambda_k^{(m)}-\bar\lambda_k)^2$, where $\bar\lambda_k=500^{-1}\sum_m \hat\lambda_k^{(m)}$.

Table 1 presents the results of BIAS, SE and $\mbox{SE}^*$ over 500 realizations for
 $k=1,\cdots, d$  and $d=2$ and 6.
It indicates that
the biases of the parameter estimates are close to 0 for any $n$ and $T$, and they become smaller as either $n$ or $T$ gets larger.
In addition, the variation of  the parameter estimate, SD, shows similar findings to those of BIAS.
Moreover, the difference between SD and $\mbox{SD}^*$ is quite small when either $n$ or $T$ is large. In sum, Table 1 demonstrates that the asymptotic results obtained in Theorem 1
are reliable and satisfactory.

We next assess the performance of the proposed EBIC criterion by considering three sizes of the full model, $d=6, 8$, and 12, while the size of the true model is
$|\mS_T|=3$. To implement the EBIC criterion, we set $\gamma=2$ in this simulation study.
Four performance measures are used: (i) the average size (AS) of the selected model
$|\hat\mS|$; (ii) the
average percentage of the correct fit (CT), $I(\hat\mS=\mS_T)$; (iii) the average true positive rate (TPR), $|\hat\mS\cap\mS_T|/|\mS_T|$;
and (iv) the average false positive rate (FPR), $|\hat\mS\cap\mS_T^c|/|\mS_T^c|$.
Since the results for all three values  of $d$ exhibit a quantitatively similar pattern, we only present the results for $d=8$.

Table 2 shows that the average percentage of
correct fit, CT, increases toward to 100\% when either $n$ or $T$ gets large. It is worth noting that the CTs are larger than 70\% even when both $n$ and $T$ are small, i.e., $n=25$ and $T=25$.
Furthermore, the average true positive
rate, TPR, is 100\%, which indicates that EBIC is unlikely to select an underfitted model even when both $n$ and $T$ are small.
In contrast, the average false positive rate, FPR, decreases toward 0 when either $n$ or $T$ becomes large.
Moreover, the average size (AS) of the
selected model, $|\hat\mS|$, approaches the true model size. The above results indicate that
EBIC performs satisfactorily in finite samples.

Lastly, we examine the performance of the proposed goodness of fit test. We consider a generative model $B_t=\lambda_1 W^{(t)}_1+\cdots+\lambda_d W^{(t)}_d+\kappa E E^\top$, where
$E\in\mR^n$ is a random normal vector of dimension $n$ with each elements that are iid simulated from a standard normal distribution.
The parameter $\kappa$ is a measure of departure from the null model of $H_0$. Specifically, $\kappa=0$ corresponds to the null model, while
$\kappa>0$ represents alternative models. Accordingly, the results for $\kappa=0$ represent empirical sizes, while
the results for $\kappa>0$ denote empirical powers.

Table 3 indicates
that the empirical sizes are slightly conservative when both $n$ and $T$ are small. However, they approach the significance level of 5\%
when either $n$ or $T$ becomes large.
Furthermore, the empirical powers increase as either
$n$ or $T$ gets larger. Moreover, they become stronger when $\kappa$ increases; in particular
the empirical power approaches 1 when either
$n$ or $T$ equals 100 and $\kappa=2$. The above findings are robust to non-normal error distributions; see Tables S.5 and S.8 in the
supplementary material. Consequently, our proposed goodness of fit test not only controls the size well, but is also consistent.
It is worth noting that the above estimation, selection and test findings are also robust to non-normal error distributions; see Tables S.3 to S.8 in the
supplementary material. For the sake of illustration, we finally conduct simulation studies for the extended models in Section 5.
Since models (5.3), (5.4) and (5.6)
can be re-expressed as a similar form of model (5.1), we only present the simulation results of model (5.1). To save space, the detailed simulation settings and
results are given in Section S.5 and Table S.9 of the supplementary material.

\begin{center}
Setting II: Endogenous Weight Matrices
\end{center}

In this setting, for any $i=1, \cdots, n$ and $t=1, \cdots, T$,
$(Z^{(t)}_{1i}, \cdots, Z^{(t)}_{di}, \epsilon_{it})^\top$ are iid
according to a multivariate normal distribution with mean 0 and covariance matrix
$(1-\rho) I_{d+1}+\rho \1\1^\top$, where $\rho=0.5$.
Since $\rho\not=0$, $\epsilon_t$ is correlated with the attributes $Z^{(t)}_{k}$ for any $k$,
and the weight matrices are endogenous.
The rest of the setting is the same as in Setting I.
The results for QMLE and EA-QMLE are presented in Table 4 with $d=6$
(the results for $d=2$ and 12 are quantitatively similar, so we have omitted them to save space).
Table 4 indicates that QMLE yields larger biases that
do not tend to 0 as $T$ increases.
In contrast, EA-QMLE performs well. The resulting biases and variations of the parameter estimates are close to 0 for any $n$ and $T$, and they become smaller as either $n$ or $T$ gets larger.
In addition, the difference between SD and $\mbox{SD}^*$ is quite small when either $n$ or $T$ is large.
 In sum, Table 4 demonstrates that the asymptotic results obtained in Theorem 5
are reliable and satisfactory.

\csection{Real Data Analysis}

\csubsection{Background and Data}

To demonstrate the practical usage of our proposed MIR model, we present an empirical  example for exploring the mechanism of spillover effects in  Chinese mutual funds.
It is known that the income and profit of a mutual fund is largely compensated from the management fees, which are charged as a fixed proportion
of the total net assets under management. As a result, the variation in cash flow across time is one of the most
influential indices closely monitored by  fund managers.
Thus, exploring the mechanism of cash flow is extremely essential (see e.g., Spitz, 1970; Nanda et al., 2004; Brown and Wu, 2016).
However, past literatures mainly
focus on
addressing the characteristics of the mutual funds that affect their cash flow from a cross-sectional prospective (see, e.g., Brown and Wu, 2016).
In this study,  we employ our proposed MIR model to
identify the  mutual fund characteristics  that can yield mutual influence on fund cash flows (i.e., a  spillover effect)
from a  network perspective.

To proceed with our study, we collect quarterly data from 2010-2017 on actively managed open ended mutual funds
through the WIND financial database,
which is one of
the most authoritative databases regarding the Chinese financial market.
After removing
funds with missing observations or existing  for less than one year, there are $n=90$ mutual funds in this empirical
study with $T=32$.
The response variable, the cash flow rate of fund $i$ at time $t$, can be calculated as follows
(Nanda et al., 2004):
\[C_{it}=\frac{TA_{it}-TA_{i,t-1}(1+r_{it})}{TA_{it}},\]
where $TA_{it}$ and  $r_{it}$ are the total net assets and the return of fund $i$ at time $t$, respectively.

We next generate the similarity matrices to explore the mechanism of spillover effects among mutual funds. To this end, we consider the following five
covariates
in the spirit of the pioneering work of Spitz (1970). To avoid endogeneity problems, all covariates were evaluated at time $t-1$.
(i). Size: the logarithm of the
total net asset of fund $i$ at time $t-1$; (ii) Age: the logarithm of the age of fund $i$ at time $t-1$;
(iii) Return; the return of fund $i$ at time $t-1$; (iv) Alpha: the risk-adjusted return of fund $i$ at time $t-1$ measured by the intercept of Carhart's (1997) four
factor model; (v) Volatility: the standard deviation of the weekly return of fund $i$ and time $t-1$.
We next generate the similarity matrices.
For the Size covariate,
we standardize the data to have zero mean and unit variance, and denote it $\mbox{SIZE}_{it}$ for $i=1,\cdots,n$ and $t=1,\cdots,T$.
Then, the similarity matrix induced by Size is $A_1^{(t)}=(a(Z_{1{j_1}}^{(t)},Z_{1{j_2}}^{(t)}))$ with zero diagonal elements and
$a(Z_{1{j_1}}^{(t)},Z_{1{j_2}}^{(t)})=\exp\{-(Z_{{1j_1}}^{(t)}-Z_{{1j_2}}^{(t)})^2\}$ when $|Z_{{1j_1}}^{(t)}-Z_{{1j_2}}^{(t)}|<\phi^{(t)}_1$ for a pre-specified
finite positive constant $\phi^{(t)}_1$, and $a(Z_{1{j_1}}^{(t)},Z_{1{j_2}}^{(t)})=0$ otherwise. As given in simulation studies,
$\phi^{(t)}_1$ is selected so that the proportion of nonzero elements of $A_1^{(t)}$
is $10/n$.
Subsequently, we obtain $W_1^{(t)}=(w(Z_{1{j_1}}^{(t)},Z_{1{j_2}}^{(t)}))_{n\times n}$
and $w(Z_{1{j_1}}^{(t)},Z_{1{j_2}}^{(t)})=a(Z_{1{j_1}}^{(t)},Z_{1{j_2}}^{(t)})/\sum_{j_2} a(Z_{1{j_1}}^{(t)},Z_{1{j_2}}^{(t)})$, which is
the row-normalized version of $A_1^{(t)}$.
Analogously, we can construct the similarity matrices $W_2^{(t)},\cdots, W_5^{(t)}$ associated with the remaining four covariates, respectively.

\csubsection{Empirical Results}

We first employ the adequacy test to
assess whether the five covariates are sufficient to explain the mutual influence matrix.
The resulting $p$-value for testing the null hypothesis of $H_0$ in (3.1) is 0.660, which is not significant under the significance level of
5\%. This indicates that one or more of the five covariates in the MIR model provide a good fit to the data.

We next employ the proposed QMLE method to estimate the model.
Table 5 presents the parameter estimates, standard errors, and their associated $p$-values. It indicates
that the covariates Return, Age and Volatility are significant and positive.
It is worth noting that these three covariates  are all related to the funds' performance and operating capacity.
Hence, we conclude that the funds' cash flows are influenced by other funds with similar performance and operating capacity.
Furthermore, the estimate of Size is positive and significant, which implies that the funds' cash flows  are influenced by other funds of similar size.
In other words, investors tend to invest in larger mutual funds.
Moreover, the estimate of Alpha is positive but not significant. Hence, investors
pay more attention to raw returns than risk-adjusted returns in judging a fund's performance. This can be due to the fact that raw returns are easier to  observe.

Subsequently, we employ EBIC to determine the most relevant covariates that are related to the cash flow with $\gamma=2$ as in the simulation studies.
The resulting model consists of the covariates Return and Size. This implies that
 fund managers tend to learn relevant information
from other  funds with a large size and good performance.
This finding is consistent
with existing studies (see, e.g., Brown et al., 1996). To check the robustness of our results against the selection of $\phi^{(t)}_k$, we also consider
 $\phi^{(t)}_k=5/n$ and $20/n$.
The results yield
similar findings to that
of $10/n$.
Finally, we take into account endogeneity in the construction of weight matrices.
The results of EA-QMLE in Table 5 show a qualitatively similar pattern to those of
QMLE. That is, except for the covariate Alpha, the other four covariates are all positive and significant.
This indicates that funds with similar return, size, age and volatility tend to influence each other's cash flow rate.
Hence, both QMLE and EA-QMLE lead to the same conclusion.
In sum, the MIR model can provide valuable insight for understanding the mechanism of mutual influence among
mutual funds.

\csection{Conclusion}

In this article, we propose
the mutual influence regression (MIR) model to explore the mechanism of
mutual influence by establishing a relationship between
the mutual influence matrix and a set of similarity matrices induced by their associated attributes among the actors.
In addition, we allow the number of similarity matrices to diverge.
The theoretical properties of the MIR model's estimations, selections, and assessments are established.
To expand the usefulness of MIR, we further extend the model to accommodate endogenous weight matrices, exogenous covariates, and both individual and time
fixed effects.
The Monte Carlo studies support the theoretical findings, and an empirical example illustrates the practical application.

To broaden the usefulness of MIR, we identify four possible
avenues for future research.
The first avenue is to allow the regression coefficients to change with
$t$ that increases model flexibility.
The second avenue is to generalize the model by accommodating discrete responses.
The third avenue is to extend the linear regression structure of MIR to the nonparametric
or semiparametric setting by changing $\lambda_k W^{(t)}_k$ to $g(\lambda_k, W^{(t)}_k)$ for some unknown smooth function $g(\cdot)$.
The last avenue is to develop a fast algorithm with theoretical justification that can implement MIR when $n$ is large.
We believe that
these efforts would further increase the application of the MIR model.

\scsection{Supplementary Material}

The supplementary material consists of five sections.
Section S.1 presents the proofs of Lemmas 3 and 4. Section S.2 gives the proofs of Theorems 3--5.
Section S.3 introduces technical conditions and lemmas that are used in proving the theoretical properties in Section 5.
Section S.4 presents the proof of Theorem 11.
Section S.5 presents five additional simulation results: (i)
the MIR model with twelve weight matrices; (ii) a comparison of the
extended BIC criterion (EBIC) with the Deviance information criterion (DIC) for the selection of weight matrices;
(iii) the MIR model with exponential errors; (iv) the MIR model with mixture errors;
 and (v) the MIR model with exogenous covariates.

%

\scsection{Appendix}

\renewcommand{\theequation}{A.\arabic{equation}}
\setcounter{equation}{0}

This Appendix includes three parts: Appendix A introduces eleven conditions and four useful Lemmas;
Appendices B--C provide the proofs of Theorems 1--2, respectively.

\scsubsection{Appendix A: Technical Conditions and  Useful Lemmas}

To study the asymptotic properties of parameter estimators and test statistics, we introduce
the following eleven technical conditions and four useful lemmas.
For the sake of convenience, we let $\varrho_{\min}(A)$ and $\varrho_{\max}(A)$
denote the smallest and largest eigenvalues of any arbitrary matrix $A$.
In addition, let $\|G\|=\{ \varrho_{\max}( G^\top G)\}^{1/2}$ be the $L_2$ (i.e., spectral) norm,
$\|G\|_F=\{\mbox{tr}( G^\top G)\}^{1/2}$ be the Frobenius norm,
and $\|G\|_R=\max_i \sum_{j}|G_{ij}|$ be the maximum absolute row-sum norm for any generic matrix $G=(G_{ij})$.
For any candidate model $\mS\subseteq \{1,\cdots, d\}$,
let $|\mS|$ be the size of $\mS$.

\begin{itemize}
\item[(C1)] Assume that $\epsilon_t=(\epsilon_{1t}, \cdots,\epsilon_{nt})^\top\in\mR^n$ is iid randomly generated with mean 0 and covariance matrix ${\sigma_0^2 I_n}$ for $t=1,\cdots,T$,
and assume that $\sup_{\psi\geq 1}\psi^{-1}\{E(|\epsilon_{it}|^{\psi})\}^{1/\psi}<\infty$ for
 $i=1,\cdots,n$ and $t=1,\cdots,T$. In addition, assume
\begin{equation*}
E(\epsilon_{i_{1}t}^{\vartheta_{1}}\epsilon_{i_{2}t}^{\vartheta _{2}}\cdots \epsilon_{i_{u}t}^{\vartheta
_{u}})=E(\epsilon_{i_{1}t}^{\vartheta_{1}})E(\epsilon_{i_{2}t}^{\vartheta_{2}})\cdots
E(\epsilon_{i_{u}t}^{\vartheta_{u}})
\end{equation*}
for a positive integer $u$ such that $\sum\nolimits_{k=1}^{u} \vartheta_{k}\leq
8$ and $i_{1}\neq i_{2}\neq \cdots \neq i_{u}$.
\item [(C2)] The parameter space of regression coefficients is
$\Lambda=\big\{\lambda=(\lambda_1,\cdots,\lambda_d)^{\top}:\sum_{k=1}^{d}|{\lambda}_k|<1-\varsigma,
\mbox{for some}\ \varsigma\in(0,1)\big\}$, and the true parameter is $\lambda_0\in \Lambda$.

\item[(C3)] For similarity matrices in $\big\{W_k^{(t)}\in \mathbb{R}^{n\times n}:k=1,\cdots, d\big\}$ and $\lambda$ in a small neighborhood
of $\lambda_0$, there exists $C_w>0$ such that
$\sup_{t\leq T}\sup_{n\geq 1}(\| W_k^{(t)}\|_R+\| W_k^{(t)\top}\|_R+\|\Delta_t^{-1}(\lambda)\|_R+\|\{\Delta_t^\top(\lambda)\}^{-1}\|_R)\leq C_w<\infty$.

\item[(C4)] Assume that
 $\mathcal{I}_{nT}(\theta_0)\to\mathcal{I}(\theta_0)$ in Frobenius norm and $\mathcal{J}_{nT}(\theta_0)\to\mathcal{J}(\theta_0)$ in $L_2$ norm as $nT\to\infty$,
where
$\mathcal{I}_{nT}(\theta_0)=-(nT)^{-1}E(\frac{\partial^2 \ell(\theta_0)}{\partial\theta\partial\theta^{\top}})$,
$\mathcal{J}_{nT}(\theta_0)=(nT)^{-1}\var(\frac{\partial \ell(\theta_0)}{\partial \theta})$,
 and $\mathcal{I}(\theta_0)$ and $\mathcal{J}(\theta_0)$ are
positive definite matrices.
In addition, assume that, for $\theta$ in a small neighborhood
of $\theta_0$, there exist two finite positive constants $c_{\min,1}$ and $c_{\min,2}$
such that
$0<c_{\min,1}< \varrho_{\min}(\mathcal{I}(\theta))\leq \varrho_{\max}(\mathcal{I}(\theta))=O(d)$ and $0<c_{\min,2}< \varrho_{\min}(\mathcal{J}(\theta))\leq \varrho_{\max}(\mathcal{J}(\theta))=O(d)$.

 \item[(C5)] Assume that $d=o\{(nT)^{1/4}\}$ as $nT\to \infty$, and $|\mS_{T}|$ is finite.

\item [(C6)] Assume that $\min_{k\in\mS_T}|\lambda_{0k}|\sqrt{nT/\log(nT)}\to\infty$ as $nT\to \infty$.

 \item[(C7)] There exist finite positive constants
${c_{\min,3}}$, ${c_{\max,3}}$, $\delta$ and $q$ such that, for sufficiently large $nT$,
$$
 c_{\min,3}<\varrho_{\min}\lsk\frac{1}{nT}\frac{\partial \ell(\tilde{\theta}_{\mS})}{\partial{\theta}_{\mS}\partial{\theta}_{\mS}^{\top}}\rsk\leq \varrho_{\max}\lsk\frac{1}{nT}\frac{\partial \ell(\tilde{\theta}_{\mS})}{\partial{\theta}_{\mS}\partial{\theta}_{\mS}^{\top}}\rsk< c_{\max,3},
 $$
where $\mS$ and $\tilde{\theta}_{\mS}$ satisfy $|\mS|\leq q$ for $q>|\mS_T|$  and $\|\tilde{\theta}_{\mS}-\theta_{0\mS}\|\leq \delta$.

\item[(E2)] For $t=1,\cdots, T$ and $i=1,\cdots,n$, define $E(\upsilon_{it}^m)=\zeta^{(m)}$ for $m>2$.
We assume that $\zeta^{(3)}=0$ and $\zeta^{(8)}<\infty$. In addition, for $i_1\neq i_2\neq i_3\neq i_4$,
 $E(\upsilon_{i_1t}^{e_1}\upsilon_{i_2t}^{e_2}\upsilon_{i_3t}^{e_3}\upsilon_{i_4t}^{e_4}|\mathbf{Z}^{(t)})=
 E(\upsilon_{i_1t}^{e_1}|\mathbf{Z}^{(t)})E(\upsilon_{i_2t}^{e_2}|\mathbf{Z}^{(t)})E(\upsilon_{i_3t}^{e_3}|\mathbf{Z}^{(t)})E(\upsilon_{i_4t}^{e_4}|\mathbf{Z}^{(t)})$
 for any non-negative constants $e_1,\cdots, e_4$ satisfying $e_1+\cdots+e_4\leq 8$.

\item [(E3)] Assume there exists a finite positive constant $C_{e_1}$ such that (i) $\sup_{t}\sup_{k} n^{-1} E\big(\|\mathcal{V}_k^{(t)}\|_s^s\big)<C_{e_1}$ and (ii)
$\sup_{t}\sup_{k} n^{-1}E\big(\|Z_k^{(t)}\|_s^s\big)<C_{e_1}$ for $s=2$ and 4.

\item [(E4)] Assume there exists a finite positive constant $C_{e_2}$ such that (i) $\sup_{t}\sup_{k,j} n^{-1}{\rm var}\big(\mathcal{V}_k^{(t)\top}\mathcal{V}_j^{(t)}\big)<C_{e_2}$; (ii) $\sup_{t}\sup_{k,j} n^{-1}{\rm var}\big(Z_k^{(t)\top}Z_j^{(t)}\big)<C_{e_2}$; (iii) $\sup_{t}\sup_{k,j} n^{-1}{\rm var}\big(\mathcal{V}_k^{(t)\top}Z_j^{(t)}\big)<C_{e_2}$
and (iv) $\sup_{t}\sup_{k,j} n^{-1}{\rm var}\big\{tr(\mathcal{U}_k^{(t)}\mathcal{U}_j^{(t)})\big\}<C_{e_2}$, where $\mathcal{U}_k^{(t)}=\frac{1}{2}\big[\Delta^{-1}_t(\lambda_0)W_k^{(t)}+W_k^{(t)\top}\{\Delta_t^{-1}(\lambda_0)\}^{\top}\big]$ and $\mathcal{V}_k^{(t)}=W_k^{(t)\top}\{\Delta_t^{-1}(\lambda_0)\}^{\top}\mathbf{Z}^{(t)}\delta_0$ were defined in Appendix F.

\item [(E5)] Assume that $\mathcal{I}^{zc}_{nT}(\theta_{0zc})\to\mathcal{I}^{zc}(\theta_{0zc})$ in
Frobenius norm and $\mathcal{J}^{zc}_{nT}(\theta_{0zc})\to\mathcal{J}^{zc}(\theta_{0zc})$ in $L_2$ norm as $nT\to\infty$, where $\mathcal{I}^{zc}(\theta_{0zc})$ and $\mathcal{J}^{zc}(\theta_{0zc})$ are
positive definite matrices.
In addition, assume that, for $\theta_{zc}$ in a small neighborhood
of $\theta_{0zc}$, there exist two finite positive constants $c_{z,1}$ and $c_{z,2}$
such that
$0<c_{z,1}< \varrho_{\min}(\mathcal{I}^{zc}(\theta_{zc}))\leq \varrho_{\max}(\mathcal{I}^{zc}(\theta_{zc}))=O(d)$ and $0<c_{z,2}< \varrho_{\min}(\mathcal{J}^{zc}(\theta_{zc}))\leq \varrho_{\max}(\mathcal{J}^{zc}(\theta_{zc}))=O(d)$.

\end{itemize}

The above conditions are mild and sensible. Condition (C1) is a moment condition,
which is much weaker than commonly used distribution assumptions; see, for example, the normal assumption in Zhou et al. (2017).
Condition (C2) specifies the parameter space of regression coefficients. A similar condition can be found in
Gupta and Robinson (2018). Condition (C3) has been carefully studied in Lee (2004) and Gupta and Robinson (2018).
Condition (C4) is used for showing
the asymptotic normality of QMLE  and it is a sufficient condition for local identification. A similar condition can be found in Gupta and Robinson (2015).
Under this condition, the quasi-loglikelihood function is strictly concave near $\theta_0$ and a local maximizer exists.
 Condition (C5) addresses the order of $d$ and the number of non-zero elements in $\lambda_0$.  The order condition of $d$ allows the number of weight matrices to diverge to
 infinity. Condition (C6) is a minimum signal assumption placed on the non-zero coefficients. Thus, if some of the nonzero coefficients
converge to zero too fast, they cannot be consistently identified. Similar conditions
have been  commonly used in extant literature such
 as Fan and Li (2001) and Chen and Chen (2012).
 Condition (C7) is used for showing the asymptotic property of EBIC.
 A similar condition can be found in Chen and Chen (2012).
Conditions (E2)--(E5) are useful in proving Theorem 5. Conditions (E1) and (E2) are moment conditions and are similar to Condition (C1).
Condition (E3) assures the variances of $\mathcal{V}_k^{(t)\top}Z_j^{(t)}$, $Z_k^{(t)\top}Z_j^{(t)}$, $\mathcal{V}_k^{(t)\top}Z_k^{(t)}$ and $tr\big\{\mathcal{U}_k^{(t)\top}\mathcal{U}_j^{(t)}\big\}$
are all of order $O(n)$ uniformly for any $k, j$ and $t$. Note that $\mathcal{V}_k^{(t)\top}Z_j^{(t)}=\sum_{l=1}^n \mathcal{V}_{kl}^{(t)}Z_{jl}^{(t)}$ (and $Z_k^{(t)\top}Z_j^{(t)}$, $\mathcal{V}_k^{(t)\top}Z_k^{(t)}$ and $tr\big\{\mathcal{U}_k^{(t)\top}\mathcal{U}_j^{(t)}\big\}$)
can be expressed as a summation of $n$ finite terms. Then, Condition (E3) is satisfied as long as the $n$ summation terms are weakly dependent.
This condition plays a similar role to the near-epoch dependence assumption used in Qu and Lee (2015).
Condition (E5) is a law of large numbers type of condition for the Hessian matrix and the variance of the score function. This condition is similar to Condition (C4).

To prove the theorems, we next introduce the following four useful lemmas. Since Lemma 1 below is directly modified from Theorem 1 of Kelejian and Prucha (2001)
and Lemma 2 is modified from Proposition A.1 of Chen et al. (2010), we only present
the proofs of Lemmas 3 and 4 in the supplementary material. It is worth mentioning that Lemma 4 is a general result for a combination of quadratic forms of random errors, and it plays a critical role
for obtaining the asymptotic variance of $T_{ql}$ (i.e., $\sigma^2_{ql}$) and proving the result of Theorem 4.

\begin{lemma} Let $\mathcal{E}=(\varepsilon_1, \cdots, \varepsilon_m)^\top$, where $\varepsilon_1,\cdots,\varepsilon_m$ are independent and identically distributed random variables
with mean 0 and finite variance $\sigma^2$.
Define
\[
Q_m=\mathcal{E}^\top B\mathcal{E}-\sigma^2{\rm tr}(B),
\]
where $B=(b_{ij})_{m\times m}\in\mathbb{R}^{m\times m}$.
Suppose the following assumptions are satisfied:

(1) for $i,j=1,\cdots,m$, $b_{ij}=b_{ji}$;

(2) $\|B\|_R<\infty$;

(3) there exists some $\eta>0$ such that $\rmE |\varepsilon_i|^{4+\eta}<\infty$.
\\Then, we have
$
 \rmE (Q_m)=0$ and
 \[
\sigma_{Q_m}^2:={\rm var}(Q_m)=4\sigma^4\sum_{i=1}^m\sum_{j=1}^{i-1}b_{ij}^2+\sum_{i=1}^m\lmk \lbk\mu^{(4)}\sigma^4-\sigma^4\rbk b_{ii}^2\rmk,
\]
where $\mu^{(k)}=E(\epsilon_i/\sigma)^{k}$.
Furthermore, suppose

(4)  $m^{-1}\sigma_{Q_m}^2\geq c_{q}$ for some finite $c_{q}>0$.
\\Then, we obtain
\[
\sigma_{Q_m}^{-1}{Q_m}\stackrel{d}{\longrightarrow} N(0,1).
\]

\end{lemma}

\begin{lemma}
Let $V=(V_1, \cdots, V_m)^\top\in\mR^m$ be a random vector distributed with mean 0 and covariance matrix $I_m$, and
\begin{equation*}
E(V_{k_{1}}^{\varsigma_{1}}V_{k_{2}}^{\varsigma _{2}}\cdots V_{k_{u}}^{\varsigma
_{u}})=E(V_{k_{1}}^{\varsigma _{1}})E(V_{k_{2}}^{\varsigma_{2}})\cdots
E(V_{k_{u}}^{\varsigma_{u}})
\end{equation*}
for a positive integer $u$ such that $\sum\nolimits_{k=1}^{u} \varsigma_{k}\leq
8$ and $k_{1}\neq k_{2}\neq \cdots \neq k_{u}$.
Then, for any arbitrary symmetric matrices $A_1$, $A_2$, $A_3$ and
$A_4$ of bounded eigenvalues,
we have
(i) $E\{(V^\top A_1 V)^2\}=tr^2(A_1)+2tr(A_1^2)+\bar\Delta tr(A_1^{\otimes2})$ with $\bar\Delta=E(V_i^4)-3$, where $A_1=(a_{j_1j_2})$ and $A_1^{\otimes
2}=(a^2_{j_1j_2})$; (ii) $E\{(V^\top A_1 V)(V^\top A_2 V)\}=tr(A_1)tr(A_2)+2tr(A_1A_2)+\bar\Delta tr(A_1\otimes A_2)$;
(iii)
there exists finite positive constant $C^*$
such that $E\big[\{V^\top A_1 V-tr(A_1)\}\{V^\top A_2 V-tr(A_2)\}\{V^\top A_3 V-tr(A_3)\}\{V^\top A_4 V-tr(A_4)\}\big]\leq C^* m^2$.
\end{lemma}

\begin{lemma}
Under Conditions (C1)-(C5) in Appendix A,   as $nT\to\infty$, we obtain the following results.
\begin{enumerate}
\item[(i)]
$$
(nTd)^{-1/2}D\frac{\partial \ell({\theta_0})}{\partial\theta}\stackrel{d}{\longrightarrow}N\lsk 0,G({\theta_0})\rsk,
$$
where $D$ is a $M\times(d+1)$ matrix  satisfying $\|D\|<\infty$ and $d^{-1}D\mathcal{J}(\theta_0)D^{\top}\to G(\theta_0)$, $M<\infty$, and $G(\theta_0)$ is a positive definite matrix.
\item[(ii)] For any $\nu>0$,
$$
P\lsk\frac{1}{\sqrt{nT}}\big|\frac{\partial {\ell(\theta_0)}}{\partial \theta_k}\big|>\nu\rsk\leq {2\exp\lbk-\min\lsk\frac{\tau_1\nu\sigma_0^2}{\|U_k\|/\sqrt{nT}},\frac{\tau_2\sigma_0^4\nu^2}{\|U_k\|_F^2/nT}\rsk\rbk,}
$$
for $k=1,\cdots, d+1$. $\tau_1$ and $\tau_2$ are two finite positive constants.
\item[(iii)]
$$
 \laak(nT)^{-1}\frac{\partial^2 {\ell(\theta_0)}}{\partial \theta\partial \theta^\top }+\mathcal{I}({\theta_0})\raak_F=o_p(1).
$$
\end{enumerate}
\end{lemma}


To state the following lemma, let $\epsilon_t=(\epsilon_{1t},\cdots,\epsilon_{nt})^\top\in\mR^n$ be independent and identically distributed
 random variables with mean 0 and covariance matrix $I_n$ for $t=1,\cdots, T$. In addition, let $\bar\Delta=E(\epsilon_{it}^4)-3$ and $W=(w_{kl})\in\mR^{d^*\times d^*}$ be a positive definite matrix, and assume that
$U_{tk}$ and $V_{tk}$ are symmetric matrices satisfying $\max_k\sup_t \|U_{tk}\|<\infty$ and $\max_k\sup_t \|V_{tk}\|<\infty$.
Moreover, define
$Z_{nT}=(nT)^{-1}\sum_t (\epsilon_t^\top\epsilon_t)^2+n^{-1}T^{-2}\sum_{t_1,t_2} \sum_{k,l}^{d^*} w_{kl}\{\epsilon_{t_1}^\top U_{t_1k}\epsilon_{t_1}-tr(U_{t_1k})\} \epsilon_{t_2}^\top V_{t_2l}\epsilon_{t_2}$
and $T_{nT}=Z_{nT}-(n+2+\bar\Delta)$.

\begin{lemma}
Let
$\sigma_{nT}^2=(8+4\bar\Delta)c+n^{-2}T^{-4}\sum_{t_1\not=t_2\not=t_3}\sum_{k_1,l_1}\sum_{k_2,l_2}w_{k_1l_1}w_{k_2l_2}\{2tr(U_{t_1k_1}U_{t_1k_2})+\bar\Delta tr(U_{t_1k_1}{\otimes}U_{t_1k_2})\} tr(V_{t_2l_1})tr(V_{t_3l_2})+
(8+4\bar\Delta)n^{-1}T^{-3}\sum_{t_1\not=t_2} \sum_{k,l}w_{kl}tr(U_{t_1k})tr(V_{t_2l})$.
Under the null hypothesis of $H_0$ and Conditions (C1)-(C5) in Appendix A,  we then have that, as $nT\to\infty$ and $n/T\rightarrow c$ for some finite constant $c>0$,
\[T_{nT}/\sigma_{nT}\rightarrow_d N(0,1).\]
\end{lemma}

\scsubsection{Appendix B: Proof of Theorem 1}

To prove this theorem, we take the following two steps:
(i) showing that $\hat\theta$ is $(nT/d)^{1/2}$-consistent;
 (ii) verifying that  $\hat\theta$ is asymptotically normal.

{\sc Step I.}
To show the consistency, it suffices to
follow the technique of Fan and Li (2001) to
demonstrate that, for an arbitrarily small positive constant
$\xi>0$, there exists a constant $C_{\xi}>0$ such that
\beq \label{eq:A1}
\rmP\lbk \sup_{u\in \mR^{d+1}:\|u\|=C_{\xi}} \ell\{ \theta_0+ (nT/d)^{-1/2}u\}<\ell(\theta_0) \rbk\geq 1-\xi
\eeq
for $nT$ sufficiently large.
To this end,  we employ a Taylor series expansion and obtain that
\bea
&&\sup_{u\in \mR^{d+1}:\|u\|=C_{\xi}} \ell\lsk \theta_0+ (nT/d)^{-1/2}u\rsk-\ell(\theta_0)\nn\\
&=&\sup_{u\in \mR^{d+1}:\|u\|=C_{\xi}} \lmk \frac{1}{(nT/d)^{1/2}}u^\top\frac{\partial\ell(\theta_0)}{\partial \theta}-\frac{d}{2nT}u^\top \lbk-\frac{\partial^2\ell(\theta_0)}{\partial\theta\partial \theta^\top}\rbk u+R_n(u) \rmk, \nonumber
\eea
where
$R_n(u)$ is a negligible term that satisfies $R_n(u)=o_p(d)$.
According to Lemma 3 (i) and Condition (C4), we have $(nT/d)^{-1/2}u^\top \frac{\partial\ell(\theta_0)}{\partial \theta}=dC_{\xi}O_p(1)$ and
$$
-\frac{d}{2nT}u^\top \Big\{-\frac{\partial^2\ell(\theta_0)}{\partial\theta\partial \theta^\top}\Big\} u=-\frac{1}{2}du^{\top}\mathcal{I}(\theta_0)u+o_p(d)\leq -\frac{1}{2}dc_{\min,1}C_{\xi}^2.
$$
Note that $dC_{\xi}O_{p}(1)-\frac{1}{2}dc_{\min,1}C_{\xi}^2$ is a quadratic function of $C_{\xi}$. Hence, as long as $C_{\xi}$ is sufficient large, we have
\beq\label{eq:supell}
\sup_{u\in \mR^{d+1}:\|u\|=C_{\xi}} \Big[ \ell\big\{ \theta_0+ (nT/d)^{-1/2}u\big\}-\ell(\theta_0)\Big]<0,
\eeq
with probability tending to 1, which demonstrates (\ref{eq:A1}).
Based on the result of (\ref{eq:supell}), there exists a local maximizer $\hat\theta$ such that $\|\hat\theta-\theta_0\|\leq {(nT/d)}^{-1/2}C_{\xi}$ for $nT$ sufficiently  large. This, in conjunction with
(\ref{eq:A1}), implies
\[
P\lsk \| \hat\theta-\theta_0\|\leq (nT/d)^{-1/2}C_{\xi}\rsk\geq  P\Big\{ \sup_{u\in \mR^{d+1}:\|u\|=C_{\xi}} \ell\lsk \theta_0+ (nT/d)^{-1/2}u\rsk<\ell(\theta_0) \Big\}\geq 1-\xi.
\]
As a result, $(nT/d)^{1/2}\|\hat\theta-\theta_0\|=O_p(1)$, which completes the proof of Step I.

{\sc Step II.}
By the result of STEP I and a Taylor series expansion,  we have that $0={\partial \ell(\hat \theta)}/{\partial \theta}={\partial \ell(\theta_0)}/{\partial \theta}+\{{\partial^2 \ell( \theta_0)}/{\partial \theta\partial \theta^\top}\}(\hat\theta-\theta_0)\{1+o_p(1)\}$.
Thus,
$$
(nT/d)^{-1/2} {\partial \ell(\theta_0)}/{\partial \theta}=-(nT/d)^{1/2} \frac{1}{nT}\{{\partial^2 \ell(\theta_0)}/{\partial \theta\partial \theta^\top}\}(\hat{\theta}-\theta_0)\{1+o_p(1)\}.
$$
By Lemma 3 (iii), we obtain
$$\Big\|(nT/d)^{1/2}\Big(\frac{1}{nT}\frac{\partial^2 \ell(\theta_0)}{\partial \theta\partial \theta^\top}+\mathcal{I}(\theta_0)\Big)(\hat{\theta}-\theta_0)\Big\| \leq (nT/d)^{1/2}\big\|\frac{1}{nT}\frac{\partial^2 \ell(\theta_0)}{\partial \theta\partial \theta^\top}+\mathcal{I}(\theta_0)\big\|\big\|\hat{\theta}-\theta_0\big\|=o_p(1).
$$
Thus, we have
$$
-(nT/d)^{1/2} \frac{1}{nT}\frac{\partial^2 \ell(\theta_0)}{\partial \theta\partial \theta^\top}(\hat{\theta}-\theta_0)=(nT/d)^{1/2}\mathcal{I}(\theta_0)(\hat{\theta}-\theta_0)+o_p(1).
$$
This, together with Lemma 3 (i), implies
$$
\sqrt{nT/d}D\mathcal{I}(\theta_0)(\hat{\theta}-\theta_0)\stackrel{d}{\longrightarrow}N\lsk 0,G(\theta_0)\rsk,
$$
which completes the entire proof.

\scsubsection{Appendix C: Proof of Theorem 2}

By definition, $\hat B_t=\hat\lambda_1W_1^{(t)}+\cdots+\hat\lambda_dW_d^{(t)}$. Thus,
$\hat B_t-B_t=(\hat\lambda_1-\lambda_{01})W_1^{(t)}+\cdots+(\hat\lambda_d-\lambda_{0d})W_d^{(t)}$.
By Theorem 1, we have $ \|\hat\lambda-\lambda_0\|=O_p\{(nT/d)^{-1/2}\}$.
By the Triangle inequality and Condition (C3), we obtain
\[\|\hat B_t-B_t\|=\|(\hat\lambda_1-\lambda_{01})W_1^{(t)}+\cdots+(\hat\lambda_d-\lambda_{0d})W_d^{(t)}\| \]
\[\leq
|\hat\lambda_1-\lambda_{01}|\times\|W_1^{(t)}\|+\cdots+|\hat\lambda_d-\lambda_{0d}|\times \|W_d^{(t)}\| \]
\[
\leq   \max_k \|W_k^{(t)}\|\times \sum_{k=1}^{d} |\hat\lambda_k-\lambda_{0k}|\leq C_w\sqrt{d}\|\hat\lambda-\lambda_0\|=O_p\{d(nT)^{-1/2}\},\]
which completes the proof.

\scsection{REFERENCES}

\begin{description}
\newcommand{\enquote}[1]{``#1''}
\expandafter\ifx\csname natexlab\endcsname\relax\def\natexlab#1{#1}\fi

\bibitem[{Anderson(1973)}]{Anderson:1973}
Anderson, T, ~W. (1973). \enquote{Asymptotically efficient estimation of covariance
matrices with linear structure,}
\textit{Annals of Statistics}, 1, 135--141.

\bibitem[{Badinger and Egger(2011)}]{Badinger:2011}
Badinger, H. and Egger, P. (2011).
\enquote{Estimation of higher order spatial autoregressive cross-section models with heteroscedastic disturbances,}
 \textit{Papers in Regional Science}, 90, 213--235.

 \bibitem[{Badinger and Egger(2015)}]{Badinger:2015}
Badinger, H. and Egger, P. (2015).
\enquote{Fixed effects and random effects estimation of higher-order spatial autoregressive models with spatial autoregressive and heteroskedastic disturbances,}
 \textit{Spatial Economic Analysis}, 10, 11--35.

\bibitem[{Brown et al.(1996)}]{Brown1996}
Brown, C., Harlow, V. and Starks, T. (1996). \enquote{Of tournaments and temptations: An analysis of managerial incentives
in the mutual fund industry,} \textit{Journal of Finance}, 51, 85--110.

\bibitem[{Brown and Wu(2016)}]{Brown:Wu:2016}
Brown, D. P. and Wu, Y. (2016). \enquote{Mutual fund flows and cross-fund learning within families,}
\textit{Journal of Finance}, 71(1), 383-424.

\bibitem[{Carhart(1997)}]{Carhart:1997}
Carhart, M. (1997). \enquote{On Persistence in Mutual Fund Performance,}
\textit{Journal of Finance}, 52(1), 57--82.

\bibitem[Chen and Chen(2008)]{Chen:2008:Extended}
Chen, J. and Chen, Z. (2008). \enquote{Extended Bayesian information criteria for model selection with large model spaces,}
\textit{Biometrika}, 95, 759--771.

\bibitem[Chen and Chen(2012)]{Chen:2012:Extended}
Chen, J. and Chen, Z. (2012). \enquote{Extended BIC for small-n-large-p sparse GLM,}
\textit{Statistica Sinica}, 22, 555--574.

\bibitem[{Chen et al.(2010)}]{Chen:Zhang:Zhong:2010}
Chen, S., Zhang, L. and Zhong, P. (2010).
\enquote{Tests for high dimensional covariance matrices,}
 \textit{Journal of
the American Statistical Association}, 105, 810--819.



\bibitem[{Dou et al.(2016)}]{Dou:Parrella:Yao:2016}
Dou, B., Parrella, M. and Yao, Q. (2016).
\enquote{Generalized Yule-Walker estimation for spatio-temporal models with unknown diagonal coefficients,}
 \textit{Journal of Econometrics}, 194, 369--382.

   \bibitem[{Elhorst et al.(2012)}]{Elhorst:2012}
Elhorst, J. P., Lacombe, D. J. and Piras, G. (2012).
\enquote{On model specification
and parameter space definitions in higher order spatial econometric models,}
 \textit{Regional Science and Urban Economics,} 42, 211--220.

 \bibitem[{Fan and Li(2001)}]{Fan:Li:2001}
Fan, J. and Li, R. (2001).
\enquote{Variable selection via nonconcave penalized likelihood and its oracle properties,}
 \textit{Journal of the American Statistical Association}, 96, 1348--1360.


\bibitem[{Gao et al.(2019)}]{Gao:Ma:Wang:Yao:2019}
Gao, Z., Ma, Y., Wang, H. and Yao, Q. (2019).
\enquote{Banded spatio-temporal autoregressions,}
 \textit{Journal of Econometrics}, 208, 211--230.

\bibitem[{Golgher and Voss(2016)}]{Golgher:Voss:2016}
Golgher, A. and Voss, P. (2016).
\enquote{How to interpret the coefficients of spatial models: spillovers, direct and indirect effects,}
 \textit{Spatial Demography}, 4, 175--205.

\bibitem[{Gupta and Robinson(2015)}]{Gupta:Robinson:2015}
Gupta, A. and Robinson, P. (2015).
\enquote{Inference on higher-order spatial autoregressive models with increasingly many parameters,}
 \textit{Journal of Econometrics}, 186, 19--31.

\bibitem[{Gupta and Robinson(2018)}]{Gupta:Robinson:2018}
Gupta, A. and Robinson, P. (2018).
\enquote{Pseudo maximum likelihood estimation of spatial autoregressive models with increasing dimension,}
 \textit{Journal of Econometrics}, 202, 92--107.

 \bibitem[{Hall and Heyde(1980)}]{Hall:Heyde:1980}
Hall, P. and Heyde, C. (1980). \textit{Martingale Limit Theory and Its Application}, New York: Academic Press.

\bibitem[{Han et al.(2021)}]{Han:2021}
Han, X., Hu, Y., Fan, L., Huang, Y., Xu, M. and Gao, S. (2021).
\enquote{Quantifying COVID-19 importation risk in a dynamic network of domestic cities and international countries,}
 \textit{Proceeding of the National Academy of Sciences}, 118(31):e2100201118.

 \bibitem[{Han et al.(2017)}]{Han:2017}
Han, X., Hsieh, C. S. and Lee, L.F. (2017).
\enquote{Estimation and model selection of higher-order spatial autoregressive
model: An efficient Bayesian approach,}
 \textit{Regional Science and Urban Economics}, 63, 97--120.

\bibitem[{Huang et al.(2019)}]{Huang:Lan:Zhang:Wang:2019}
Huang, D., Lan, W., Zhang, H. and Wang, H. (2019).
\enquote{Least squares estimation for social autocorrelation in large-scale networks,}
 \textit{Electronic Journal of Statistics}, 13, 1135--1165.

\bibitem[{Jenish and Prucha(2012)}]{Jenish:2012}
Jenish, N. and Prucha, I. R. (2012).
\enquote{On spatial processes and asymptotic inference
under near-epoch dependence,}
 \textit{Journal of Econometrics,} 170, 178--190.

\bibitem[{Kelejian and Prucha(2001)}]{Kelejian:Prucha:2001}
Kelejian, H. and Prucha, I. (2001). \enquote{On the asymptotic distribution of the Moran I test statistic with applications,}\textit{Journal of Econometrics}, 104, 219--257.

\bibitem[{Kelejian and Prucha(2010)}]{Kelejian:Prucha:2010}
Kelejian, H. and Prucha, I. (2010). \enquote{Specification and estimation of spatial autoregressive models with
autoregressive and heteroskedastic disturbances,}\textit{Journal of Econometrics}, 157, 53--67.

\bibitem[{Lam and Souza (2020)}]{Lam:Souza:2020}
Lam, C. and Souza, P. (2020).
\enquote{Estimation and selection of spatial weight matrix in a spatial lag model,}
 \textit{Journal of Business \& Economic Statistics}, 38, 693--710.


\bibitem[{Ledoit and Wolf(2002)}]{Ledoit:Wolf:2002}
Ledoit, O. and Wolf, M. (2002).
\enquote{Some hypotheses tests for the covariance matrix when the dimension is
  large compare to the sample size,}
\textit{ Annals of Statistics}, 30, 1081--1102.

\bibitem[{Lee(2004)}]{Lee:2004}
Lee, L.~F. (2004). \enquote{Asymptotic distributions of quasi-maximum likelihood estimators for
spatial autoregressive models,} \textit{Econometrica}, 72, 1899--1925.

\bibitem[{Lee and Liu(2010)}]{Lee:2015}
Lee, L.~F. and Liu, X. (2010). \enquote{Efficient GMM estimation of high order spatial
autoregressive models with autoregressive disturbances,} \textit{Econometric Theory}, 26, 187-230.

\bibitem[Lee and Yu(2010a)]{Lee:2010}
Lee, L.~F. and Yu, J. (2010a). \enquote{Estimation of spatial autoregressive panel data models with fixed effects,}
\textit{Journal of Econometrics}, 154, 165--185.

\bibitem[Lee and Yu(2010b)]{Lee:20102}
Lee, L.~F. and Yu, J. (2010b). \enquote{A spatial dynamic panel data model with both time and individual effects,}
\textit{Econometric Theory}, 26, 564--597.

\bibitem[Leenders(2002)]{Leenders:2002}
Leenders, R.~T. (2002). \enquote{Modeling social influence through network autocorrelation: constructing the weight matrix,}
\textit{Social Networks}, 24, 21--47.

 \bibitem[{LeSage and Pace(2009)}]{LeSage:Pace:2009}
LeSage, J. and Pace, R.~K. (2009). \textit{Introduction to Spatial
  Econometrics}, New York: Chapman \& Hall.

   \bibitem[{Liu et al.(2014)}]{Liu:2014}
Liu, X., Patacchini, E. and Zenou, Y. (2014). \enquote{Endogenous peer effects: local
aggregate of global average?,} \textit{Journal of Economic Behavior and Organization}, 103, 39--59.


\bibitem[{Nanda et al.(2004)}]{Nanda2004}
Nanda, N., Wang, J. and Zheng, L. (2004). \enquote{Family values and the star phenomenon: Strategies of mutual fund families,} \textit{Review of Financial Studies}, 17, 667--698.

\bibitem[{Peng et al.(2017)}]{Peng:Yang:Cao:Yu:Xie:2017}
Peng, S., Yang, A., Cao, L., Yu, S. and Xie, D. (2017). \enquote{Social influence modeling using information theory in mobile social networks,} \textit{Information Sciences}, 379, 146--159.

\bibitem[{Qu et al.(2000)}]{Qu:Lindsay:Li:2000}
Qu, A., Lindsay, G. and Li, B. (2000). \enquote{Improving generalised estimating equations using quadratic inference functions,} \textit{Biometrika}, 87, 823--836.


 \bibitem[{Qu and Lee(2015)}]{Qu:2015}
Qu, X. and Lee, L.~F. (2015). \enquote{Estimating a spatial autoregressive model with an endogenous spatial weight matrix,} \textit{Journal of Econometrics}, 184, 209-232.

 \bibitem[{Qu et al.(2017)}]{Qu:2017}
Qu, X., Lee, L.~F. and Yu, J. (2017). \enquote{QML Estimation of spatial dynamic panel data models with endogenous time varying spatial weights matrices,} \textit{Journal of Econometrics}, 197, 173-201.


 \bibitem[{Spitz(1970)}]{Spitz:1970}
Spitz, E. (1970). \enquote{Mutual fund performance and cash inflows,} \textit{Applied Economics}, 2, 141--145.


 \bibitem[{Trusov et al.(2010)}]{Trusov:Bodapati:Bucklin:2010}
Trusov, M., Bodapati, A. and Bucklin, R. (2010).
\enquote{Determining influential users in internet social networks,}
 \textit{Journal of Marketing Research}, 47, 643--558.

\bibitem[{Wooldridge(2002)}]{Wooldridge:2002}
Wooldridge, J. (2002). \textit{Econometric Analysis of Cross Section and Panel Data}, MIT Press, Cambridge, Mass.


\bibitem[{Zhang and Yu(2018)}]{Zhang:Yu:2018}
Zhang, X. and Yu, J. (2018).
\enquote{Spatial weights matrix selection and model averaging for spatial
autoregressive models,}
 \textit{Journal of Econometrics}, 203, 1--18.

\bibitem[{Zheng et al.(2019)}]{Zheng:Chen:Cui:Li:2019}
Zheng, S., Chen, Z., Cui, H. and Li, R. (2019).
\enquote{Hypothesis testing on linear structures of high dimensional covariance matrix,}\textit{Annals of Statistics},
47, 3300--3334.

 \bibitem[{Zhou et al.(2017)}]{Zhou:Tu:Chen:Wang:2017}
Zhou, J., Tu, Y., Chen, Y. and Wang, H. (2017). \enquote{Estimating spatial autocorrelation with sampled network data,}
\textit{Journal of Business \& Economics Statistics}, 35, 130--138.

\bibitem[{Zhu et al.(2017)}]{Zhu:Pan:Li:Liu:Wang:2017}
Zhu, X., Pan, R., Li, G., Liu, Y. and Wang, H. (2017).
\enquote{Network vector autoregression,}
 \textit{Annals of Statistics}, 45, 1096--1123.

 \bibitem[{Zou et al.(2017)}]{Zou:Lan:Wang:Tsai:2017}
Zou, T., Lan, W., Wang, H. and Tsai, C. L. (2017).
\enquote{Covariance regression analysis,}
 \textit{Journal of the American Statistical Association}, 112, 266--281.

 \end{description}

 \newpage

\begin{table}[!h]
\begin{center} \caption{\label{tab:t1} The bias and standard error of the parameter estimates when the true parameters are
$\lambda_k=0.2$ for $k=1, \cdots, d$, and the random errors follow a normal distribution.
BIAS: the average bias; SE: the average of the estimated standard errors via Theorem 1; SE$^*$: the standard error of parameter estimates calculated from 500 realizations.}
\vspace{0.18 cm}

\begin{tabular}{ccc|rr|rrrrrr}
\hline
\hline
&&&\multicolumn{2}{|c}{$d=2$ } & \multicolumn{6}{|c}{$d=6$}\\
$n$ & $T$ &&  $\lambda_1$ & $\lambda_2$ & $\lambda_1$ & $\lambda_2$ & $\lambda_3$ & $\lambda_4$ & $\lambda_5$ & $\lambda_6$\\
\hline
 25 &  25   &BIAS        &   -0.003 &   -0.004 &   -0.015 &   -0.017 &   -0.014 &   -0.017 &   -0.012 &   -0.009 \\
           &&SE          &    0.050 &    0.050 &    0.047 &    0.048 &    0.048 &    0.048 &    0.047 &    0.047 \\
           &&SE$^*$      &    0.049 &    0.050 &    0.051 &    0.051 &    0.052 &    0.055 &    0.054 &    0.057 \\
\hline
 25   &  50   &BIAS        &   -0.006 &   -0.002 &   -0.003 &   -0.004 &   -0.002 &   0.003 &   0.004 &   -0.002 \\
            &&SE          &    0.035 &    0.035 &    0.034 &    0.034 &    0.034 &    0.034 &    0.034 &    0.034 \\
            &&SE$^*$      &    0.034 &    0.033 &    0.040 &    0.044 &    0.041 &    0.041 &    0.040 &    0.041 \\
\hline
 25   &  100  &BIAS        &   -0.001 &    0.001 &   -0.001 &   -0.001 &   0.002 &   -0.002 &    0.003 &   -0.002 \\
            &&SE          &    0.025 &    0.025 &    0.024 &    0.025 &    0.026 &    0.024 &    0.026 &    0.027 \\
            &&SE$^*$      &    0.026 &    0.024 &    0.026 &    0.027 &    0.027 &    0.026 &    0.028 &    0.026 \\
\hline
 50 &  25   &BIAS        &   -0.002 &    0.002 &   -0.002 &   -0.006 &   -0.003 &    0.004 &   -0.002 &   -0.005 \\
            &&SE          &    0.035 &    0.034 &    0.032 &    0.032 &    0.032 &    0.032 &    0.032 &    0.032 \\
            &&SE$^*$      &    0.035 &    0.034 &    0.035 &    0.036 &    0.034 &    0.036 &    0.034 &    0.034 \\
\hline
 50   &  50   &BIAS        &   -0.002 &   -0.000 &   -0.003 &   0.003 &   -0.001 &   -0.003 &   -0.004 &   -0.004 \\
            &&SE          &    0.024 &    0.024 &    0.023 &    0.023 &    0.023 &    0.023 &    0.023 &    0.023 \\
            &&SE$^*$      &    0.023 &    0.025 &    0.025 &    0.026 &    0.026 &    0.025 &    0.025 &    0.026 \\
\hline
 50   &  100  &BIAS        &   -0.001 &    0.000 &   -0.002 &   -0.001 &   -0.001 &   0.001 &   -0.002 &   0.002 \\
            &&SE          &    0.017 &    0.017 &    0.016 &    0.016 &    0.016 &    0.016 &    0.016 &    0.016 \\
            &&SE$^*$      &    0.016 &    0.018 &    0.018 &    0.018 &    0.018 &    0.018 &    0.018 &    0.017 \\
\hline
 100 &  25  &BIAS        &    0.001 &   -0.001 &   -0.001 &   -0.001 &   -0.002 &    0.001 &   -0.000 &   -0.002 \\
            &&SE          &    0.024 &    0.024 &    0.022 &    0.022 &    0.022 &    0.022 &    0.022 &    0.022 \\
            &&SE$^*$      &    0.024 &    0.025 &    0.024 &    0.023 &    0.023 &    0.023 &    0.023 &    0.023 \\
\hline
 100    &  50  &BIAS        &    0.000 &   -0.000 &    0.000 &   -0.001 &   -0.001 &    0.000 &   -0.003 &   -0.000 \\
            &&SE          &    0.017 &    0.017 &    0.016 &    0.016 &    0.016 &    0.016 &    0.016 &    0.016 \\
            &&SE$^*$      &    0.018 &    0.017 &    0.017 &    0.018 &    0.016 &    0.017 &    0.016 &    0.017 \\
\hline
 100   &  100 &BIAS        &    0.000 &    0.001 &   -0.001 &    0.000 &   -0.001 &   -0.000 &   -0.000 &   -0.001 \\
            &&SE          &    0.012 &    0.012 &    0.011 &    0.011 &    0.011 &    0.011 &    0.011 &    0.011 \\
             &&SE$^*$      &    0.012 &    0.012 &    0.012 &    0.012 &    0.012 &    0.012 &    0.013 &    0.012 \\
 \hline\hline
\end{tabular}
\end{center}
\end{table}

 \begin{table}[!h]
\begin{center} \caption{\label{tab:t1} Model selections via EBIC when
$d=8$ and the random errors are normally distributed. AS: the average size of the selected model; CT: the average percentage of the correct fit;
TPR: the average true positive rate; FPR: the average false positive rate.} \vspace{0.28 cm}
\begin{tabular}{cc|cccc}
\hline
\hline
$n$ & $T$ &  AS & CT & TPR & FPR    \\
\hline
25  &     25&      3.3 &    74.5 &    92.6  &    9.7 \\
    &     50&      3.2 &    79.2 &    96.8  &    8.2 \\
    &    100&      3.1 &    82.1 &    100.0 &    5.5 \\
    \hline
50  &     25&      3.2 &    79.3 &    95.1  &    7.2 \\
    &     50&      3.1 &    82.5 &    98.4  &    5.3 \\
    &    100&      3.1 &    85.7 &    100.0 &    4.4 \\
    \hline
100 &     25&      3.1 &    83.1 &    100.0 &    6.2 \\
    &     50&      3.1 &    85.2 &    100.0 &    4.5 \\
    &    100&      3.0 &    88.4 &    100.0 &    3.9 \\
\hline
\end{tabular}
\end{center}
\end{table}

\begin{table}[!h]
\begin{center} \caption{\label{tab:t1} The empirical sizes and powers of the goodness of fit test. The $\kappa=0$ corresponds to the null model and
$\kappa>0$ represents alternative models. The random errors are normally distributed, and the full model sizes are $d=2$ and 6. } \vspace{0.28 cm}
\begin{tabular}{cc|ccc|ccc}
\hline\hline
&&\multicolumn{3}{|c}{$d$=2} & \multicolumn{3}{|c}{$d$=6}  \\
$n$ & $T$ &  $\kappa$=0 & $\kappa$=0.1 & $\kappa$=0.2 & $\kappa$=0 & $\kappa$=0.1 & $\kappa$=0.2  \\
\hline
25  &     25&      0.027 &     0.326 &     0.685  &      0.022 &     0.259 &     0.604\\
    &     50&      0.032 &     0.544 &     0.846  &      0.031 &     0.453 &     0.774\\
    &    100&      0.040 &     0.679 &     0.922  &      0.042 &     0.584 &     0.849\\
    \hline
50  &     25&      0.030 &     0.464 &     0.791  &      0.025 &     0.368 &     0.672\\
    &     50&      0.034 &     0.601 &     0.908  &      0.034 &     0.496 &     0.802\\
    &    100&      0.045 &     0.711 &     1.000  &      0.047 &     0.672 &     0.973\\
    \hline
100 &     25&      0.032 &     0.528 &     0.991  &      0.032 &     0.469 &     0.976\\
    &     50&      0.039 &     0.751 &     1.000  &      0.036 &     0.604 &     1.000\\
    &    100&      0.044 &     0.928 &     1.000  &      0.046 &     0.855 &     1.000\\
\hline
\end{tabular}
\end{center}
\end{table}

\begin{landscape}
\begin{table}[!h]
\begin{center}
\caption{\label{tab:t1} The bias and standard error of the parameter estimates,
QMLE and EA-QMLE,
when the true parameters are
$\lambda_k=0.2$ for $k=1, \cdots, d$, and the random errors follow a normal distribution.
BIAS: the average bias; SE: the average of the estimated standard errors via Theorems 1 and 5; SE$^*$: the standard error of parameter estimates calculated from 500 realizations.}
\vspace{0.18 cm}

\begin{tabular}{ccc|rrrrrr|rrrrrr}
\hline
\hline
&&&\multicolumn{6}{|c}{QMLE} & \multicolumn{6}{|c}{EA-QMLE}\\
$n$ & $T$ &&  $\lambda_1$ & $\lambda_2$ & $\lambda_3$ & $\lambda_4$ & $\lambda_5$ & $\lambda_6$ & $\lambda_1$ & $\lambda_2$ & $\lambda_3$ & $\lambda_4$ & $\lambda_5$ & $\lambda_6$\\
\hline
  25 &   25   &BIAS        &    0.045 &    0.044 &    0.046 &    0.046 &    0.042 &    0.047 &   -0.002 &   -0.004 &   -0.002 &   -0.002 &   -0.002 &   -0.001  \\
     &        &SE          &    0.053 &    0.053 &    0.053 &    0.053 &    0.053 &    0.053 &    0.043 &    0.043 &    0.043 &    0.043 &    0.043 &    0.043  \\
     &        &SE$^*$      &    0.044 &    0.045 &    0.043 &    0.045 &    0.043 &    0.047 &    0.044 &    0.045 &    0.043 &    0.045 &    0.043 &    0.047  \\
\hline
  25 &   50   &BIAS        &    0.047 &    0.043 &    0.046 &    0.048 &    0.043 &    0.045 &    0.001 &   -0.004 &   -0.001 &    0.000 &   -0.003 &   -0.001  \\
     &        &SE          &    0.037 &    0.037 &    0.037 &    0.037 &    0.037 &    0.037 &    0.030 &    0.030 &    0.030 &    0.030 &    0.030 &    0.030  \\
     &        &SE$^*$      &    0.030 &    0.032 &    0.032 &    0.032 &    0.030 &    0.032 &    0.030 &    0.032 &    0.032 &    0.032 &    0.030 &    0.032  \\
\hline
  25 &  100   &BIAS        &    0.046 &    0.045 &    0.046 &    0.047 &    0.044 &    0.045 &    0.001 &   -0.001 &    0.000 &   -0.000 &   -0.000 &   -0.002  \\
     &        &SE          &    0.026 &    0.026 &    0.026 &    0.026 &    0.026 &    0.026 &    0.021 &    0.021 &    0.021 &    0.021 &    0.021 &    0.021  \\
     &        &SE$^*$      &    0.021 &    0.022 &    0.021 &    0.021 &    0.022 &    0.022 &    0.021 &    0.022 &    0.021 &    0.021 &    0.022 &    0.022 \\
\hline
  50 &   25   &BIAS        &    0.060 &    0.058 &    0.057 &    0.055 &    0.058 &    0.056 &    0.001 &   -0.001 &   -0.003 &   -0.003 &   -0.001 &   -0.002  \\
     &        &SE          &    0.034 &    0.034 &    0.034 &    0.034 &    0.034 &    0.034 &    0.029 &    0.029 &    0.029 &    0.029 &    0.029 &    0.029  \\
     &        &SE$^*$      &    0.029 &    0.029 &    0.029 &    0.028 &    0.030 &    0.029 &    0.029 &    0.029 &    0.029 &    0.028 &    0.030 &    0.029  \\
\hline
  50 &   50   &BIAS        &    0.059 &    0.058 &    0.056 &    0.056 &    0.059 &    0.057 &    0.001 &   -0.000 &   -0.001 &   -0.003 &   -0.001 &   -0.001  \\
     &        &SE          &    0.024 &    0.024 &    0.024 &    0.024 &    0.024 &    0.024 &    0.020 &    0.020 &    0.020 &    0.020 &    0.020 &    0.020  \\
     &        &SE$^*$      &    0.021 &    0.019 &    0.021 &    0.021 &    0.021 &    0.020 &    0.021 &    0.019 &    0.021 &    0.021 &    0.021 &    0.020  \\
\hline
  50 &  100   &BIAS        &    0.059 &    0.058 &    0.057 &    0.056 &    0.059 &    0.057 &    0.000 &    0.000 &   -0.001 &   -0.002 &    0.001 &   -0.001  \\
     &        &SE          &    0.017 &    0.017 &    0.017 &    0.017 &    0.017 &    0.017 &    0.014 &    0.014 &    0.014 &    0.014 &    0.014 &    0.014  \\
     &        &SE$^*$      &    0.015 &    0.014 &    0.015 &    0.015 &    0.015 &    0.014 &    0.015 &    0.014 &    0.015 &    0.015 &    0.015 &    0.014  \\
\hline
 100 &   25   &BIAS        &    0.064 &    0.064 &    0.063 &    0.066 &    0.065 &    0.066 &   -0.002 &   -0.000 &   -0.002 &    0.000 &    0.000 &    0.001  \\
     &        &SE          &    0.023 &    0.023 &    0.023 &    0.023 &    0.023 &    0.023 &    0.020 &    0.020 &    0.020 &    0.020 &    0.020 &    0.020  \\
     &        &SE$^*$      &    0.020 &    0.021 &    0.020 &    0.020 &    0.021 &    0.020 &    0.020 &    0.021 &    0.020 &    0.020 &    0.021 &    0.020 \\
\hline
 100 &   50   &BIAS        &    0.064 &    0.064 &    0.066 &    0.065 &    0.065 &    0.065 &   -0.002 &   -0.001 &    0.001 &   -0.000 &   -0.000 &    0.000  \\
     &        &SE          &    0.016 &    0.016 &    0.016 &    0.016 &    0.016 &    0.016 &    0.014 &    0.014 &    0.014 &    0.014 &    0.014 &    0.014  \\
     &        &SE$^*$      &    0.014 &    0.015 &    0.014 &    0.014 &    0.015 &    0.015 &    0.014 &    0.015 &    0.014 &    0.014 &    0.015 &    0.015  \\
\hline
 100 &  100   &BIAS        &    0.064 &    0.064 &    0.066 &    0.066 &    0.065 &    0.065 &   -0.001 &   -0.001 &    0.000 &    0.000 &    0.000 &    0.001  \\
     &        &SE          &    0.012 &    0.012 &    0.012 &    0.012 &    0.012 &    0.012 &    0.010 &    0.010 &    0.010 &    0.010 &    0.010 &    0.010  \\
     &        &SE$^*$      &    0.009 &    0.010 &    0.010 &    0.009 &    0.011 &    0.010 &    0.009 &    0.010 &    0.010 &    0.009 &    0.011 &    0.010  \\

 \hline\hline
\end{tabular}
\end{center}
\end{table}
\end{landscape}

\begin{table}[htbp!]
\begin{center}
\caption
{The QMLE and EA-QMLE parameter estimates and associated standard errors and p-values for the five covariates.}
%
\label{tb:4}
\begin{tabular}{l|ccc|cccc}
\hline
&\multicolumn{3}{|c}{QMLE} & \multicolumn{3}{|c}{EA-QMLE}  \\
\hline
& Estimate  & Standard-Error   &  $p$-Value  & Estimate  & Standard-Error    &  $p$-Value\\
\hline
   Alpha   &   0.005 & 0.027 & 0.853 & 0.042 & 0.030 & 0.162 \\
  Return   &   0.569 & 0.019 & 0.000 & 0.116 & 0.027 & 0.000\\
  Size     &   0.330 & 0.014 & 0.000 & 0.112 & 0.023 & 0.000\\
   Age     &   0.036 & 0.018 & 0.046 & 0.174 & 0.026 & 0.000\\
  Volatility &  0.209 & 0.020 & 0.000 & 0.176 & 0.022 & 0.000\\
\hline
\end{tabular}
\end{center}
\end{table}

\end{document}


\begin{center}
{\bf\Large Mutual Influence Regression Model}\\
{Supplementary Material} \\
\bigskip
\end{center}
\begin{center}
{Xinyan Fan, Wei Lan, Tao Zou and Chih-Ling Tsai}

{\it Renmin University of China, Southwestern University of Finance and Economics, Australian National University
and University of California, Davis}
\end{center}

\renewcommand{\theequation}{S.\arabic{equation}}
\renewcommand{\thela}{S.\arabic{la}}

This supplementary material consists of five sections.
Section S.1 presents the proofs of Lemmas 3 and 4. Section S.2 gives the proofs of Theorems 3--5.
Section S.3 introduces technical conditions and lemmas that are used in proving the theoretical properties in Section 5.
Section S.4 presents the proof of Theorem 11.
Section S.5 presents five additional simulation results: (i)
the MIR model with twelve weight matrices; (ii) a comparison of the
extended BIC criterion (EBIC) with the Deviance information criterion (DIC) for the selection of weight matrices;
(iii) the MIR model with exponential errors; (iv) the MIR model with mixture errors;
 and (v) the MIR model with exogenous covariates.

\csection{Proofs of Lemmas 3--4}

\noindent {\textbf{Proof of Lemma 3:}}
To show part (i) in the above lemma, we  define $\epsilon=(\epsilon_1^\top, \cdots, \epsilon_T^\top)^\top\in\mR^{nT}$.
After simple calculation, we obtain that
\[\frac{\partial \ell(\theta_0)}{\partial \theta}=\big(\frac{1}{\sigma_0^2}\epsilon^\top U_1\epsilon-tr(U_1), \cdots, \frac{1}{\sigma_0^2}\epsilon^\top U_d\epsilon-tr(U_d),\frac{1}{2\sigma_0^4}\epsilon^\top\epsilon-\frac{nT}{2\sigma_0^2}\big),\]
where $U_k$ for $k=1,\cdots, d$ is defined below Theorem 1. Denote $U_{d+1}=\frac{1}{2\sigma_0^2}I_{nT}$.
By Cram\'er's theorem, it suffices to show that $d^{-1/2}c^\top D\frac{\partial \ell(\theta_0)}{\partial \theta}$ is asymptotic normal for any
finite vector $c=(c_1, \cdots, c_{M})^\top\in\mR^{M}$.
Define $U_c=d^{-1/2}\sum_{k=1}^{d+1}(\sum_{m=1}^{M}c_mD_{mk})U_k$ with $D=(D_{mk})$, then we have
$d^{-1/2}c^\top D\frac{\partial \ell(\theta_0)}{\partial \theta}=(\sigma_0^2)^{-1}\big\{\epsilon^\top U_c \epsilon-\sigma_0^2tr(U_c)\big\}$.
By Lemma 1, we only need to  verify that $U_c$ satisfies Assumptions (1)--(4)
listed in Lemma 1.
Using the fact that $U_k$ is symmetric for any $k=1, \cdots, d$, Assumption (1) is satisfied.
By Conditions (C2) and (C3), we have that $\|U_k\|_R\leq \sup_t\|W_k^{(t)}\Delta_t(\lambda_0)\|_R\leq \sup_t\|W_k^{(t)}\|_R\|\Delta_t(\lambda_0)\|_R<\infty$.
This implies that
\[\|U_c\|_R\leq \sum_{k=1}^{d+1}d^{-1/2}\sum_{m=1}^{M}|c_mD_{mk}|\|U_{k}\|_R\leq c_{u,1}d^{-1/2}\sum_{k=1}^{d+1}\sum_{m=1}^{M}|D_{mk}|\leq c_{u,2}\|D\|< \infty\]
for two finite positive constants $c_{u,1}$ and $c_{u,2}$. Thus, $U_c$ satisfies Assumption (2). By Condition (C1),
Assumption (3) holds. In addition, since
$(nT)^{-1}\text{var} (d^{-1/2}c^\top D\frac{\partial \ell(\theta_0)}{\partial \theta})=d^{-1}c^{\top}D\mathcal{J}_{nT}(\theta_0)D^{\top}c>2^{-1}c^{\top} G(\theta_0)c>c_{g}$
for some positive constant $c_{g}$ with probability approaching 1, Assumption (4) holds, which completes the first part of the proof.

We next very part (ii). Note that
$
P\lsk (nT)^{-1/2}|\frac{\partial \ell(\theta_0)}{\partial \theta_k}|>\nu\rsk=P\big\{ (nT)^{-1/2}|\epsilon^{\top} U_k\epsilon/\sigma_0^2-tr(U_k)|>\nu\big\}.
$
By Condition (C1) and the Hanson-Wright inequality (Hanson and Wright, 1971; Wright, 1973), we have
$$P\Big\{ (nT)^{-1/2}|\epsilon^{\top} U_k\epsilon/\sigma_0^2-tr(U_k)|>\nu\Big\}\leq 2\exp\lmk-\min\lsk\frac{\tau_1\nu\sigma_0^2}{\|U_k\|/\sqrt{nT}},\frac{\tau_2\sigma_0^4\nu^2}{\|U_k\|_F^2/nT}\rsk\rmk,$$
which completes the second part of the proof.

We lastly prove (iii).
By Condition (C4), it suffices to show that
 $\|(nT)^{-1}\frac{\partial^2 \ell(\theta_0)}{\partial \theta\partial \theta^\top }+ \mathcal{I}_{nT}(\theta_0)\|_F=o_p(1)$.
 After simple calculation, we have that
\bda
-(nT)^{-1}\frac{\partial^2 \ell(\theta_0)}{\partial \theta\partial \theta^\top }&=&
\begin{pmatrix}
-(nT)^{-1}\frac{\partial^2 \ell(\theta_0)}{\partial \lambda\partial \lambda^\top }& -(nT)^{-1}\frac{\partial^2 \ell(\theta_0)}{\partial \lambda \partial \sigma^2 }\\
-(nT)^{-1}\frac{\partial^2 \ell(\theta_0)}{\partial \sigma^2\partial \lambda }& -(nT)^{-1}\frac{\partial^2 \ell(\theta_0)}{\partial^2 \sigma^2 }\\
\end{pmatrix},\textrm{ where }
\eda
$-(nT)^{-1}\frac{\partial^2 \ell(\theta_0)}{\partial^2 \sigma^2 }=\frac{1}{nT\sigma_0^6}\epsilon^\top\epsilon-\frac{1}{2\sigma_0^4}$,
$-(nT)^{-1}\frac{\partial^2 \ell(\theta_0)}{\partial\lambda_k\partial\sigma^2 }=\frac{1}{nT\sigma_0^4}\epsilon^\top U_k \epsilon$ and
$-(nT)^{-1}\frac{\partial^2 \ell(\theta_0)}{\partial\lambda_k\partial\lambda_l}=\frac{1}{nT\sigma_0^2}\epsilon^\top U_kU_l\epsilon+(nT)^{-1}tr(U_kU_l)$ for  $k=1, \cdots, d$ and $l=1, \cdots, d$.
Note that  $\var\big(\frac{1}{nT\sigma_0^6}\epsilon^\top\epsilon\big)=O({(nT)^{-1}})$. Accordingly, we only need to show that $\var(\frac{1}{nT\sigma_0^4}\epsilon^\top U_k \epsilon)=O({(nT)^{-1}})$ and $\var(\frac{1}{nT\sigma_0^2}\epsilon^\top U_kU_l\epsilon)=O({(nT)^{-1}})$ and they are given below.

By Lemma 2 (i) and the fact that $(nT)^{-1}E\big(\epsilon^\top U_k\epsilon\big)=(nT)^{-1}\sigma_0^2tr(U_k)$, we have
$(nT)^{-2}\var\big(\epsilon^\top U_k\epsilon\big)=(nT)^{-2}\sigma_0^4\{2tr(U_k^2)+(\mu^{(4)}-3)tr(U_k^{\otimes 2})\}$.
By Condition (C3), we further have $tr(U_k^2)=O(nT)$ and
$tr(U_k^{\otimes 2})\leq tr(U_k^2)=O(nT)$. This leads to $(nT\sigma_0^4)^{-2}\var\big(\epsilon^\top U_k\epsilon\big)=O((nT)^{-1})$.
Analogously, by Conditions (C3) and Lemma 2 (i),  we can show that
$\var(\frac{1}{nT\sigma_0^2}\epsilon^\top U_kU_l\epsilon)=O({(nT)^{-1}})$.

Consequently, for any $\tau>0$, by Condition (C5) and employ Chebyshev's inequality, we have
\begin{align*}
&P\lsk\Big\|(nT)^{-1}\frac{\partial^2 \ell(\theta_0)}{\partial \theta\partial \theta^\top }+ \mathcal{I}_{nT}(\theta_0)\Big\|_F>\tau/d\rsk  \leq d^2/\tau^2 \sum_{k=1}^{d+1}\sum_{l=1}^{d+1}\var\Big\{(nT)^{-1}\frac{\partial^2 \ell(\theta_0)}{\partial \theta_k\partial \theta_l }\Big\}\\
& \leq d^2/\tau^2\sum_{k=1}^{d+1}\sum_{l=1}^{d+1}O\big\{(nT)^{-1}\big\}=O\big\{d^4/(nT\tau^2)\big\}=o(1).
\end{align*}
The above result, together with Condition (C4), leads to
\[
\Big\|(nT)^{-1}\frac{\partial^2 \ell(\theta_0)}{\partial \theta\partial \theta^\top }+ \mathcal{I}(\theta_0)\Big\|_F\leq \Big\|(nT)^{-1}\frac{\partial^2 \ell(\theta_0)}{\partial \theta\partial \theta^\top }+ \mathcal{I}_{nT}(\theta_0)\Big\|_F+ \Big\|\mathcal{I}(\theta_0)- \mathcal{I}_{nT}(\theta_0)\Big\|_F =o_p(1),\]
which completes the proof of the lemma.

\noindent {\textbf{Proof of Lemma 4}}:
Let $Z_{nT}= Z_{nT,1}+Z_{nT,2}$, where
$Z_{nT,1}=(nT)^{-1}\sum_t (\epsilon_t^\top\epsilon_t)^2$ and $Z_{nT,2}=n^{-1}T^{-2}\sum_{t_1,t_2} \sum_{k,l}^{d^*} w_{kl}\{\epsilon_{t_1}^\top U_{t_1k}\epsilon_{t_1}-tr(U_{t_1k})\} \epsilon_{t_2}^\top V_{t_2l}\epsilon_{t_2}$.
By Condition (C4), after tedious calculations, we obtain that
$E(Z_{nT,1})=n+2+\bar\Delta$, $E(Z_{nT,2})=o(1)$,
$\var(Z_{nT,1})=(8+4\bar\Delta)c\{1+o(1)\}$,
\[\var(Z_{nT,2})=n^{-2}T^{-4}\sum_{t_1\not=t_2\not=t_3}\sum_{k_1,l_1}\sum_{k_2,l_2}w_{k_1l_1}w_{k_2l_2}\]
\[\times\{2tr(U_{t_1k_1}U_{t_1k_2})+\bar\Delta tr(U_{t_1k_1}{\otimes}U_{t_1k_2})\}
 tr(V_{t_2l_1})tr(V_{t_3l_2})\{1+o(1)\}\]
and $\cov(Z_{nT,1},Z_{nT,2})=(4+2\bar\Delta)n^{-1}T^{-3}\sum_{t_1\not=t_2} \sum_{k,l}w_{kl}tr(U_{t_1k})tr(V_{t_2l})\{1+o(1)\}$.
Then,
one can straightforwardly demonstrate that
$E(Z_{nT})=n+2+\bar\Delta+o(1)$ and $\var(Z_{nT})=O\{(d^*)^2\}$.

We next define $\mF_r=\{\epsilon_t, t\leq r\}$ to be the $\sigma$-field generated by $\{\epsilon_t\}$ for $t\leq r$. In addition,
define $T_{nT,r}=(nT)^{-1}\sum_{t=1}^r (\epsilon_t^\top\epsilon_t)^2+n^{-1}T^{-2}\sum_{t_1\leq r,t_2\leq r}\sum_{k,l}^{d^*} w_{kl}\{\epsilon_{t_1}^\top U_{t_1k}\epsilon_{t_1}-tr(U_{t_1k})\} \epsilon_{t_2}^\top V_{t_2l}\epsilon_{t_2}-T^{-1}(n+2+\bar\Delta)r$.
Obviously, we have $T_{nT,r}\in\mF_r$.
Then, set $\Delta_{nT,r}=T_{nT,r}-T_{nT,r-1}$ with $\Delta_{nT,0}=0$. One can easily verify that
$E(\Delta_{nT,r}|\mF_q)=0$ and $E(T_{nT,r}|\mF_q)=T_{nT,q}$ for any $q<r$. This implies that,
for an arbitrary fixed $T$, $\{\Delta_{nT,r}, 0\leq r\leq T\}$ is a martingale difference sequence with
respect to $\{\mF_r, 0\leq r\leq T\}$ and $\mF_0=\emptyset$. Accordingly, by the Martingale Central
Limit Theorem (Hall and Heyde, 1980), it suffices to show that
\beq (d^*)^{-4}\var\big(\sum_{r=1}^T \sigma^*_{nT,r})\rightarrow_p 0, \mbox{~and~} (d^*)^{-4}\sum_{r=1}^T E(\Delta_{nT,r}^4)\rightarrow 0,\eeq
where $\sigma^*_{nT,r}=E(\Delta_{nT,r}^2|\mF_{r-1})$. We next verify the above two parts separately via the following two steps.
Without loss of generality, we assume that $w_{kk}=1$ and $w_{kl}=0$ for any $k\not= l$ to ease the calculation.

{\sc STEP I.} We begin by showing the first term of (A.1). After simple calculation, we have
$\Delta_{nT,r}=(nT)^{-1}(\epsilon_r^\top\epsilon_r)^2-T^{-1}(n+2+\bar\Delta)+n^{-1}T^{-2}\sum_k\sum_{t<r}\{\epsilon_{t}^\top U_{tk}\epsilon_t-tr(U_{tk})\} \epsilon_r^\top V_{rk}\epsilon_r+
n^{-1}T^{-2}\sum_k\sum_{t<r}\{\epsilon_r^\top U_{rk}\epsilon_r-tr(U_{rk})\} \epsilon_t^\top V_{tk}\epsilon_t$.
Based on the above three components of $\Delta_{nT,r}$, we further obtain
$\sigma^*_{nT,r}=E(\Delta_{nT,r}^2|\mF_{r-1})=\sum_{k=1}^6\Pi_{nT,r(k)}$,
where
\[\Pi_{nT,r(1)}=(nT)^{-2}E[\{(\epsilon_r^\top\epsilon_r)^2-n(n+2+\bar\Delta)\}^2],\]
\[\Pi_{nT,r(2)}=n^{-2}T^{-4}\sum_{k_1,k_2}^{d^*}\sum_{t_1<r}\sum_{t_2<r}\{tr(V_{rk_1})tr(V_{rk_2})+2tr(V_{rk_1}V_{rk_2})+\bar\Delta tr(V_{rk_1}\otimes V_{rk_2})\}\]
\[\times\{\epsilon_{t_1}^\top U_{t_1k_1}\epsilon_{t_1}-tr(U_{t_1k_1})\}
\{\epsilon_{t_2}^\top U_{t_2k_2}\epsilon_{t_2}-tr(U_{t_2k_2})\},\]
\[\Pi_{nT,r(3)}=n^{-2}T^{-4}\sum_{k_1,k_2}^{d^*}\sum_{t_1<r}\sum_{t_2<r}\{2tr(U_{rk_1}U_{rk_2})+\bar\Delta tr(U_{rk_1}\otimes U_{rk_2})\}\epsilon_{t_1}^\top V_{t_1k_1}\epsilon_{t_1}
\epsilon_{t_2}^\top V_{t_2k_2}\epsilon_{t_2},\]
\[\Pi_{nT,r(4)}=n^{-1}T^{-2}\sum_k\sum_{t<r}Q_{r(1)k}\{\epsilon_t^\top U_{tk}\epsilon_t-tr(U_{tk})\},\quad \Pi_{nT,r(5)}=n^{-1}T^{-2}\sum_k\sum_{t<r}Q_{r(2)k}\epsilon_t^\top V_{tk}\epsilon_t \]
\[\mbox{~~and~~}
\Pi_{nT,r(6)}=n^{-2}T^{-4}\sum_{k_1,k_2}\sum_{t_1<r}\sum_{t_2<r}\{2tr(U_{rk_1}V_{rk_2})+\bar\Delta tr(U_{rk_1}\otimes V_{rk_2})\}\{\epsilon_{t_1}^\top U_{t_1k_1}\epsilon_{t_1}-tr(U_{t_1k_1})\}\epsilon_{t_2}^\top V_{t_2k_2}\epsilon_{t_2},\]
where $Q_{r(1)k}=(nT)^{-1}E[\epsilon_r^\top V_{rk}\epsilon_r\{(\epsilon_r^\top\epsilon_r)^2-n(n+2+\bar\Delta)\}]$ and
$Q_{r(2)k}=(nT)^{-1}E[\{\epsilon_r^\top U_{rk}\epsilon_r-tr(U_r)\}\{(\epsilon_r^\top\epsilon_r)^2-n(n+2+\bar\Delta)\}]$.
Accordingly,
to prove the first term of (A.1), it suffices to show that $(d^*)^{-4}\var\{\sum_{r=1}^T\Pi_{nT,r(k)}\}\rightarrow 0$ for $k=1, \cdots, 6$. It is worth noting that
 $\var\{\sum_{r=1}^T\Pi_{nT,r(1)}\}=0$. Hence, we only need to verify five of them.

We first consider $\Pi_{nT,r(2)}$, and decompose it
into two terms $\Pi_{nT,r(2)}\triangleq\Pi_{nT,r(2)}^{(1)}+\Pi_{nT,r(2)}^{(2)}$, where
\[\Pi_{nT,r(2)}^{(1)}=n^{-2}T^{-4}\sum_{k_1, k_2}\sum_{t_1\not=t_2<r}\{tr(V_{rk_1})tr(V_{rk_2})+2tr(V_{rk_1}V_{rk_2})+\bar\Delta tr(V_{rk_1}\otimes V_{rk_2})\}\]
\[\times\{\epsilon_{t_1}^\top U_{t_1k_1}\epsilon_{t_1}-tr(U_{t_1k_1})\}
\{\epsilon_{t_2}^\top U_{t_2k_2}\epsilon_{t_2}-tr(U_{t_2k_2})\}\]
\[\mbox{~and~} \Pi_{nT,r(2)}^{(2)}=n^{-2}T^{-4}\sum_{k_1, k_2}\sum_{t<r}\{tr(V_{rk_1})tr(V_{rk_2})+2tr(V_{rk_1}V_{rk_2})+\bar\Delta tr(V_{rk_1}\otimes V_{rk_2})\}\]
\[\times \{\epsilon_{t}^\top U_{tk_1}\epsilon_{t}-tr(U_{tk_1})\}
\{\epsilon_{t}^\top U_{tk_2}\epsilon_{t}-tr(U_{tk_2})\}.\]
Using the fact that $E(\Pi_{nT,r(2)}^{(1)})=0$, we can have
\[\var\big(\sum_{r=1}^T\Pi_{nT,r(2)}^{(1)}\big)=E\big(\sum_{r=1}^T\Pi_{nT,r(1)}^{(1)}\big)^2\]
\[=n^{-4}T^{-8}\sum_{k_1, k_2, k_3, k_4}\sum_{r_1=1}^T\sum_{r_2=1}^T \sum_{t_1\not=t_2<r_1}
\sum_{t_3\not=t_4<r_2}\{tr(V_{r_1k_1})tr(V_{r_1k_2})+2tr(V_{r_1k_1}V_{r_1k_2})+\bar\Delta tr(V_{r_1k_1}\otimes V_{r_1k_2})\}\]
\[\times\{tr(V_{r_2k_3})tr(V_{r_2k_4})+2tr(V_{r_2k_3}V_{r_2k_4})+\bar\Delta tr(V_{r_2k_3}\otimes V_{r_2k_4})\}\]
\[\times E\big[\{\epsilon_{t_1}^\top U_{t_1k_1}\epsilon_{t_1}-tr(U_{t_1k_1})\}\{\epsilon_{t_2}^\top U_{t_2k_2}\epsilon_{t_2}-tr(U_{t_2k_2})\}
\{\epsilon_{t_3}^\top U_{t_3k_3}\epsilon_{t_3}-tr(U_{t_3k_3})\}\{\epsilon_{t_4}^\top U_{t_4k_4}\epsilon_{t_4}-tr(U_{t_4k_4})\}\big]\]
\[
=n^{-4}T^{-8}\sum_{k_1, k_2, k_3, k_4}\sum_{r_1=1}^T\sum_{r_2=1}^T\sum_{t_1\not=t_2<\max\{r_1,r_2\}}\{tr(V_{r_1k_1})tr(V_{r_1k_2})+2tr(V_{r_1k_1}V_{r_1k_2})+\bar\Delta tr(V_{r_1k_1}\otimes V_{r_1k_2})\}\]
\[\times \{tr(V_{r_2k_3})tr(V_{r_2k_4})+2tr(V_{r_2k_3}V_{r_2k_4})+\bar\Delta tr(V_{r_2k_3}\otimes V_{r_2k_4})\}\]
\[\times E\big[\{\epsilon_{t_1}^\top U_{t_1k_1}\epsilon_{t_1}-tr(U_{t_1k_1})\}\{\epsilon_{t_1}^\top U_{t_1k_3}\epsilon_{t_1}-tr(U_{t_1k_3})\}\big] E\big[\{\epsilon_{t_2}^\top U_{t_2k_2}\epsilon_{t_2}-tr(U_{t_2k_2})\}\{\epsilon_{t_2}^\top U_{t_2k_4}\epsilon_{t_2}-tr(U_{t_2k_4})\}\big] \]
\[+n^{-4}T^{-8}\sum_{k_1, k_2, k_3, k_4}\sum_{r_1=1}^T\sum_{r_2=1}^T\sum_{t_1\not=t_2<\max\{r_1,r_2\}}\{tr(V_{r_1k_1})tr(V_{r_1k_2})+2tr(V_{r_1k_1}V_{r_1k_2})+\bar\Delta tr(V_{r_1k_1}\otimes V_{r_1k_2})\}\]
\[\times \{tr(V_{r_2k_3})tr(V_{r_2k_4})+2tr(V_{r_2k_3}V_{r_2k_4})+\bar\Delta tr(V_{r_2k_3}\otimes V_{r_2k_4})\}\]
\[E\big[\{\epsilon_{t_1}^\top U_{t_1k_1}\epsilon_{t_1}-tr(U_{t_1k_1})\}\{\epsilon_{t_1}^\top U_{t_1k_4}\epsilon_{t_1}-tr(U_{t_1k_4})\}\big] E\big[\{\epsilon_{t_2}^\top U_{t_2k_2}\epsilon_{t_2}-tr(U_{t_2k_2})\}\{\epsilon_{t_2}^\top U_{t_2k_3}\epsilon_{t_2}-tr(U_{t_2k_3})\}\big]\]
\[
=n^{-4}T^{-8}\times O\{(d^*)^4\}\times T^4\times O(n^4)\times O(n^2)=O\{(d^*)^4n^2T^{-4}\}=o(1),\]
where the last equality follows from the lemma's assumption that $n/T\rightarrow c$ for some finite positive constant $c$ and Condition (C5).
In addition,
\[\var\big(\sum_{r=1}^T\Pi_{nT,r(2)}^{(2)}\big)=n^{-4}T^{-8}\var\big[\sum_{k_1, k_2}\sum_r\sum_{t<r}\{tr(V_{rk_1})tr(V_{rk_2})+2tr(V_{rk_1}V_{rk_2})+\bar\Delta tr(V_{rk_1}\otimes V_{rk_2})\}\]
\[\times \{\epsilon_{t}^\top U_{tk_1}\epsilon_{t}-tr(U_{tk_1})\}\{\epsilon_{t}^\top U_{tk_2}\epsilon_{t}-tr(U_{tk_2})\}\big] \]
\[
\leq n^{-4}T^{-8}\times (d^*)^2T^2 \sum_{k_1, k_2}\sum_r\sum_{t<r}\var\big[\{tr(V_{rk_1})tr(V_{rk_2})+2tr(V_{rk_1}V_{rk_2})+\bar\Delta tr(V_{rk_1}\otimes V_{rk_2})\}\]
\[\times \{\epsilon_{t}^\top U_{tk_1}\epsilon_{t}-tr(U_{tk_1})\}\{\epsilon_{t}^\top U_{tk_2}\epsilon_{t}-tr(U_{tk_2})\}\big].\]
By Lemma 2 (iii), we have
$\var\big[\{\epsilon_{t}^\top U_{tk_1}\epsilon_{t}-tr(U_{tk_1})\}\{\epsilon_{t}^\top U_{tk_2}\epsilon_{t}-tr(U_{tk_2})\}\big]=O(n^2)$ uniformly for any $k_1$ and $k_2$.
As a result, $\var\big(\sum_{r=1}^T\Pi_{nT,r(1)}^{(2)}\big)=n^{-4}T^{-8} \times (d^*)^4\times T^2 \times T^2\times O(n^4)O(n^2)=O\{(d^*)^4 n^2T^{-4}\}=o(1)$.
This, together with the above result, implies that
$\var\{\sum_{r=1}^T\Pi_{nT,r(2)}\}\rightarrow0$.

We next consider $\Pi_{nT,r(3)}$, and analogously decompose $\Pi_{nT,r(3)}$ into two parts
$\Pi_{nT,r(3)}=\Pi_{nT,r(3)}^{(1)}+\Pi_{nT,r(3)}^{(2)}$,
where $\Pi_{nT,r(3)}^{(1)}=
n^{-2}T^{-4}\sum_{k_1, k_2}\sum_{t_1\not=t_2<r}\{2tr(U_{rk_1}U_{rk_2})+\bar\Delta tr(U_{rk_1}\otimes U_{rk_2})\}\epsilon_{t_1}^\top V_{t_1k_1}\epsilon_{t_1}\epsilon_{t_2}^\top V_{t_2k_2}\epsilon_{t_2}$ and
$\Pi_{nT,r(3)}^{(2)}=n^{-2}T^{-4}\sum_{k_1, k_2}\{2tr(U_{rk_1}U_{rk_2})+\bar\Delta tr(U_{rk_1}\otimes U_{rk_2})\}\epsilon_{t}^\top V_{tk_1}\epsilon_{t}\epsilon_{t}^\top V_{tk_2}\epsilon_{t}$.
Note that
\[\var\Big(\sum_{r=1}^T\Pi_{nT,r(3)}^{(1)}\Big)=n^{-4}T^{-8}\var\Big[\sum_{k_1, k_2}\sum_r \sum_{t_1\not=t_2<r}\{2tr(U_{rk_1}U_{rk_2})+\bar\Delta tr(U_{rk_1}\otimes U_{rk_2})\}\epsilon_{t_1}^\top V_{t_1k_1}\epsilon_{t_1}\epsilon_{t_2}^\top V_{t_2k_2}\epsilon_{t_2}\Big] \]
\[\leq n^{-4}T^{-8}\times (d^*)^2T^3\sum_{k_1, k_2}\sum_r \sum_{t_1\not=t_2<r}\{2tr(U_{rk_1}U_{rk_2})+\bar\Delta tr(U_{rk_1}\otimes U_{rk_2})\}^2\var\big(\epsilon_{t_1}^\top V_{t_1k_1}\epsilon_{t_1}
\epsilon_{t_2}^\top V_{t_2k_2}\epsilon_{t_2}\big).\]
By
$\var\big(\epsilon_{t_1}^\top V_{t_1k_1}\epsilon_{t_1}
\epsilon_{t_2}^\top V_{t_2k_2}\epsilon_{t_2}\big)=\var\big(\epsilon_{t_1}^\top V_{t_1k_1}\epsilon_{t_1}\big)
\var\big(
\epsilon_{t_2}^\top V_{t_2k_2}\epsilon_{t_2}\big)=O(n)\times O(n)=O(n^2)$, we further have
$\var\big(\sum_{r=1}^T\Pi_{nT,r(3)}^{(1)}\big)\leq n^{-4}T^{-8} (d^*)^4\times T^3\times T^3 \times O(n^2)O(n^2)=O\{(d^*)^4T^{-2}\}=o(1)$.

By Lemma 2 (iii), we obtain that
$\var(\epsilon_{t}^\top V_{tk_1}\epsilon_{t}\epsilon_{t}^\top V_{tk_2}\epsilon_{t})\leq E\{(\epsilon_{t}^\top V_{tk_1}\epsilon_{t})^2(\epsilon_{t}^\top V_{tk_2}\epsilon_{t})^2\}=O(n^4)$.
Then using Condition (C5), we can have
\[\var\big(\sum_{r=1}^T\Pi_{nT,r(3)}^{(2)}\big)\leq n^{-4}T^{-8}\times (d^*)^2T^2 \sum_{k_1, k_2}\sum_r\sum_{t<r}\{2tr(U_{rk_1}U_{rk_2})+\bar\Delta tr(U_{rk_1}\otimes U_{rk_2})\}^2 \]
\[\times\var(\epsilon_t^\top V_{tk_1}\epsilon_t\epsilon_t^\top V_{tk_2}\epsilon_t)
=n^{-4}T^{-8}\times (d^*)^4\times T^2\times T^2 \times O(n^2)O(n^4)=O\{(d^*)^4n^2T^{-4}\}=o(1)\]
This, in conjunction with the above result, leads to
$\var\{\sum_{r=1}^T\Pi_{nT,r(3)}\}\rightarrow0$.

Applying techniques  similar  to those used in the above proofs, one can verify that
$\var\{\sum_{r=1}^T\Pi_{nT,r(6)}\}\rightarrow0$. Finally, using the results of Bao and Ullah (2010), we can
obtain $Q_{r(1)k}\leq (nT)^{-1}E\{(\epsilon_r^\top\epsilon_t)^3\}=O(nT^{-1})=O(1)$
and $Q_{r(2)k}=O(1)$ uniformly for any $k$, one can also employ techniques similar to those used in the above proofs to demonstrate that
$\var\{\sum_{r=1}^T\Pi_{nT,r(4)}\}\rightarrow0$ and
$\var\{\sum_{r=1}^T\Pi_{nT,r(5)}\}\rightarrow0$.
Consequently, we have completed the proof of the first term of (A.1).

{\sc STEP II.} To verify the second term of (A.1), denote $\Delta_{nT,r}\triangleq\Delta_{nT,r(1)}+\Delta_{nT,r(2)}+\Delta_{nT,r(3)}$,
where $\Delta_{nT,r(1)}=(nT)^{-1}(\epsilon_r^\top\epsilon_r)^2-T^{-1}(n+2+\bar\Delta)$,
$\Delta_{nT,r(2)}=n^{-1}T^{-2}\sum_{k}\sum_{t<r}\{\epsilon_{t}^\top U_{tk}\epsilon_t-tr(U_{tk})\} \epsilon_r^\top V_{rk}\epsilon_r$
and $\Delta_{nT,r(3)}=n^{-1}T^{-2}\sum_k\sum_{t<r}\{\epsilon_r^\top U_{rk}\epsilon_r-tr(U_{rk})\} \epsilon_t^\top V_{tk}\epsilon_t$.
Accordingly,  it  suffices to show that, for $k=1, 2$ and 3,
$\sum_{r=1}^T E(\Delta_{nT,r(k)}^4)\rightarrow 0$.

We first consider $k=1$.
Note that $E\big\{(\epsilon_r^\top\epsilon_r)^2-n(n+2+\bar\Delta)\big\}^4=O(n^6)$ by the results of Bao and Ullah (2010). This, together with the lemma's assumption that
$n/T\rightarrow c$ for some finite positive constant $c$,
implies that
$\sum_{r=1}^T E(\Delta_{nT,r(1)}^4)=(nT)^{-4}\sum_{r=1}^T E\big\{(\epsilon_r^\top\epsilon_r)^2-n(n+2+\bar\Delta)\big\}^4=(nT)^{-4}T\times O(n^6)=o(1)$.

We next consider $k=2$. After algebraic simplification, we have that
 $\sum_{r=1}^T E(\Delta_{nT,r(2)}^4)=n^{-4}T^{-8}\sum_{k_1, k_2, k_3, k_4}\sum_{r=1}^T \sum_{t_1<r}\sum_{t_2<r}\sum_{t_3<r}\sum_{t_4<r}E\big[\{\epsilon_{t_1}^\top U_{t_1k_1}\epsilon_{t_1}-tr(U_{t_1k_1})\} \{\epsilon_{t_2}^\top U_{t_2k_2}\epsilon_{t_2}-tr(U_{t_2k_2})\}\{\epsilon_{t_3}^\top U_{t_3k_3}\epsilon_{t_3}-tr(U_{t_3k_3})\}\{\epsilon_{t_4}^\top U_{t_4k_4}\epsilon_{t_4}-tr(U_{t_4k_4})\} \big]
 E\big[(\epsilon_r^\top V_{rk_1}\epsilon_r)(\epsilon_r^\top V_{rk_2}\epsilon_r)(\epsilon_r^\top V_{rk_3}\epsilon_r)(\epsilon_r^\top V_{rk_4}\epsilon_r)\big]$.
Under the assumption that
$\sup_t\|V_t\|<\infty$, we obtain $E\big[(\epsilon_r^\top V_{rk_1}\epsilon_r)(\epsilon_r^\top V_{rk_2}\epsilon_r)(\epsilon_r^\top V_{rk_3}\epsilon_r)(\epsilon_r^\top V_{rk_4}\epsilon_r)\big]=O(n^4)$, which immediately leads to
\[\sum_{r=1}^T E(\Delta_{nT,r(2)}^4)=T^{-8}\sum_{k_1, k_2, k_3, k_4}\sum_{r=1}^T\sum_{t_1<r}\sum_{t_2<r}\sum_{t_3<r}\sum_{t_4<r}E\big[\{\epsilon_{t_1}^\top U_{t_1k_1}\epsilon_{t_1}-tr(U_{t_1k_1})\} \{\epsilon_{t_2}^\top U_{t_2k_2}\epsilon_{t_2}-tr(U_{t_2k_2})\}\]
\[\times \{\epsilon_{t_3}^\top U_{t_3k_3}\epsilon_{t_3}-tr(U_{t_3k_3})\}\{\epsilon_{t_4}^\top U_{t_4k_4}\epsilon_{t_4}-tr(U_{t_4k_4})\} \big]\times O(1).\]
Furthermore,
\[T^{-8}\sum_{k_1, k_2, k_3, k_4}\sum_{t=1}^T\sum_{t_1<r}\sum_{t_2<r}\sum_{t_3<r}\sum_{t_4<r}E\big[\{\epsilon_{t_1}^\top U_{t_1k_1}\epsilon_{t_1}-tr(U_{t_1k_1})\} \{\epsilon_{t_2}^\top U_{t_2k_2}\epsilon_{t_2}-tr(U_{t_2k_2})\}\{\epsilon_{t_3}^\top U_{t_3k_3}\epsilon_{t_3}-tr(U_{t_3k_3})\}\]
\[\times\{\epsilon_{t_4}^\top U_{t_4k_4}\epsilon_{t_4}-tr(U_{t_4k_4})\} \big]=T^{-8}\sum_{k_1, k_2, k_3, k_4}\sum_{t=1}^T\sum_{t<r}E\big[\{\epsilon_{t}^\top U_{tk_1}\epsilon_{t}-tr(U_{tk_1})\}\{\epsilon_{t}^\top U_{tk_2}\epsilon_{t}-tr(U_{tk_2})\}\]
\[\times
\{\epsilon_{t}^\top U_{tk_3}\epsilon_{t}-tr(U_{tk_3})\}\{\epsilon_{t}^\top U_{tk_4}\epsilon_{t}-tr(U_{tk_4})\}\big]\]
\[+T^{-8}\sum_{k_1, k_2, k_3, k_4}\sum_{t=1}^T\sum_{t_1<r, t_2<r, t_1\not=t_2}E\big[\{\epsilon_{t_1}^\top U_{t_1k_1}\epsilon_{t_1}-tr(U_{t_1k_1})\}\{\epsilon_{t_1}^\top U_{t_1k_2}\epsilon_{t_1}-tr(U_{t_1k_2})\}\big]\]
\[\times E\big[\{\epsilon_{t_2}^\top U_{t_2k_3}\epsilon_{t_2}-tr(U_{t_2k_3})\}\{\epsilon_{t_2}^\top U_{t_2k_4}\epsilon_{t_2}-tr(U_{t_2k_4})\}\big] \]
\[+T^{-8}\sum_{k_1, k_2, k_3, k_4}\sum_{t=1}^T\sum_{t_1<r, t_2<r, t_1\not=t_2}E\big[\{\epsilon_{t_1}^\top U_{t_1k_1}\epsilon_{t_1}-tr(U_{t_1k_1})\}\{\epsilon_{t_1}^\top U_{t_1k_2}\epsilon_{t_1}-tr(U_{t_1k_2})\}\big]\]
\[\times E\big[\{\epsilon_{t_2}^\top U_{t_2k_3}\epsilon_{t_2}-tr(U_{t_2k_3})\}\{\epsilon_{t_2}^\top U_{t_2k_4}\epsilon_{t_2}-tr(U_{t_2k_4})\}\big] \]
\[+T^{-8}\sum_{k_1, k_2, k_3, k_4}\sum_{t=1}^T\sum_{t_1<r, t_2<r, t_1\not=t_2}E\big[\{\epsilon_{t_1}^\top U_{t_1k_1}\epsilon_{t_1}-tr(U_{t_1k_1})\}\{\epsilon_{t_1}^\top U_{t_1k_2}\epsilon_{t_1}-tr(U_{t_1k_2})\}\big]\]
\[\times E\big[\{\epsilon_{t_2}^\top U_{t_2k_3}\epsilon_{t_2}-tr(U_{t_2k_3})\}\{\epsilon_{t_2}^\top U_{t_2k_4}\epsilon_{t_2}-tr(U_{t_2k_4})\}\big]
\triangleq
\Pi_1+\Pi_2+\Pi_3+\Pi_4.\]
By Lemma 2 (iii), we have
$E\big[\{\epsilon_{t}^\top U_{tk_1}\epsilon_{t}-tr(U_{tk_1})\}\{\epsilon_{t}^\top U_{tk_2}\epsilon_{t}-tr(U_{tk_2})\}
\{\epsilon_{t}^\top U_{tk_3}\epsilon_{t}-tr(U_{tk_3})\}\{\epsilon_{t}^\top U_{tk_4}\epsilon_{t}-tr(U_{tk_4})\}\big]=O(n^2)$. Thus,
$\Pi_1=T^{-8}\times T(T-1)\times O(n^2)\times (d^*)^4=o(1)$.
By Lemma 2 (ii), we have $E\big[\{\epsilon_{t_1}^\top U_{t_1k_1}\epsilon_{t_1}-tr(U_{t_1k_1})\}\{\epsilon_{t_1}^\top U_{t_1k_2}\epsilon_{t_1}-tr(U_{t_1k_2})\}\big]=O(n)$. As a result,
$\Pi_2=T^{-8}\times T(T-1)^2\times O(n^2)\times (d^*)^2=o(1)$. This, in conjunction with the above results, implies that
$\sum_{r=1}^T E(\Delta_{nT,r(2)}^4)\rightarrow 0$.
Analogously, we can demonstrate that $\sum_{r=1}^T E(\Delta_{nT,r(3)}^4)\rightarrow 0$, which completes the entire proof.

\csection{Proofs of Theorems 3--5}

\noindent {\textbf{Proof of Theorem 3}}:
To prove the theorem, we consider the following two cases:
(i) $\mS$ is underfitted; (ii) $\mS$ is overfitted.

 \textbf{CASE I: Underfitted models (i.e., $\mS\in \mA_1$). }
In this case, we need to prove $P\big\{\min_{\mS\in \mA_1}\mbox{EBIC}(\mS)\leq \mbox{EBIC}(\mS_T)\big\}\to 0$. Note that
$|\mS|-|\mS_T|\geq -|\mS_T|\geq -q$ and $\ell(\hat{\theta}_{\mS_T})\geq \ell({\theta}_{0\mS_T})$. Then, we have
\begin{align*}
&P\big\{\min_{\mS\in \mA_1}\mbox{EBIC}(\mS)\leq \mbox{EBIC}(\mS_T)\big\}\\
=& P\big\{\min_{\mS\in \mA_1}\ell(\hat\theta_{\mS})-\ell(\hat{\theta}_{\mS_T})\geq \{\log(nT)+\gamma\log(d)\}(|\mS|-|\mS_T|)/2 \big\}\\
\leq & P\big\{\min_{\mS\in \mA_1}\ell(\hat\theta_{\mS})-\ell(\hat{\theta}_{\mS_T}) \geq -q/2(\log(nT)+\gamma\log(d)) \big\}\\
\leq & P\big\{\min_{\mS\in \mA_1}\ell(\hat\theta_{\mS})-\ell({\theta}_{0\mS_T}) \geq -q/2(\log(nT)+\gamma\log(d)) \big\}.
\end{align*}
Accordingly, it suffices to show
$P\big\{\min_{\mS\in \mA_1}\ell(\hat\theta_{\mS})-\ell({\theta}_{0\mS_T}) \geq -q/2(\log(nT)+\gamma\log(d)) \big\}\to 0$.

For any $\mS\in \mA_1$, let $\bar{\mS}=\mS\cup \mS_T$. We then
consider those $\tilde{\theta}_{\bar{\mS}}$ such that $\|\tilde{\theta}_{\bar{\mS}}-\theta_{0\bar{\mS}}\|=\rho_{nT}$ with $\rho_{nT}\to 0$ and $\rho_{nT}\sqrt{nT/\log(nT)}\to \infty$.
By a Taylor series expansion,
we have
\begin{align*}
\ell(\tilde{\theta}_{\bar{\mS}})-\ell(\theta_{0\bar{\mS}})=(\tilde{\theta}_{\bar{\mS}}-\theta_{0\bar{\mS}})^{\top}\frac{\partial\ell(\theta_{0\bar{\mS}})}{\partial\theta_{\bar{\mS}}}-\frac{1}{2}(\tilde{\theta}_{\bar{\mS}}-\theta_{0\bar\mS})^{\top} \Big\{-\frac{\partial^2 \ell(\tilde{\theta}^{*}_{\bar{\mS}})}{\partial\tilde{\theta}^{*}_{\bar{\mS}}\partial\tilde{\theta}_{\bar{\mS}}^{*\top}}\Big\}(\tilde{\theta}_{\bar{\mS}}-\theta_{0\bar{\mS}}),
\end{align*}
where $\tilde{\theta}^{*}_{\bar{\mS}}$ lies between $\tilde{\theta}_{\bar\mS}$ and $\theta_{0\bar\mS}$. By Condition (C7),
$$
\ell(\tilde{\theta}_{\bar{\mS}})-\ell(\theta_{0\bar{\mS}})\leq (\tilde{\theta}_{\bar{\mS}}-\theta_{0\bar{\mS}})^{\top}\frac{\partial\ell(\theta_{0\bar{\mS}})}{\partial\theta_{\bar{\mS}}} -\frac{c_{\min,3}}{2}nT\|\tilde{\theta}_{\bar{\mS}}-\theta_{0\bar{\mS}}\|^2\leq \rho_{nT} \big\|\frac{\partial\ell(\theta_{0\bar{\mS}})}{\partial\theta_{\bar{\mS}}} \big\|- \frac{c_{\min,3}}{2}nT\rho_{nT}^2.
$$

We next prove that $\sup_{\bar{\mS}}\|{\partial\ell(\theta_{0\bar{\mS}})}/{\partial\theta_{\bar{\mS}}}\|=O_p\big\{\sqrt{nT\log(nT)}\big\}$. According to Lemma 3 (ii),
for sufficiently large constant $c_l$,
\begin{align}
& P\Big(\sup_{\bar{\mS}}\big\|\frac{\partial\ell(\theta_{0\bar{\mS}})}{\partial\theta_{\bar{\mS}}}\big\|>c_l\sqrt{nT\log(nT)}\Big)\leq \sum_{\bar{\mS}}P\Big(\big\|\frac{\partial\ell(\theta_{0\bar{\mS}})}{\partial\theta_{\bar{\mS}}}\big\|>c_l\sqrt{nT\log(nT)}\Big) \nonumber\\
& \leq \sum_{\bar{\mS}}\sum_{k=1}^{|\bar\mS|}P\Big(\big|\frac{\partial\ell(\theta_{0\bar{\mS}})}{\partial\theta_{\bar{\mS},k}}\big|>c_l\sqrt{nT\log(nT)/|\bar{\mS}|}\Big)\nonumber\\
& \leq \sum_{\bar{\mS}}\sum_{k=1}^{|\bar\mS|}2\exp\Bigg[{-\min\Big({\frac{\tau_1c_l\sigma_0^2\sqrt{\log(nT)/|\bar{\mS}|}}{\|U_k\|/\sqrt{nT}},\frac{\tau_2c_l^2\sigma_0^4\log(nT)/|\bar\mS|}
{\|U_k\|_F^2/nT}}\Big)}\Bigg].
\end{align}
By Conditions (C2) and (C3),  we have $\|U_k\|=\sup_{t}\|W^{(t)}_k\Delta_{t}^{-1}(\lambda_{0\bar{\mS}})\|\leq \sup_{t}\|W^{(t)}_k\|\|\Delta_{t}^{-1}(\lambda_{0\bar{\mS}})\|<\infty$ and $\|U_k\|_F^2=O(nT)$ uniformly for any $k\leq |\bar\mS|$.
Hence,
$$\frac{\tau_1c_l\sigma_0^2\sqrt{\log(nT)/|\bar{\mS}|}}{\|U_k\|/\sqrt{nT}}=O\Big\{\sqrt{nT\log(nT)}\Big\}>\frac{\tau_2c_l^2\sigma_0^4\log(nT)/|\bar\mS|}{\|U_k\|_F^2/nT}=O\{\log(nT)\}.$$
Accordingly, for sufficiently large constants $c_l$ and $c_L$, we have
$$
P\lsk\sup_{\bar{\mS}}\big\|\frac{\partial\ell(\theta_{0\bar{\mS}})}{\partial\theta_{\bar{\mS}}}\big\|>c_l\sqrt{nT\log(nT)}\rsk\leq 2q\exp\Big\{-c_L\log(nT)+q\log(d)\Big\}\to 0,
$$
which immediately leads to $\sup_{\bar{\mS}}\|{\partial\ell(\theta_{0\bar{\mS}})}/{\partial\theta_{\bar{\mS}}}\|=O_p\big\{\sqrt{nT\log(nT)}\big\}$.

Using the above result, we obtain that, for $\|\tilde{\theta}_{\bar{\mS}}-\theta_{0\bar{\mS}}\|=\rho_{nT}$,
$$
\ell(\tilde{\theta}_{\bar{\mS}})-\ell(\theta_{0\bar{\mS}})\leq \rho_{nT}\Big\{c_l\sqrt{nT\log(nT)}-\frac{c_{\min,3}}{2}\rho_{nT}{nT}\Big\}
$$
uniformly over $\bar{\mS}$ and $\tilde{\theta}_{\bar\mS}$ with probability  tending to 1. Since $\rho_{nT}\sqrt{nT/\log(nT)}\to \infty$,  we have $\sqrt{nT\log(nT)}=o(\rho_{nT}nT)$. Then $\ell(\tilde{\theta}_{\bar{\mS}})-\ell(\theta_{0\bar{\mS}})\leq -c_l^{'}\rho_{nT}^2nT$,
where $c_l^{'}$ is a positive constant.
In addition, the concavity of $\ell(\cdot)$ in a neighborhood of $\theta_{0\bar{S}}$ implies that
$$
\sup\Big\{\ell(\tilde{\theta}_{\bar{\mS}})-\ell(\theta_{0\bar{\mS}}): \|\tilde{\theta}_{\bar{\mS}}-\theta_{0\bar{\mS}}\|\geq \rho_{nT} \Big\}\leq
\sup\Big\{\ell(\tilde{\theta}_{\bar{\mS}})-\ell(\theta_{0\bar{\mS}}): \|\tilde{\theta}_{\bar{\mS}}-\theta_{0\bar{\mS}}\|=\rho_{nT}\Big\}\leq  -c_l^{'}\rho_{nT}^2nT
$$
with probability approaching 1. Let $\breve{\theta}_{\bar\mS}$ be $\hat{\theta}_{\mS}$ augmented with zeros corresponding to the elements in $\bar\mS/\mS$, and
let $\rho_{nT}=\min_{k\in \mS_{T}}|\lambda_{0k}|$. It can be shown that $\|\breve{\theta}_{\bar\mS}-\hat{\theta}_{\mS}\|\geq\rho_{nT}$.
Therefore, $\ell(\hat\theta_{\mS})-\ell(\theta_{0\mS_T})\leq -c_l^{'}\rho_{nT}^2nT$ uniformly over $\mS\in \mA_{1}$. Since
 $\rho_{nT}\sqrt{nT/\log(nT)}\to\infty$ and $-c_l^{'}\rho_{nT}^2nT< -q/2\big\{\log(nT)+\gamma\log(d)\big\}$ for large $nT$.
 The above results imply $P\big\{\min_{\mS\in \mA_1}\ell(\hat\theta_{\mS})-\ell({\theta}_{0\mS_T}) \geq -q/2(\log(nT)+\gamma\log(d)) \big\}\to 0$, which completes the first part of the proof.

 \textbf{CASE II: Overfitted models (i.e., $\mS\in \mA_0,\mS\neq \mS_T$).}
For any $\mS\in\mA_{0}$, let $m=|\mS|-|\mS_{T}|$.
 By definition, $\mbox{EBIC}(\mS)\leq  \mbox{EBIC}(\mS_T)$ if and only if $\ell(\hat\theta_{\mS})-\ell(\hat\theta_{\mS_T})\geq m/2\big\{\log(nT)+\gamma
\log(d)\big\}$. For sufficiently large $nT$,
 \begin{align*}
\ell(\hat{\theta}_{\mS})-\ell(\hat{\theta}_{\mS_T})\leq \ell(\hat{\theta}_{\mS})-\ell(\theta_{0\mS})=
 (\hat{\theta}_{\mS}-\theta_{0\mS})^{\top}\frac{\partial\ell(\theta_{{0\mS}})}{\partial\theta_{{\mS}}}-\frac{1}{2}(\hat{\theta}_{{\mS}}-\theta_{0\mS})^{\top} \Big(-\frac{\partial^2 \ell({\theta}^{*}_{{\mS}})}{\partial{\theta}^{*}_{{\mS}}\partial{\theta}_{{\mS}}^{*\top}}\Big)(\hat{\theta}_{{\mS}}-\theta_{{0\mS}}),
  \end{align*}
where ${\theta}^{*}_{{\mS}}$ lies between $\hat{\theta}_{\mS}$ and $\theta_{0\mS}$.
Define $H(\theta^{*}_{\mS})=-\frac{1}{nT}\frac{\partial^2 \ell({\theta}^{*}_{{\mS}})}{\partial{\theta}^{*}_{{\mS}}\partial{\theta}_{{\mS}}^{*\top}}$. By Condition (C7), we have
$$
\ell(\hat{\theta}_{\mS})-\ell({\theta}_{0\mS_T})=\frac{1}{2nT}\frac{\partial\ell^{\top}(\theta_{{0\mS}})}{\partial\theta_{\mS}}(H(\theta^{*}_{\mS}))^{-1}
\frac{\partial\ell(\theta_{{0\mS}})}{\partial\theta_{\mS}}\leq \frac{1}{2nTc_{\min,3}}\|\frac{\partial\ell(\theta_{{0\mS}})}{\partial\theta_{\mS}}\|^2.
$$
Then, employing similar techniques to those used for obtaining (A.4) in the proof of CASE I,
we have
\begin{align*}
 & P\Bigg\{ \frac{1}{2nTc_{\min,3}}\sup_{{\mS}\in \mA_{0},\mS\neq \mS_T}\|\frac{\partial\ell(\theta_{0\mS})}{\partial\theta_{\mS}}\|^2\geq {m/2\big(\log(nT)+\gamma\log(d)\big)}\Bigg\}\\
 \leq & \sum_{{\mS}}\sum_{k=1}^{|\mS|} 2\exp\lmk{-\frac{\tau_2c_{\min,3}m\sigma_0^4(\log(nT)+\gamma\log(d))/|\mS|}{\|U_k\|_F^2/nT}}\rmk.
\end{align*}
Since $\|U_k\|_F^2/nT\leq\|U_k\|^2\leq \big(\sup_{t\leq T}\sup_{n\geq1}\|W_k^{(t)}\|\|\Delta^{-1}(\lambda_0)\|\big)^2\leq C_w^2$, we further have that
\begin{align*}
&2\exp\lmk{-\tau_2c_{\min,3}m\sigma_0^4(\log(nT)+\gamma\log(d))/(|\mS|\|U_k\|_F^2/nT)}\rmk
\\
\leq &2\exp\lmk{-\tau_2c_{\min,3}m\sigma_0^4(\log(nT)+\gamma\log(d))/qC_w^2}\rmk.
\end{align*}
The above results imply that
\begin{align*}
& P\Big\{ \frac{1}{2nTc_{\min,3}}\sup_{{\mS}\in \mA_{0},\mS\neq \mS_T}\|\frac{\partial\ell(\theta_{0\mS})}{\partial\theta_{\mS}}\|^2\geq {m/2\big(\log(nT)+\gamma\log(d)\big)}\Big\}\\
& \leq 2q\exp\lmk{-\tau_2c_{\min,3}m\sigma_0^4(\log(nT)+\gamma\log(d))/qC_w^2+m\log(d)}\rmk\\
& \leq 2q \exp\lmk-\tau_2c_{\min,3}m\sigma_0^4/(qC_w^2)\{4+\gamma-qC_w^2/(\tau_2c_{\min,3}\sigma_0^4)\}\log(d)  \rmk\to 0
\end{align*}
for $\gamma>qC_w^2/(\tau_2c_{\min,3}\sigma_0^4)-4$. Accordingly, we have
\begin{align*}
&P\Big\{\mbox{EBIC}(\mS)\leq \mbox{EBIC}(\mS_T)\Big\}=
P\Big\{ \sup_{\mS\in \mA_{0}, \mS\neq \mS_T}\ell(\hat\theta_{\mS})-\ell(\hat\theta_{\mS_T})\geq m/2\big\{\log(nT)+\gamma
\log(d)\big\}\Big\}\\
& \leq P\Big\{ \frac{1}{2nTc_{\min,3}}\sup_{{\mS}\in \mA_{0},\mS\neq \mS_T}\|\frac{\partial\ell(\theta_{0\mS})}{\partial\theta_{\mS}}\|^2\geq {m/2\big(\log(nT)+\gamma\log(d)\big)}\Big\}\to 0,
\end{align*}
which completes the entire proof.

\noindent {\textbf{Proof of Theorem 4}}:
To prove the theorem, we decompose $T_{ql}$ into the following three parts,
\[T_{ql}=(nT)^{-1}\sum_{t=1}^T tr(Y_t Y_t^\top\Sigma_t^{-1}-I_n)^2+2(nT)^{-1}\sum_{t=1}^T tr \big\{(Y_t Y_t^\top\Sigma_t^{-1}-I_n)(Y_tY_t^\top\hat\Sigma_t^{-1}-Y_tY_t^\top
\Sigma_t^{-1})\big\} \]\[+(nT)^{-1}\sum_{t=1}^T tr\big\{Y_tY_t^\top\hat\Sigma_t^{-1}-Y_tY_t^\top\Sigma_t^{-1})\big\}^2\triangleq T_{ql}^{(1)}+2T_{ql}^{(2)}+T_{ql}^{(3)}.\]
Then we consider the following three steps: (i) STEP I, showing that $T_{ql}^{(3)}$ is negligible;
(ii) STEP II, showing that $T_{ql}^{(2)}$ is of order $O_p(d)$
and then employing a Taylor series expansion to derive the leading term of $T_{ql}^{(2)}$ as a combination of quadratic forms of the random error $\epsilon_t$,
which directly connects to the result of Lemma 4; (iii) STEP III, demonstrating the asymptotic normality of $T_{ql}$ based on the results of Lemma 4.

{STEP I:} Note that
$T_{ql}^{(3)}=(nT)^{-1}\sum_{t=1}^T \{Y_t^\top(\hat\Sigma_t^{-1}-\Sigma_t^{-1})Y_t\}^2\leq \max_t \|\hat\Sigma_t^{-1}-\Sigma_t^{-1}\|^2\times (nT)^{-1}\sum_{t=1}^T (Y_t^\top Y_t)^2$.
By Condition (C3), we obtain $Y_t^\top Y_t=\epsilon_t^\top \{\Delta_t^{-1}({\lambda_0})\}^\top \Delta_t^{-1}({\lambda_0}) \epsilon_t\leq \max_t \|\{\Delta_t^{-1}({\lambda_0})\}^\top \Delta_t^{-1}({\lambda_0})\|\times \epsilon_t^\top \epsilon_t
=O(1)\times \epsilon_t^\top \epsilon_t=O_p(n)$.
Then $(nT)^{-1}\sum_{t=1}^T (Y_t^\top Y_t)^2=O_p(n)$.
In addition,
$\hat\Sigma_t^{-1}=(\hat\sigma^2)^{-1}\Delta_t^\top(\hat\lambda)\Delta_t(\hat\lambda)$
and $\Sigma_t^{-1}=({\sigma^2_0})^{-1}\Delta_t^\top({\lambda_0})\Delta_t({\lambda_0})$. Thus,
$\hat\Sigma_t^{-1}-\Sigma_t^{-1}=({\sigma^2_0})^{-1}\big\{\Delta_t^\top(\hat\lambda)\Delta_t(\hat\lambda)-\Delta_t^\top({\lambda_0})\Delta_t({\lambda_0})\big\}+
\{(\hat\sigma^2)^{-1}-({\sigma^2_0})^{-1}\}\Delta_t^\top(\hat\lambda)\Delta_t(\hat\lambda)
\triangleq \Delta_{11t}+\Delta_{12t}$.
Note that
${\|}\Delta_{11t}{\|}={\|}({\sigma^2_0})^{-1}\big\{\Delta_t^\top(\hat\lambda)\Delta_t(\hat\lambda)-\Delta_t^\top(\hat\lambda)\Delta_t({\lambda_0})+\Delta_t^\top(\hat\lambda)\Delta_t({\lambda_0})-\Delta_t^\top({\lambda_0})\Delta_t({\lambda_0})\big\}{\|}
\leq ({\sigma^2_0})^{-1}\|\Delta_t^\top(\hat\lambda)\|\times \|{\Delta_t(\hat\lambda)}-\Delta_t({\lambda_0})\|+({\sigma^2_0})^{-1}\|\Delta_t({\lambda_0})\|\times \|\Delta_t(\hat\lambda)-\Delta_t({\lambda_0})\|\leq
({\sigma^2_0})^{-1}(\|\Delta_t^\top(\lambda)\|+\|\Delta_t(\hat\lambda)-\Delta_t({\lambda_0})\|)\times \|{\Delta_t(\hat\lambda)}-\Delta_t({\lambda_0})\|+({\sigma^2_0})^{-1}\|\Delta_t({\lambda_0})\|\times \|\Delta_t(\hat\lambda)-\Delta_t({\lambda_0})\|$.
By Condition (C3), we have $\max_t\|\Delta_t^\top(\lambda)\|=O(1)$.
Moreover, by the results of Theorem 2, $\max_t\|\Delta_t(\hat\lambda)-\Delta_t({\lambda_0})\|=O_p\{d(nT)^{-1/2}\}$,
which leads to
$\max_t{\|}\Delta_{11t}{\|}\leq ({\sigma^2_0})^{-1}\max_t(\|\Delta_t^\top(\lambda)\|+\|\Delta_t(\hat\lambda)-\Delta_t({\lambda_0})\|)\times \|{\Delta_t(\hat\lambda)}-\Delta_t({\lambda_0})\|+({\sigma^2_0})^{-1}\max_t\|\Delta_t({\lambda_0})\|\times \|\Delta_t(\hat\lambda)-\Delta_t({\lambda_0})\|=O_p\big\{\|\Delta_t(\hat\lambda)-\Delta_t({\lambda_0})\|\big\}=O_p\{d(nT)^{-1/2}\}$.
Analogously, we can show that $\max_t\Delta_{12t}=O_p\{d(nT)^{-1/2}\}$.
Combining the above results, we have $\max_t\|\hat\Sigma_t-\Sigma_t\|=O_p\{d(nT)^{-1/2}\}$, which leads to
$T_{ql}^{(3)}\leq \max_t \|\hat\Sigma_t^{-1}-\Sigma_t^{-1}\|^2\times (nT)^{-1}\sum_{t=1}^T (Y_t^\top Y_t)^2=O_p\{d(nT)^{-1}\}\times O_p(n)=o_p(1)$. The proof is complete.

{STEP II:}
Note that $T_{ql}^{(2)}=(nT)^{-1}\sum_{t=1}^T Y_t^\top(\hat\Sigma_t^{-1}-\Sigma_t^{-1})Y_t\{Y_t^\top\Sigma_t^{-1}Y_t-1\}$. In addition, $Y_t^\top\Sigma_t^{-1}Y_t=\epsilon_t^\top\epsilon_t=n{\sigma^2_0}\{1+o_p(1)\}=O_p(n)$. By the theorem's assumption that $n/T\rightarrow c>0$, we  have
$Y_t^\top(\hat\Sigma_t^{-1}-\Sigma_t^{-1})Y_t\leq \varrho_{\max}(\hat\Sigma_t^{-1}-\Sigma_t^{-1})Y_t^\top Y_t=O_p\{d(nT)^{-1/2}\}\times O_p(n)=O_p(dn^{1/2}T^{-1/2})=O_p(d)$.  Accordingly,
$T_{ql}^{(2)}=O_p(d)$, and it is non-negligible. Hence, we need to evaluate the mean
and variance of $T_{ql}^{(2)}$ for the bias-correction. However, $T_{ql}^{(2)}$ involves the
estimator $\hat\lambda$, which does not have a closed form. As a result,  it is not plausible to
obtain explicit formulas for $E\{T_{ql}^{(2)}\}$ and $\var\{T_{ql}^{(2)}\}$. Thus,
we apply a Taylor series expansion with respect to $\hat\theta=(\hat\lambda^{\top},\hat\sigma^2)^{\top}$  given below to obtain the leading term of $T_{ql}^{(2)}$.

Let $\widetilde T_{ql}^{(2)}=T^{-1}\sum_{t=1}^T Y_t^\top(\hat\Sigma_t^{-1}-\Sigma_t^{-1})Y_t.$ We then obtain
$T_{ql}^{(2)}=\widetilde T_{ql}^{(2)}
+o_p(1)$. Thus, we only need to consider $\widetilde T_{ql}^{(2)}$
in the rest of proof.
Define $\Sigma_t^{-1}\triangleq\Sigma_t^{-1}({\theta_0})=({\sigma^2_0})^{-1}\Delta_t^\top({\lambda_0})\Delta_t({\lambda_0})$ and
$\hat\Sigma_t^{-1}\triangleq\Sigma_t^{-1}(\hat\theta)=(\hat\sigma^2)^{-1}\Delta_t^\top(\hat\lambda)\Delta_t(\hat\lambda)$.
By Theorem 1, we have
$\hat\theta-\theta_0=\mathcal{I}^{-1}({\theta_0})(nT)^{-1}\frac{\partial \ell(\theta_0)}{\partial \theta} \{1+o_p(1)\}$. Furthermore, employing a Taylor series expansion, we have
\[\widetilde T_{ql}^{(2)}=T^{-1}\sum_{t=1}^T \mbox{vec}^\top\{\Sigma_t^{-1}(\hat\theta)-\Sigma_t^{-1}({\theta_0})\} \mbox{vec}(Y_t Y_t^\top)\]
\[=T^{-1}\sum_{t=1}^T(\hat\theta-\theta_0)^\top \frac{\partial \mbox{vec}^\top\{\Sigma_t^{-1}({\theta_0})\}}{\partial \theta}\mbox{vec}(Y_t Y_t^\top)+o_p(1)\]
\[=n^{-1}T^{-2}\sum_{t=1}^T \frac{\partial^\top\ell(\theta_0)}{\partial \theta} \mathcal{I}^{-1}({\theta_0}) \Lambda_t^\top \mbox{vec}(Y_t Y_t^\top)+o_p(1),\]
where $\Lambda_t=\partial \mbox{vec}\{\Sigma_t^{-1}({\theta_0})\}/\partial \theta=(\partial \mbox{vec}\{\Sigma_t^{-1}({\theta_0})\}/\partial \lambda_1, \cdots, \partial \mbox{vec}\{\Sigma_t^{-1}({\theta_0})\}/\partial \sigma^2)\in\mR^{n^2\times (d+1)}$.
It is worth noting that
\[\Lambda_t^\top \mbox{vec}(Y_t Y_t^\top)=(Y_t^\top\widetilde\Lambda_{t1} Y_t, \cdots, Y_t^\top\widetilde\Lambda_{t(d+1)}Y_t)^\top
\]
\[
=(\epsilon_t^\top\{\Delta_t^{-1}({\lambda_0})\}^\top\widetilde\Lambda_{t1} \Delta_t^{-1}({\lambda_0}) \epsilon_t, \cdots, \epsilon_t^\top\{\Delta_t^{-1}({\lambda_0})\}^\top\widetilde\Lambda_{t(d+1)}\Delta_t^{-1}({\lambda_0}) \epsilon_t)^\top\in\mR^{d+1},\]
where $\widetilde\Lambda_{tk}$ is the matrix form of $\partial \mbox{vec}\{\Sigma_t^{-1}({\theta_0})\}/\partial \theta_k$ for $k=1,\cdots, d+1$.
In addition,
$\frac{\partial \ell(\theta_0)}{\partial \theta}=\sum_{t=1}^T (\epsilon_t^\top U_{t1}\epsilon_t/{\sigma^2_0}-tr(U_{t1}), \cdots, \epsilon_t^\top U_{t(d+1)}\epsilon_t/{\sigma^2_0}-tr(U_{t(d+1)}))^\top$,
where $U_{tk}=s\{W_k^{(t)}\Delta_t^{-1}({\lambda_0})\}$, for $k=1,\cdots, d$, and $U_{t(d+1)}=I_n/(2{\sigma^2_0})$.
Accordingly,
\[\widetilde T_{ql}^{(2)}=n^{-1}T^{-2}\sum_{t_1=1}^T\sum_{t_2=1}^T \Big(\epsilon_{t_1}^\top U_{t_11}\epsilon_{t_1}/{\sigma^2_0}-tr(U_{t_11}), \cdots, \epsilon_{t_1}^\top U_{t_1(d+1)}\epsilon_{t_1}/{\sigma^2_0}-tr(U_{t_1(d+1)})\Big)\mathcal{I}^{-1}({\theta_0}) \]
\[\times (\epsilon_{t_2}^\top V_{t_21} \epsilon_{t_2}, \cdots, \epsilon_{t_2}^\top V_{t_2(d+1)} \epsilon_{t_2})^\top+o_p(1),\]
where $V_{tk}=\{\Delta_t^{-1}({\lambda_0})\}^\top\widetilde\Lambda_{tk} \Delta_t^{-1}({\lambda_0})$ for $k=1,\cdots, d+1$.
Consequently, the leading term of $T_{ql}^{(2)}$ is a combination of quadratic forms
of the random errors $\epsilon_t$, which allows us to directly apply the results from
Lemma 4 and complete this part of the proof.

{STEP III:}
Note that
$T_{ql}^{(1)}=(nT)^{-1}\sum_{t=1}^T tr(\epsilon_t\epsilon_t^\top/{\sigma^2_0}-I_n)^2=(nT)^{-1}\sum_t (\epsilon_t^\top\epsilon_t/{\sigma^2_0})^2+(nT)^{-1}\sum_t(n-2\epsilon_t^\top\epsilon_t/{\sigma^2_0})\triangleq
T_{ql}^{(11)}+T_{ql}^{(12)}$.
Since $E\{T_{ql}^{(12)}\}=-1$ and $\var\{T_{ql}^{(12)}\}=(nT)^{-2}\sum_t \var(\epsilon_t^\top\epsilon_t/{\sigma^2_0})=O\{(nT)^{-1}\}=o(1)$, we obtain that  $T_{ql}^{(1)}=T_{ql}^{(11)}-1+o_p(1)$.
This, together with the results from STEPs I--II, leads to
\[T_{ql}=T_{ql}^{(11)}+2\widetilde T_{ql}^{(2)}-1+o_p(1)
=(nT)^{-1}\sum_{t=1}^T (\epsilon_t^\top\epsilon_t/{\sigma^2_0})^2-1\]
\[+2n^{-1}T^{-2}\sum_{t_1=1}^T\sum_{t_2=1}^T \Big(\epsilon_{t_1}^\top U_{t_11}\epsilon_{t_1}/{\sigma^2_0}-tr(U_{t_11}), \cdots, \epsilon_{t_1}^\top U_{t_1(d+1)}\epsilon_{t_1}/{\sigma^2_0}-tr(U_{t_1(d+1)})\Big)\times\]\[ \mathcal{I}^{-1}({\theta_0})
(\epsilon_{t_2}^\top V_{t_21} \epsilon_{t_2}, \cdots, \epsilon_{t_2}^\top V_{t_2(d+1)} \epsilon_{t_2})^\top+o_p(1)\]
\[=(nT)^{-1}\sum_{t=1}^T (\epsilon_t^\top\epsilon_t/{\sigma^2_0})^2-1+2n^{-1}T^{-2}\sum_{t_1=1}^T\sum_{t_2=1}^T \sum_{k,l}^{d+1} \mathcal{I}_{kl}^{-1}({\theta_0})\{\epsilon_{t_1}^\top U_{t_1k}\epsilon_{t_1}/{\sigma^2_0}-tr(U_{t_1k})\}\epsilon_{t_2}^\top V_{t_2l} \epsilon_{t_2}+o_p(1)\]
with $\mathcal{I}^{-1}({\theta_0})=(\mathcal{I}_{kl}^{-1}({\theta_0}))$.
One can further apply Condition (C3) to verify that $\max_t\max_k\|U_{tk}\|<\infty$ and $\max_t\max_k\|V_{tk}\|<\infty$. Moreover, $\mathcal{I}^{-1}({\theta_0})$ is a positive definite matrix
with bounded upper eigenvalues by Condition (C4). Finally,
we apply Lemma 4 and demonstrate that $T_{ql}$ is asymptotic normal, which completes the entire proof.

\noindent {\textbf{Proof of Theorem 5}}:
Analogously to the proof of Theorem 1, we first show
\begin{equation}
(nTd)^{-1/2}D_z\frac{\partial \ell_{zc}({\theta_{0zc}})}{\partial\theta_{zc}}\stackrel{d}{\longrightarrow}N\lsk 0,G_z(\theta_{0zc})\rsk.
\end{equation}
After some simple calculations, we obtain that
\[\frac{\partial \ell_{zc}(\theta_{0zc})}{\partial \lambda_k}=\sum_{t=1}^T \big[\upsilon_t^\top \mathcal{U}_k^{(t)}\upsilon_t-\sigma_{0z}^2 tr\big\{\mathcal{U}^{(t)}_k\big\}+\mathcal{V}_k^{(t)\top}\upsilon_t\big],\]
\[ \frac{\partial \ell_{zc}(\theta_{0zc})}{\partial \sigma_z^2}=-\frac{nT}{2\sigma_{0z}^2}+\frac{1}{2\sigma_{0z}^4} \sum_{t=1}^T \upsilon_t^\top\upsilon_t,
\mbox{~and~}\frac{\partial \ell_{zc}(\theta_{0zc})}{\partial \delta}=\frac{1}{\sigma_{0z}^2}\sum_{t=1}^T \mathbf{Z}^{(t)\top}\upsilon_t,\]
where $\mathcal{U}_k^{(t)}=\frac{1}{2}\big[\Delta^{-1}_t(\lambda_0)W_k^{(t)}+W_k^{(t)\top}\{\Delta_t^{-1}(\lambda_0)\}^{\top}\big]$ and $\mathcal{V}_k^{(t)}=W_k^{(t)\top}\{\Delta_t^{-1}(\lambda_0)\}^{\top}\mathbf{Z}^{(t)}\delta_0$.
By Cram\'er's theorem, in order to prove (A.5),
it suffices to show that $d^{-1/2}c^\top D_z\frac{\partial \ell_{zc}(\theta_{0zc})}{\partial \theta_{zc}}$ is asymptotically normal for any
finite vector $c=(c_1, \cdots, c_{M})^\top\in\mR^{M}$.

Define $\mathbf{A}_t=(\mathbf{A}_{t, ij})=d^{-1/2}\big\{\sum_{k=1}^{d}(\sum_{m=1}^{M}c_mD_{z,mk})\mathcal{U}_k^{(t)}+{ (\sum_{m=1}^{M}c_mD_{z,m(d+1)})I_n/\sigma_{0z}^4}\big\}$ and $\mathbf{B}_t=d^{-1/2}\sum_{k=1}^{d}(\sum_{m=1}^{M}c_mD_{z,mk})\mathcal{V}_k^{(t)}+d^{-1/2}\sum_{k=1}^{d}\big(\sum_{m=1}^{M}c_mD_{z,m(k+d+1)}\big)Z_k^{(t)}/\sigma_{0z}^2$.
Then, we have
$d^{-1/2}c^\top D_z\frac{\partial \ell_{zc}(\theta_{0zc})}{\partial \theta_{zc}}=\sum_{t=1}^T \big[\upsilon_t^\top \mathbf{A}_t \upsilon_t-\sigma_{0z}^2 tr(\mathbf{A}_t)+\upsilon_t^\top \mathbf{B}_t\big]$.
In addition, define $\Xi=\sum_{t=1}^T \big[\upsilon_t^\top \mathbf{A}_t \upsilon_t-\sigma_{0z}^2 tr(\mathbf{A}_t)+\upsilon_t^\top \mathbf{B}_t\big]=
\sum_{t=1}^T \sum_{i\not=j\leq n} \upsilon_{it}\upsilon_{jt} \mathbf{A}_{t, ij}+\sum_{j=1}^{n}\upsilon_{jt}^\top \mathbf{B}_{tj}+\sum_{t=1}^T \sum_{i=1}^n (\upsilon_{it}^2-\sigma_{0z}^2) \mathbf{A}_{t, ii}$,
which is a linear-quadratic form of $\upsilon_t$s.
Moreover, define
$\Xi_r=\sum_{t=1}^T \big[\sum_{i\not=j\leq r} \upsilon_{it}\upsilon_{jt} \mathbf{A}_{t, ij}+\sum_{j=1}^{r}
\upsilon_{jt} \mathbf{B}_{tj}\big]+\sum_{t=1}^T \sum_{i=1}^r (\upsilon_{it}^2-\sigma_{0z}^2) \mathbf{A}_{t, ii}$ and
$\Psi_r=\Xi_{r}-\Xi_{r-1}$. Then, one can verify that $\Psi_r$ is a martingale difference sequence with respect to the sigma-field
$\tilde{\mF}_r$, where $\tilde{\mF}_r=\sigma\{\upsilon_{it}, \mathbf{Z}^{(t)}, 1\leq i\leq r, 1\leq t\leq T\}$ and $\tilde{\mF}_0=\emptyset$.
Accordingly, by the martingale central limit theorem (Hall and Heyde, 1980), it suffices to show
\beq \frac{\sum_{r=1}^n \sigma_{n,r}^{*2}}{\var(\Xi)}\rightarrow_p 1 \mbox{~and~} \frac{\sum_{r=1}^n E(\Psi_r^4)}{\var^2(\Xi)}\rightarrow_p 0,\eeq
where $\sigma_{n,r}^{*2}=E(\Psi_r^2|\tilde{\mF}_{r-1})$. By Condition (E5), $(nT)^{-1}{\rm var}\big\{\frac{\partial\ell_{zc}(\theta_{0zc})}{\partial \theta_{zc}}\big\}$ is positive definite for large $nT$.
As a result, with probability approaching 1, we can have that
$(nT)^{-1}{\rm var}\{d^{-1/2}c^\top D_z\frac{\partial \ell_{zc}(\theta_{0zc})}{\partial \theta_{zc}}\}=d^{-1}c^\top D_z \mathcal{J}^{zc}_{nT}(\theta_{zc0}) D_z^\top c\rightarrow_p
c^\top G_z(\theta_{0zc}) c=O(1)$. This implies that $\var(\Xi)$ is of order $O(nT)$.

We next prove the first part of (A.6).
Note that $E(\sum_{r=1}^n \sigma_{n,r}^{*2})=\var(\Xi)$. Then, it suffices to show that
${\rm var}(\sum_{r=1}^n \sigma_{n,r}^{*2})=o(n^2T^2)$. By definition, we have
$\sigma_{n,r}^{*2}=E(\Psi_r^2|\tilde{\mF}_{r-1})$.
By Conditions (E1) and (E2), after some simple calculations, we have
\[
\sigma_{n,r}^{*2}=\sum_{t=1}^{T}\sigma_{0z}^2\Big[4\sum_{j=1}^{r-1}\mathbf{A}_{t,rj}^2\upsilon_{jt}^2+4\sum_{j\neq k}^{r-1}\upsilon_{jt}\upsilon_{kt}\mathbf{A}_{t,rj}\mathbf{A}_{t,rk}+{4}\sum_{j=1}^{r-1}\upsilon_{jt}\mathbf{A}_{t,rj}\mathbf{B}_{tr}+\mathbf{B}_{tr}^2\Big]
\]
\[
 + (\zeta^{(4)}-\sigma_{0z}^4)\sum_{t=1}^T \mathbf{A}_{t, rr}^2. \]
 Define $I_t^{(1)}=\sigma_{0z}^2(4\sum_{r=1}^n \sum_{j=1}^{r-1}\mathbf{A}_{t,rj}^2\upsilon_{jt}^2{+2\sigma_{0z}^2\sum_{r=1}^{n}\mathbf{A}_{t,rr}^2})$, $I_t^{(2)}=4\sigma_{0z}^2\sum_{r=1}^n \sum_{j\neq k}^{r-1}\upsilon_{jt}\upsilon_{kt}\mathbf{A}_{t,rj}\mathbf{A}_{t,rk}$,
 $I_t^{(3)}={4\sigma_{0z}^2}\sum_{r=1}^n \sum_{j=1}^{r-1}\upsilon_{jt}\mathbf{A}_{t,rj}\mathbf{B}_{tr}$, $I_t^{(4)}=\sigma_{0z}^2\sum_{r=1}^n \mathbf{B}_{tr}^2$,
 and $I_t^{(5)}=\sum_{r=1}^n (\zeta^{(4)}-3\sigma_{0z}^4) \mathbf{A}_{t, rr}^2$.
Then, $\sum_{r=1}^n \sigma_{n,r}^{*2}=\sum_{t=1}^T \big\{I_t^{(1)}+I_t^{(2)}+I_t^{(3)}+I_t^{(4)}+I_t^{(5)}\big\}$
and ${\rm var}(\sum_{r=1}^n \sigma_{n,r}^{*2})=\sum_{t=1}^T \big\{{\rm var}(I_t^{(1)})+{\rm var}(I_t^{(2)})+{\rm var}(I_t^{(3)})+{\rm var}(I_t^{(4)})+{\rm var}(I_t^{(5)})\big\}$.
Consequently, to prove ${\rm var}(\sum_{r=1}^n \sigma_{n,r}^{*2})=o(n^2T^2)$, it suffices to
show that $\var(I_t^{(l)})=o(n^2T)$ uniformly for any $l=1, \cdots, 5$ and $t=1,\cdots, T$.

For $I_t^{(1)}$, we have ${\rm var}(I_t^{(1)})={\rm var}\big\{E(I_t^{(1)}\big|\mathbf{Z}^{(t)})\big\}
+E\big\{{\rm var}(I_t^{(1)}\big|\mathbf{Z}^{(t)})\big\}=I_t^{(1,1)}+I_t^{(1,2)}$. 
Note that
$$
I_t^{(1,1)}={\rm var}\big(4\sigma_{0z}^4\sum_{r=1}^{n}\sum_{j=1}^{r-1}\mathbf{A}_{t,rj}^2+2\sigma_{0z}^4\sum_{r=1}^n\mathbf{A}_{t,rr}^2\big)= 4\sigma_{0z}^8{\rm var}(\|\mathbf{A}_t\|_F^2).
$$
By the definition of $\mathbf{A}_t$,  we obtain that
\begin{align*}
{\rm var}(\|\mathbf{A}_t\|_F^2)
= &{\rm var}\Big[d^{-1}\sum_{j=1}^d\sum_{k=1}^d\big\{(\sum_{m=1}^Mc_mD_{z,mj})(\sum_{m=1}^Mc_mD_{z,mk})tr(\mathcal{U}_j^{(t)}\mathcal{U}_k^{(t)})\big\}\Big]\\
\leq & \sum_{j=1}^{d}\sum_{k=1}^{d}(\sum_{m=1}^{M}c_mD_{z,mj})^2(\sum_{m=1}^{M}c_mD_{z,mk})^2{\rm var}\big\{tr(\mathcal{U}_j^{(t)}\mathcal{U}_k^{(t)})\big\}=O(d^2n),
\end{align*}
where the last equality is due to Condition (E4) (iv).
This, together with the  Theorem's assumption,
$d^2/n\rightarrow 0$ as $n\rightarrow\infty$, leads to $I_t^{(1,1)}=o(n^2 T)$.
In addition,
after some algebraic simplification, we have
\begin{align*}
I_t^{(1,2)}=&16\sigma_{0z}^4E\Big[E\big\{(\sum_{r=1}^{n}\sum_{j=1}^{r}\mathbf{A}_{t,rj}^2(\upsilon_{jt}^2-\sigma_{0z}^2))^2\big|\mathbf{Z}^{(t)}\big\}\Big]
=16\sigma_{0z}^4E\Big\{\sum_{j=1}^{n}(\sum_{r=j+1}^{n}\mathbf{A}_{t,rj}^2)^2E\{(\upsilon_{jt}^2-\sigma_{0z}^2))^2\big|\mathbf{Z}^{(t)}\}\Big\}\\
=&16\sigma_{0z}^4(\zeta^{(4)}-\sigma_{0z}^4)E\Big\{\sum_{j=1}^{n}(\sum_{r=j+1}\mathbf{A}_{t,rj}^2)^2\Big\}\leq 16\sigma_{0z}^4(\zeta^{(4)}-\sigma_{0z}^4) n E\big\{\sup_t \|\mathbf{A}_t\|^4_R\big\}=O(n)=o(n^2 T).
\end{align*}
Consequently, $I_t^{(1)}=o(n^2 T)$.

For $I_t^{(2)}$, let $|\mathbf{A}_t|=(|\mathbf{A}_{t,ij}|)$. Then, by Condition (C3), we have $\sup_t \lambda_{\max}(|\mathbf{A}_t|)<\infty$.
After some algebraic simplification, we have
\begin{align*}
{\rm var}(I_t^{(2)})&=E\Big\{{\rm var}(I_t^{(2)}\big|\mathbf{Z}^{(t)})\Big\}=16\sigma_{0z}^4E\Big[E\big\{(\sum_{r=1}^{n}\sum_{j\neq k}\upsilon_{jt}\upsilon_{kt}\mathbf{A}_{t,rj}\mathbf{A}_{t,rk})^2|\mathbf{Z}^{(t)}\big\}\Big]\\
&=64\sigma_{0z}^4E\Big[E\big\{(\sum_{j=1}^{n}\sum_{ k=j+1}^n\upsilon_{jt}\upsilon_{kt}\sum_{r>k}\mathbf{A}_{t,rj}\mathbf{A}_{t,rk})^2|\mathbf{Z}^{(t)}\big\}\Big]\\
& =64\sigma_{0z}^4E\Big[E\big\{\sum_{j=1}^{n}\sum_{ k=j+1}^n\upsilon_{jt}^2\upsilon_{kt}^2(\sum_{r>k}\mathbf{A}_{t,rj}\mathbf{A}_{t,rk})^2|\mathbf{Z}^{(t)}\big\}\Big]\\
& = 64\sigma_{0z}^8E\Big\{\sum_{j=1}^{n}\sum_{k=j+1}^n(\sum_{r>k}\mathbf{A}_{t,rj}\mathbf{A}_{t,rk})^2\Big\}\\
&\leq 64\sigma_{0z}^8E\Big\{\sum_{j=1}^{n}\sum_{k=1}^n(\sum_{r=1}^n |\mathbf{A}_{t,rj}||\mathbf{A}_{t,rk}|)^2\Big\}\leq 64\sigma_{0z}^8 E\big[tr\big\{(|\mathbf{A}_t||\mathbf{A}_t^\top|)^2\big\}\big]=O(n).
\end{align*}
 As a result, $I_t^{(2)}=o(n^2 T)$.

For $I_t^{(3)}$, we have
\begin{align*}
{\rm var}(I_t^{(3)})&=16\sigma_{0z}^4E\Big\{{\rm var}(\sum_{r=1}^{n}\sum_{j=1}^{r-1}\upsilon_{jt}\mathbf{A}_{t,rj}\mathbf{B}_{tr}|\mathbf{Z}^{(t)})\Big\}=16\sigma_{0z}^6E\Big\{\sum_{j=1}^{n}(\sum_{r>j}\mathbf{A}_{t,rj}\mathbf{B}_{tr})^2\Big\}\\
&\leq 16\sigma_{0z}^6E\big\{tr(\mathbf{B}_t^\top \mathbf{A}_{t}\mathbf{A}_{t}^\top\mathbf{B}_t)\big\}\leq C_B\sigma_{0z}^6E( \|\mathbf{B}_t\|^2)
\end{align*}
for some finite positive constant $C_B$.
We then evaluate the order of $E(\|\mathbf{B}_t\|^2)$ given below.
\begin{align*}
E(\|\mathbf{B}_t\|^2)&\leq E\Big\{2\big\|d^{-1/2}\sum_{k=1}^{d}(\sum_{m=1}^{M}c_mD_{z,mk})\mathcal{V}_k^{(t)}\big\|^2+2\big\|d^{-1/2}\sum_{k=1}^{d}(\sum_{m=1}^{M}c_mD_{z,m(k+d+1)})Z_k^{(t)}\big\|^2\Big\}\\
&\leq 2E\Big\{d^{-1/2}\sum_{k=1}^{d}\sum_{m=1}^{M}|c_mD_{z,mk}|\|\mathcal{V}_k^{(t)}\|\Big\}^2+2E\Big\{d^{-1/2}\sum_{k=1}^{d}\sum_{m=1}^{M}|c_mD_{z,m(k+d+1)}|\|{Z}_k^{(t)}\|\Big\}^2\\
&={O(nd)},
\end{align*}
where the last equality is due to the fact that $E\big\{\|\mathcal{V}_k^{(t)}\|^2\big\}=O(n)$, $E\big\{\|{Z}_k^{(t)}\|^2\big\}=O(n)$
by Condition (E3),
and {$d^{-1/2}\sum_{k=1}^{2d+1}\sum_{m=1}^{M}|c_mD_{z,mk}|=O(d^{1/2}\|D_z\|)=O(d^{1/2})$}.
Accordingly, ${\rm var}(I_t^{(3)})=O(nd)$, which implies $I_t^{(3)}=o(n^2 T)$.

For $I_t^{(4)}$, we have ${\rm var}(I_t^{(4)})={\rm var}(\|\mathbf{B}_t\|^2)$. By Condition (E4),
\begin{align*}
{\rm var}(\|\mathbf{B}_t\|^2)=d^{-2}{\rm var}\Big\{&\sum_{j,k}^{d}(\sum_{m=1}^{M}c_mD_{z,mj})(\sum_{m=1}^{M}c_mD_{z,mk})\mathcal{V}_j^{(t)\top}\mathcal{V}_k^{(t)}\\
&+\sum_{j,k}^{d}(\sum_{m=1}^{M}c_mD_{z,mj})(\sum_{m=1}^{M}c_mD_{z,m(k+d+1)})\mathcal{V}_j^{(t)\top}Z_k^{(t)}\\
&+\sum_{j,k}^{d}(\sum_{m=1}^{M}c_mD_{z,m(j+d+1)})(\sum_{m=1}^{M}c_mD_{z,m(k+d+1)})Z_j^{(t)\top}Z_k^{(t)}\Big\}\\
\leq 3\Big\{&\sum_{j,k}^{d}(\sum_{m=1}^{M}c_mD_{z,mj})^2(\sum_{m=1}^{M}c_mD_{z,mk})^2{\rm var}(\mathcal{V}_j^{(t)\top}\mathcal{V}_k^{(t)})\\
&+\sum_{j,k}^{d}(\sum_{m=1}^{M}c_mD_{z,mj})^2(\sum_{m=1}^{M}c_mD_{z,m(k+d+1)})^2{\rm var}(\mathcal{V}_j^{(t)\top}Z_k^{(t)})
\end{align*}
\begin{align*}
&+\sum_{j,k}^{d}(\sum_{m=1}^{M}c_mD_{z,m(j+d+1)})^2(\sum_{m=1}^{M}c_mD_{z,m(k+d+1)})^2{\rm var}(Z_j^{(t)\top}Z_k^{(t)})
\Big\}\\
=O(d^2n).&
\end{align*}
Thus, $I_t^{(4)}=O(d^2n)$, which leads to $I_t^{(4)}=o(n^2 T)$.
Finally, by Condition (C3), we have $\|\mathbf{A}_t\|_R<\infty$.
Then, one can verify that
${\rm var}(I_t^{(5)})=O(n)$, which implies $I_t^{(5)}=o(n^2 T)$.
Combining the above results, we have
${\rm var}\big\{\sum_{t=1}^{T}I_t^{(m)}/{\rm var}(\Xi)\big\}=O\{d^2/(nT)\}\to 0$ for $m=1,\cdots,5$, and $\frac{\sum_{r=1}^n \sigma_{n,r}^{*2}}{\var(\Xi)}\rightarrow_p 1$.
This completes the first part of the proof.

We next verify the second part of (A.6). Define $\Psi_r^{(t)}=(\sum_{j=1}^{r-1}2\mathbf{A}_{t,rj}\upsilon_{jt}+\mathbf{B}_{tr})\upsilon_{rt}+(\upsilon_{rt}^2-\sigma_{0z}^2)\mathbf{A}_{t,rr}$.
Then, $\Psi_r=\sum_{t=1}^{T}\Psi_r^{(t)}$ and
$\sum_{r=1}^{n}E(\Psi_r^4)=E\big\{\sum_{r=1}^{n}E(\Psi_r^4|\tilde{\mF}_{r-1})\big\}$.
By Condition (E2),
it can be shown that
\begin{align*}
 \sum_{r=1}^{n}E(\Psi_r^4|\tilde{\mF}_{r-1})=&\sum_{r=1}^{n}E\Big\{\big(\sum_{t=1}^{T}\Psi_r^{(t)}\big)^4|\tilde{\mF}_{r-1}\Big\}\\
=& E\Big\{\sum_{r=1}^n\sum_{t=1}^{T}(\Psi_r^{(t)})^4|\tilde{\mF}_{r-1}\Big\}+3E\Big\{\sum_{r=1}^{n}\sum_{t_1\neq t_2}(\Psi_r^{(t_1)})^2(\Psi_r^{(t_2)})^2|\tilde{\mF}_{r-1}\Big\}\\
\leq & \sum_{t=1}^{T}E\Big\{\sum_{r=1}^{n}(\Psi_r^{(t)})^4|\tilde{\mF}_{r-1}\Big\}+\frac{3}{2}\sum_{t_1\neq t_2}E\Big\{\sum_{r=1}^n\big((\Psi_r^{(t_1)})^4+(\Psi_r^{(t_2)})^4\big)|\tilde{\mF}_{r-1}\Big\}.
\end{align*}

After some simple calculations, we obtain
\begin{align*}
E\Big\{\sum_{r=1}^{n}(\Psi_r^{(t)})^4|\tilde{\mF}_{r-1}\Big\}\leq &27\sum_{r=1}^nE\big[\{(\sum_{j=1}^{r-1}2\mathbf{A}_{t,rj}\upsilon_{jt})^4+\mathbf{B}_{tr}^4\}\upsilon_{rt}^4+(\upsilon_{rt}^2-\sigma_{0z}^2)^4\mathbf{A}_{t,rr}^4|\tilde{\mF}_{r-1}\big]\\
& \leq 27\sum_{r=1}^n\big[E(\upsilon_{rt}^4|\tilde{\mF}_{r-1})\{(\sum_{j=1}^{r-1}2\mathbf{A}_{t,rj}\upsilon_{jt})^4+\mathbf{B}_{tr}^4\}+E(\upsilon_{rt}^8|\tilde{\mF}_{r-1})\mathbf{A}_{t,rr}^4\big].
\end{align*}
This, in conjunction with Condition (E2), leads to
\begin{align*}
E\Big\{\sum_{r=1}^{n}(\Psi_r^{(t)})^4|\mathbf{Z}^{(t)}\Big\}\leq 27\sum_{r=1}^{n}\big[\zeta^{(4)}E\{(\sum_{j=1}^{r-1}2\mathbf{A}_{t,rj}\upsilon_{jt})^4|\mathbf{Z}^{(t)}\}+\zeta^{(4)}\mathbf{B}_{tr}^4+\zeta^{(8)}\mathbf{A}_{t,rr}^4\big],
\end{align*}
where
\begin{align*}
\sum_{r=1}^{n}E\Big\{(\sum_{j=1}^{r-1}2\mathbf{A}_{t,rj}\upsilon_{jt})^4|\mathbf{Z}^{(t)}\Big\}=16\sum_{r=1}^n\sum_{j=1}^{r-1}\mathbf{A}_{t,rj}^4\zeta^{(4)}+48\sum_{r=1}^{n}\sum_{k\neq j}^{r}\mathbf{A}_{t,rj}^2\mathbf{A}_{t,rk}^2\sigma_{0z}^4.
\end{align*}
Since $E(\sum_{r=1}^{n}\sum_{j=1}^{r-1}\mathbf{A}_{t,rj}^4)\leq E\big[tr\{(|\mathbf{A}_t||\mathbf{A}_t|^{\top})^2\}\big]=O(n)$,  $E(\sum_{r=1}^{n}\sum_{k\neq j}^{r}\mathbf{A}_{t,rj}^2\mathbf{A}_{t,rk}^2) \leq E\big[tr\{(|\mathbf{A}_t||\mathbf{A}_t|^{\top})^2\}\big]=O(n)$, and $E(\sum_{r=1}^{n}\mathbf{B}_{tr}^4)=O(nd^2)$ by Condition (E3), we have $E\big\{E(\sum_{r=1}^{n}(\Psi_r^{(t)})^4|\mathbf{Z}^{(t)})\big\}=O(nd^2)$.

Based on the above results, we obtain that $\sum_{r=1}^{n}E(\Psi_r^4)=E\{\sum_{r=1}^{n}E(\Psi_r^4|\tilde{\mF}_{r-1})\}=E\{\sum_{r=1}^{n}E(\Psi_r^4|\mathbf{Z}^{(t)})\}=O(nd^2T^2)$.
Accordingly, $\sum_{r=1}^{n}E(\Psi_r^4)/\big\{{\rm var}(\Xi)\big\}^2=O(d^2n^{-1})\to 0$,
which completes the second part of (A.6). In sum, we have demonstrated (A.5).

Finally, under Condition (E5), we can employ similar techniques to those used in the proof of Lemma 3 to verify that
$-(nT)^{-1}\frac{\partial^2 \ell_{zc}(\theta_{0zc})}{\partial \theta_{zc}\partial\theta_{zc}^{\top}}$ converges to $\mathcal{I}^{zc}(\theta_{0zc})$ in Frobenius form.
Furthermore, applying similar techniques to those used in the proof of Theorem 1 (Step II) and Condition (E5), we
complete the entire proof of Theorem 5.

\csection{Additional Conditions and Lemmas}

We present six technical conditions in this section that are useful for proving the theoretical results in Section 5 of the manuscript.

\begin{itemize}
\item[(C4-I)] Assume that $\mathcal{I}^e_{nT}(\theta_{e0})\to\mathcal{I}^e(\theta_{e0})$ in Frobenius norm and $\mathcal{J}^e_{nT}(\theta_{e0})\to\mathcal{J}^e(\theta_{e0})$ in $L_2$ norm
 as $nT\to\infty$, where
$\mathcal{I}^e_{nT}(\theta_{e0})=-(nT)^{-1}E(\frac{\partial^2 \ell_{e}(\theta_{e0})}{\partial\theta_e\partial\theta_e^\top})$,
$\mathcal{J}^e_{nT}(\theta_{e0})=(nT)^{-1}\mbox{var}(\frac{\partial \ell_{e}(\theta_{e0})}{\partial \theta_e})$, and $\mathcal{I}^e(\theta_{e0})$ and $\mathcal{J}^e(\theta_{e0})$
are positive definite matrices.
In addition, assume that
there exist finite positive constants $c_{e,1}$ and $c_{e,2}$ such that
$0<c_{e,1}< \varrho_{\min}(\mathcal{I}^e(\theta_{e}))\leq \varrho_{\max}(\mathcal{I}^e(\theta_{e}))=O(d)$ and $0<c_{e,2}< \varrho_{\min}(\mathcal{J}^e(\theta_{e}))\leq \varrho_{\max}(\mathcal{J}^e(\theta_{e}))=O(d)$ for $\theta_e$ in a small neighborhood of
$\theta_{e0}$.

\item[(C4-II)] Assume that $\mathcal{I}^s_{nT}(\theta_{s0})\to\mathcal{I}^s(\theta_{s0})$ in Frobenius norm, and $\mathcal{J}^s_{nT}(\theta_{s0})\to\mathcal{J}^s(\theta_{s0})$
in $L_2$ norm as $nT\to\infty$,
where
$\mathcal{I}^s_{nT}(\theta_{s0})=-(nT)^{-1}E(\frac{\partial^2 \ell_{s}(\theta_{s0})}{\partial\theta_s\partial\theta_s^\top})$
and $\mathcal{J}^s_{nT}(\theta_{s0})=(nT)^{-1}\mbox{var}(\frac{\partial \ell_{s}(\theta_{s0})}{\partial \theta_{s}})$, and
$\mathcal{I}^s(\theta_{s0})$ and $\mathcal{J}^s(\theta_{s0})$ are  positive definite matrices. In addition, assume that there exist finite positive constants $c_{s,1}$ and $c_{s,2}$ such that
$0<c_{s,1}< \varrho_{\min}(\mathcal{I}^s(\theta_{s}))\leq \varrho_{\max}(\mathcal{I}^s(\theta_{s}))=O(d)$ and $0<c_{s,2}< \varrho_{\min}(\mathcal{J}^s(\theta_{s}))\leq \varrho_{\max}(\mathcal{J}^s(\theta_{s}))=O(d)$ for $\theta_s$ in a small neighborhood of
$\theta_{s0}$.

\item[(C4-III)] Assume that
 $\mathcal{I}^f_{nT}(\theta_{fT})\to\mathcal{I}^f(\theta_{f0})$ in Frobenius norm, and $\mathcal{J}^f_{nT}(\theta_{fT})\to\mathcal{J}^f(\theta_{f0})$ in $L_2$ norm as $nT\to\infty$,
 where
$\mathcal{I}^f_{nT}(\theta_{fT})=-(nT)^{-1}E(\frac{\partial^2 \ell_{f}(\theta_{fT})}{\partial\theta_f\partial\theta_f^\top})$,
$\mathcal{J}^f_{nT}(\theta_{fT})=(nT)^{-1}\mbox{var}(\frac{\partial \ell_{f}(\theta_{fT})}{\partial \theta_f})$, and  $\mathcal{I}^f(\theta_{f0})$ and $\mathcal{J}^f(\theta_{f0})$
are positive definite matrices. In addition, assume that  there exist finite positive constants $c_{f,1}$ and $c_{f,2}$ such that
$0<c_{f,1}< \varrho_{\min}(\mathcal{I}^f(\theta_{f}))\leq \varrho_{\max}(\mathcal{I}^f(\theta_{f}))=O(d)$ and $0<c_{f,2}< \varrho_{\min}(\mathcal{J}^f(\theta_{f}))\leq \varrho_{\max}(\mathcal{J}^f(\theta_{f}))=O(d)$ for $\theta_f$ in a small neighborhood of
$\theta_{f0}$.

\item[(C4-IV)] Assume that
 $\mathcal{I}^g_{nT}(\theta_{gT})\to\mathcal{I}^g(\theta_{g0})$ in Frobenius norm, and $\mathcal{J}^g_{nT}(\theta_{gT})\to\mathcal{J}^g(\theta_{g0})$ in $L_2$ norm as $nT\to\infty$,
 where
$\mathcal{I}^g_{nT}(\theta_{gT})=-(nT)^{-1}E(\frac{\partial^2 \ell_{g}(\theta_{gT})}{\partial\theta_g\partial\theta_g^\top})$,
$\mathcal{J}^g_{nT}(\theta_{gT})=(nT)^{-1}\mbox{var}(\frac{\partial \ell_{g}(\theta_{gT})}{\partial \theta_g})$, and  $\mathcal{I}^g(\theta_{g0})$ and $\mathcal{J}^g(\theta_{g0})$
are positive definite matrices. In addition, assume that  there exist finite positive constants $c_{g,1}$ and $c_{g,2}$ such that
$0<c_{g,1}< \varrho_{\min}(\mathcal{I}^g(\theta_{g}))\leq \varrho_{\max}(\mathcal{I}^g(\theta_{g}))=O(d)$ and $0<c_{g,2}< \varrho_{\min}(\mathcal{J}^g(\theta_{g}))\leq \varrho_{\max}(\mathcal{J}^g(\theta_{g}))=O(d)$ for $\theta_g$ in a small neighborhood of
$\theta_{g0}$.


 \item[(C8-I)]  Assume $p<\infty$ and $\lim_{nT\to \infty} (nT)^{-1}\sum_{t=1}^{T}X_t^\top X_t=Q$, where $Q$ is finite and positive definite. In addition, assume $\sup_{i,j,t}|X_{j,i}^{(t)}|$ is finite,
 where $X_{j,i}^{(t)}$ is the $i$-th element of $X_{tj}$ and $X_t=(X_{t1}, \cdots, X_{tp})\in\mR^{n\times p}$.

  \item[(C8-II)]  Assume $p<\infty$ and $\lim_{nT\to \infty} (nT)^{-1}\sum_{t=1}^{T}\tilde{X}^{(t)\top}\tilde{X}^{(t)}=\tilde{Q}$, where $\tilde{Q}$ is finite and positive definite.
  In addition,  assume $\sup_{i,t}|\tilde{X}_{i,\cdot}^{(t)}\tilde{\beta}|$ is finite, where $\tilde{X}^{(t)}_{i,\cdot}$ is the $i$-th row of $\tilde{X}^{(t)}$.


\end{itemize}

The above conditions are mild and sensible. Conditions (C4-I)--(C4-IV) are parallel to Condition (C4) of the main paper.
Conditions (C8-I)--(C8-II) are  standard conditions used in linear regression models to avoid multicollinearity.

To prove the theorems in Section 5, we next introduce the following seven useful lemmas. Note that Lemma S.2 given
below is directly modified from Theorem 1 of Kelejian and Prucha (2001), and the proof of Lemma S.4 is similar to that of Lemma S.3. Thus, we
only present the proofs of Lemmas S.1, S.3, S.5, S.6 and S.7.

\begin{la}
Let $\xi_t\in \mathbb{R}^{n}$  be independent and identically distributed random variables
with mean 0 and covariance $\sigma^2I_n$ for $t=1, \cdots, T$. Then, for any vector $\nu_t$, there exists a constant $c_{\xi}$ such that
$\mbox{\rm var}\big\{\sum_{t=1}^{T}\nu_t^{\top}(\xi_{t}-\bar{\xi})\big\}\leq c_{\xi}\mbox{\rm var}(\sum_{t=1}^{T}\nu_t^{\top}\xi_{t})$, where $\bar{\xi}=T^{-1}\sum_{t=1}^{T}\xi_{t}$.
\end{la}
\noindent {\textbf{Proof:}} Using the fact that $\mbox{\rm var}\big\{\sum_{t=1}^{T}\nu_t^{\top}(\xi_{t}-\bar{\xi})\big\}\leq 2\mbox{var}(\sum_{t=1}^{T}\nu_t^{\top}\xi_{t})+2\mbox{var}(\sum_{t=1}^{T}\nu_t^{\top}\bar{\xi})$, it suffices to show that $\mbox{var}(\sum_{t=1}^{T}\nu_t^{\top}\xi_{t})\geq\mbox{var}(\sum_{t=1}^{T}\nu_t^{\top}\bar{\xi})$.
 Denote  $\bar{\nu}=T^{-1}\sum_{t=1}^{T}\nu_t$. We then have
\begin{align*}
\mbox{var}(\sum_{t=1}^{T}\nu_t^{\top}\xi_{t})= \sum_{t=1}^{T}\sigma^2\nu_t^{\top}\nu_t
\quad and\quad
\mbox{var}(\sum_{t=1}^{T}\nu_t^{\top}\bar{\xi})=T\sigma^2\bar{\nu}^{\top}\bar{\nu}.
\end{align*}
The above results, together with $ \sum_{t=1}^{T}\sigma^2\nu_t^{\top}\nu_t-T\sigma^2\bar{\nu}^{\top}\bar{\nu}=\sigma^2\sum_{t=1}^{T}(\nu_t-\bar{\nu})^{\top}(\nu_t-\bar{\nu})\geq 0$, complete the proof.

\begin{la} Let $\mathcal{E}=(\varepsilon_1, \cdots, \varepsilon_m)^\top$, where $\varepsilon_1,\cdots,\varepsilon_m$ are independent and identically distributed random variables
with mean 0 and finite variance $\sigma^2$.
Define
\[
\bar Q_m=\mathcal{E}^\top \bar B\mathcal{E}+h^{\top}\mathcal{E}-\sigma^2{\rm tr}(\bar B),
\]
where $\bar B=(\bar b_{ij})_{m\times m}\in\mathbb{R}^{m\times m}$ and $h=(h_1,\cdots, h_m)^{\top}$.
Suppose the following assumptions are satisfied:

(1) for $i,j=1,\cdots,m$, $\bar b_{ij}=\bar b_{ji}$;

(2) $\|\bar B\|_R<\infty$;

(3) $\sum_{i=1}^{m}|h_{i}|^{2+\eta}/m<\infty$ for some $\eta>0$;

(4) there exists some $\eta_2>0$ such that $\rmE |\varepsilon_i|^{4+\eta_2}<\infty$.
\\Then, we have
$
 \rmE (\bar Q_m)=0$ and
 \[
\sigma_{\bar Q_m}^2:={\rm var}(\bar Q_m)=4\sigma^4\sum_{i=1}^m\sum_{j=1}^{i-1}\bar b_{ij}^2+\sum_{i=1}^m\lmk \lbk\mu^{(4)}\sigma^4-\sigma^4\rbk \bar b_{ii}^2 +2\mu^{(3)}\sigma^3 \bar b_{ii}h_{i}+h_i^2\sigma^2\rmk,
\]
where $\mu^{(k)}=E(\epsilon_{i}/\sigma)^k$.
Furthermore, suppose

(5)  $m^{-1}\sigma_{\bar Q_m}^2\geq c_{\min,4}$ for some finite $c_{\min,4}>0$.
\\Then, we obtain
\[
\sigma_{\bar Q_m}^{-1}{\bar Q_m}\stackrel{d}{\longrightarrow} N(0,1).
\]
\end{la}

\begin{la}
Under Conditions (C1)-(C3) and (C5) in Appendix A
and (C4-I) and (C8-I),
as $nT\to\infty$, we have the following results.
\begin{enumerate}
\item[(i)]
$$
(nTd)^{-1/2}D_1\frac{\partial \ell_{e}(\theta_{e0})}{\partial\theta_{e}}\stackrel{d}{\longrightarrow}N\lsk 0,G_1(\theta_{e0})\rsk,
$$
where $D_1$ is an arbitrary $M\times(d+p+1)$ matrix
satisfying $d^{-1}D_1\mathcal{J}^e(\theta_{e0})D_1^{\top}\to G_1(\theta_{e0})$,   $M<\infty$, and $G_1(\theta_{e0})$ is a positive definite matrix.
\item[(ii)]
$$
 \laak(nT)^{-1}\frac{\partial^2 \ell_{e}(\theta_{e0})}{\partial \theta_{e}\partial \theta_{e}^\top }+\mathcal{I}^e(\theta_{e0})\raak_F=o_p(1).
$$
\end{enumerate}
\end{la}
\noindent {\textbf{Proof:}}
To show part (i) in the above lemma, we  define $\epsilon=(\epsilon_1^\top, \cdots, \epsilon_T^\top)^\top\in\mR^{nT}$.
After some simple calculations, we obtain that
\begin{align*}
& \frac{\partial\ell_{e}(\theta_{e0})}{\partial\lambda_k} =\frac{1}{\sigma_0^2}\epsilon^{\top}U_k\epsilon-tr(U_k)+\frac{1}{\sigma_0^2}\sum_{t=1}^{T}\beta_0^{\top}X_t^\top\{\Delta_t^{\top}(\lambda_0)\}^{-1}W_k^{(t)\top}\epsilon_t,  \\
&\frac{\partial\ell_{e}(\theta_{e0})}{\partial\sigma^2}    = -\frac{nT}{2\sigma_0^2}+\frac{1}{2\sigma_0^4}\epsilon^{\top}\epsilon,\ \mbox{and}\quad
\frac{\partial\ell_{e}(\theta_{e0})}{\partial\beta_j}       = \frac{1}{\sigma_0^2}\sum_{t=1}^{T}X_{tj}^\top\epsilon_t,
\end{align*}
 for $k=1,\cdots, d$ and $j=1,\cdots, p$, where $U_k$ is defined below Theorem 1.
By  Cram\'er's theorem, it suffices to show that
 $d^{-1/2}c^\top D_1\frac{\partial \ell_{e}(\theta_{e0})}{\partial \theta_{e}}$ is asymptotically normal for any
finite vector $c=(c_1, \cdots, c_{M})^\top\in\mR^{M}$. Denote $U_{d+1}=\frac{1}{2\sigma_0^2}I_{nT}$ and $h_{x,tk}^{\top}=\beta_0^{\top}X_t^\top\{\Delta_t^{\top}(\lambda_0)\}^{-1}W_k^{(t)\top}$.
Further define $U_c=d^{-1/2}\sum_{j=1}^{d+1}(\sum_{m=1}^{M}c_mD^{(1)}_{mj})U_j$ with $D_1=(D^{(1)}_{mj})$, and
\begin{align*}
h_c= d^{-1/2}\Big\{\sum_{j=1}^{d}(\sum_{m=1}^{M}c_m D^{(1)}_{mj})h_{x,1j}^{\top}+\sum_{j=d+2}^{d+p+1}(\sum_{m=1}^{M}c_mD^{(1)}_{mj})X_{1j}^\top,\cdots,\\
\sum_{j=1}^{d}(\sum_{m=1}^{M}c_mD^{(1)}_{mj})h_{x,Tj}^{\top}+\sum_{j=d+2}^{d+p+1}(\sum_{m=1}^{M}c_mD^{(1)}_{mj})X_{Tj}^\top \Big\}^{\top}\in\mR^{nT}.
\end{align*}
Then, we have
$d^{-1/2}c^\top D_1\frac{\partial \ell_{e}(\theta_{e0})}{\partial \theta_{e}}=(\sigma_0^2)^{-1}\big\{\epsilon^\top U_c \epsilon-\sigma_0^2tr(U_c)+h_c^{\top}\epsilon\big\}$.
By Lemma S.2, we only need to  verify that $U_c$ and $h_c$ satisfy Assumptions (1)--(5) listed in Lemma S.2.

Using the fact that $U_k$ is symmetric for any $k=1, \cdots, d$, Assumption (1) is satisfied.
By Condition (C3), we have that $\|U_k\|_R\leq \sup_t\|W_k^{(t)}\Delta_t(\lambda_0)\|_R\leq \sup_t\|W_k^{(t)}\|_R\|\Delta_t(\lambda_0)\|_R<\infty$.
Applying the Cauchy-Schwartz inequality, we have
\[\|U_c\|_R\leq \sum_{j=1}^{d+1}d^{-1/2}\sum_{m=1}^{M}|c_mD^{(1)}_{mj}|\|U_{j}\|_R\leq c_{u,3}d^{-1/2}\sum_{j=1}^{d+1}\sum_{m=1}^{M}|D^{(1)}_{mj}|\leq c_{u,4}\|D_1\|< \infty,\]
for two finite positive constants $c_{u,3}$ and $c_{u,4}$. Thus, $U_c$ satisfies Assumption (2). Furthermore,
the $i$-th element of $h_{x,tk}$ satisfies that
$$\max_{i}|h_{x,tk,i}|=\max_{i}|\beta_0^{\top}X_t^\top\{\Delta_t^{\top}(\lambda_0)\}^{-1}W_{k,\cdot i}^{(t)\top}|\leq \max_{i} (|X_t\beta_0|)_i\|\{\Delta^\top_t(\lambda_0)\}^{-1}\|_R\|W_k^{(t)\top}\|_R,$$
where $W_{k,\cdot i}^{(t)\top}$ is the $i$-th column of $W_k^{(t)\top}$.
By Conditions (C3) and (C8-I), we immediately
have $\max_{i}|h_{x,tk,i}|\leq \tau_{h1}$  and $\max_{i}|X_{j,i}^{(t)}|\leq\tau_{x1}$, where  $\tau_{h1}$ and $\tau_{x1}$ are positive constants. Thus, the $i$-th element of $h_c$ satisfies that
\[|h_{c,i}|\leq\max\{\tau_{h1},\tau_{x1}\}d^{-1/2}\sum_{m=1}^{M}|c_m|\sum_{j=1}^{d+p+1}|D^{(1)}_{mj}|\leq \tau_1^{*}\|D_1\|< \infty\]
 for some positive constant $\tau_1^{*}>0$.
Accordingly, $(nT)^{-1}\sum_{i=1}^{nT}|h_{c,i}|^{2+\eta}<\infty$ for any finite constant $\eta$, and Assumption (3) holds. Note that  Assumption (4) holds under Condition (C1).
Lastly, using the fact that  $d^{-1}D_1\mathcal{J}^e(\theta_{e0})D_1^{\top}\to G_1(\theta_{e0})$  for sufficiently large $nT$, and  $G_1(\theta_{e0})$ is a positive definite matrix, we have
 $(nT)^{-1}\text{var} (d^{-1/2}c^\top D_1\frac{\partial \ell_{e}(\theta_{e0})}{\partial \theta_{e}})=d^{-1}c^{\top}D_1\mathcal{J}^e_{nT}(\theta_{e0})D_1^{\top}c>2^{-1}c^\top c\varrho_{\min}\{G_1(\theta_{e0})\}$.
Hence,  Assumption (5) holds, which completes the first part of the proof.

We next verify part (ii). By Condition (C4-I), it suffices to show that
 $\|(nT)^{-1}\frac{\partial^2 \ell_{e}(\theta_{e0})}{\partial \theta_{e}\partial \theta_{e}^\top }+ \mathcal{I}^e_{nT}(\theta_{e0})\|_F=o_p(1)$.
 After some simple calculations, we have that
\bda
-(nT)^{-1}\frac{\partial^2\ell_{e}(\theta_{e0}) }{\partial \theta_{e}\partial \theta_{e}^\top }&=&
\begin{pmatrix}
-(nT)^{-1}\frac{\partial^2 \ell_{e}(\theta_{e0})}{\partial \lambda\partial \lambda^\top }& -(nT)^{-1}\frac{\partial^2 \ell_{e}(\theta_{e0})}{\partial \lambda \partial \sigma^2 }&  -(nT)^{-1}\frac{\partial^2 \ell_{e}(\theta_{e0})}{\partial \lambda \partial \beta^{\top} } \\
-(nT)^{-1}\frac{\partial^2 \ell_{e}(\theta_{e0})}{\partial \sigma^2\partial \lambda }& -(nT)^{-1}\frac{\partial^2 \ell_{e}(\theta_{e0})}{\partial^2 \sigma^2 }&  -(nT)^{-1}\frac{\partial^2 \ell_{e}(\theta_{e0})}{\partial \sigma^2\partial \beta^{\top} }\\
-(nT)^{-1}\frac{\partial^2 \ell_{e}(\theta_{e0})}{\partial \beta\partial \lambda^\top }& -(nT)^{-1}\frac{\partial^2 \ell_{e}(\theta_{e0})}{\partial \beta \partial \sigma^2 }&  -(nT)^{-1}\frac{\partial^2 \ell_{e}(\theta_{e0})}{\partial \beta \partial \beta^{\top} }
\end{pmatrix},
\eda
where
$-(nT)^{-1}\frac{\partial^2 \ell_{e}(\theta_{e0})}{\partial^2 \sigma^2 }=\frac{1}{nT\sigma_0^6}\epsilon^\top\epsilon-\frac{1}{2\sigma_0^4}$,
$-(nT)^{-1}\frac{\partial^2 \ell_{e}(\theta_{e0})}{\partial\lambda_k\partial\sigma^2 }=\frac{1}{nT\sigma_0^4}(\epsilon^\top U_k \epsilon+\sum_{t=1}^{T}h_{x,tk}\epsilon_t)$,
$-(nT)^{-1}\frac{\partial^2 \ell_{e}(\theta_{e0})}{\partial \beta_j \partial \sigma^2}=\frac{1}{nT\sigma_0^4}\sum_{t=1}^{T}X_{tj}^\top\epsilon_t $,
$$
-(nT)^{-1}\frac{\partial^2 \ell_{e}(\theta_{e0})}{\partial\lambda_k\partial\lambda_l}=\frac{1}{2nT\sigma_0^2}\sum_{t=1}^{T}\big\{\epsilon_t^\top\Omega_{kl}^{(t)}\epsilon_t+2\beta_0^{\top}X_t^\top\Omega_{kl}^{(t)}\epsilon_t
+\beta_0^{\top}X_t^\top\Omega_{kl}^{(t)}X_t\beta_0\big\}+\frac{1}{2nT}\sum_{t=1}^{T}\mbox{tr}(\Omega_{kl}^{(t)})
$$
with $\Omega_{kl}^{(t)}=\{\Delta^{\top}_t(\lambda_0)\}^{-1}W_k^{(t)\top}W_l^{(t)}\Delta_t^{-1}(\lambda_0)+\{\Delta^{\top}_t(\lambda_0)\}^{-1}W_l^{(t)\top}W_k^{(t)}\Delta_t^{-1}(\lambda_0)$,	
$-(nT)^{-1}\frac{\partial^2 \ell_{e}(\theta_{e0})}{\partial \beta_j \partial \lambda^\top_{k} }=\frac{1}{nT\sigma_0^2}\sum_{t=1}^TX_{tj}^\top W_k^{(t)}\Delta_t^{-1}(\lambda_0)(X_t\beta_0+\epsilon_t)$,
 and
$-(nT)^{-1}\frac{\partial^2 \ell_{e}(\theta_{e0})}{\partial \beta_j \partial \beta_m}=\frac{1}{nT\sigma_0^2}\sum_{t=1}^{T}X_{tj}^\top X_{tm}$, for any $k=1,\cdots,d$, $l=1,\cdots,d$, $j=1,\cdots,p$, and $m=1,\cdots,p$.

We then show that the variance of each element in $-(nT)^{-1}\frac{\partial^2 \ell_{e}(\theta_{e0})}{\partial \theta_{e}\partial \theta_{e}^\top }$ is $O\{(nT)^{-1}\}$.
Following similar techniques to those used in the proof of Lemma 3, we have $\mbox{var}\big\{-\frac{1}{nT}\frac{\partial^2 \ell_{e}(\theta_{e0})}{\partial^2 \sigma^2 }\big\}=O\{(nT)^{-1}\}$.
By Condition (C8-I), $\mbox{var}\big\{-\frac{1}{nT}\frac{\partial^2 \ell_{e}(\theta_{e0})}{\partial \beta_{j} \partial \sigma^2}\big\}=\frac{1}{(nT)^2\sigma_0^6}\sum_{t=1}^{T}X_{tj}^\top X_{tj}=O\{(nT)^{-1}\}$.
Applying Conditions (C2)--(C3) and (C8-I), we  have
\[\mbox{var}\big\{-\frac{1}{nT}\frac{\partial^2 \ell_{e}(\theta_{e0})}{\partial \beta_j \partial \lambda^\top_{k}}\big\}=\frac{1}{(nT)^2\sigma_0^2}\sum_{t=1}^{T}X_{tj}^\top W_k^{(t)}\Delta_t^{-1}(\lambda_0)
\{\Delta_t^{-1}(\lambda_0)\}^{\top}W_k^{(t)\top}X_{tj} \]
\[\leq \frac{1}{(nT)^2\sigma_0^2}c_{w\delta}\sum_{t=1}X_{tj}^\top X_{tj}=O\{(nT)^{-1}\}\] for some finite constant $c_{w\delta}>0$. By
Lemma S.2,
the variance of $-(nT)^{-1}\frac{\partial^2 \ell_{e}(\theta_{e0})}{\partial\lambda_k\partial\sigma^2}$ is
$$
\frac{1}{(nT)^2\sigma_0^8}\Big[\sigma_0^4\big\{2tr(U_k^2)+(\mu^{(4)}-3)tr(U_k^{\otimes 2})\big\}+\sum_{i=1}^{nT}\{2\mu^{(3)}\sigma_0^3(U_k)_{ii}h_{x,i}^{(k)}+h_{x,i}^{(k)2}\sigma_0^2\}
\Big],
$$
with $h_x^{(k)}=(h_{x,1}^{(k)},\cdots,h_{x,nT}^{(k)})^{\top}=(h_{x,1k}^{\top},\cdots,h_{x,Tk}^{\top})^{\top}$.
 Employing similar techniques to those used in the proof of Lemma 3,
 we have $tr(U_k^{\otimes 2})\leq tr(U_k^2)=O(nT)$.
By the proof of the first part of Lemma S.3, we have $|h_{x,i}^{(k)}|<\infty$. Thus,
the variance of $ -(nT)^{-1}\frac{\partial^2 \ell_{e}(\theta_{e0})}{\partial\lambda_k\partial\sigma^2}$ is $O\{(nT)^{-1}\}$.
Define $\Omega_{kl}=\mbox{diag}\{\Omega_{kl}^{(1)},\cdots,\Omega_{kl}^{(T)}\}$ and $h_{\beta,kl,t}=2\beta_0^{\top}X_t^\top\Omega_{kl}^{(t)}$.
Then, $-(nT)^{-1}\frac{\partial^2 \ell_{e}(\theta_{e0})}{\partial\lambda_k\partial\lambda_l}$ can be rewritten as
$\frac{1}{2nT\sigma_0^2}\epsilon^{\top}\Omega_{kl}\epsilon+\sum_{t=1}^{T}(h_{\beta,kl,t}\epsilon_t+\beta_0^{\top}X_t^\top\Omega_{kl}^{(t)}X_t\beta_0)$.
By Conditions (C3) and (C8-I), we have $tr(\Omega^2_{kl})=O(nT)$ and $\max_{i}|(h_{\beta,kl,t})_i|<\max_{i}{(|X_t\beta_0|)_{i}}\sup_{t}\|\Omega^{(t)}_{kl}\|_R\leq 2\max_{i}{(|X_t\beta_0|)_{i}}C_w^4<\infty$.
Consequently,  $ \mbox{var}\big\{\frac{1}{nT}\frac{\partial^2 \ell_{e}(\theta_{e0})}{\partial\lambda_k\partial\lambda_l}\big\}$ is $O\{(nT)^{-1}\}$.

The above results, together with Chebyshev's inequality, imply that, for any $\tau>0$,
\begin{align*}
&\mbox{P}\Big(\Big\|(nT)^{-1}\frac{\partial^2 \ell_{e}(\theta_{e0})}{\partial \theta_{e}\partial \theta_{e}^\top }+ \mathcal{I}^e_{nT}(\theta_{e0})\Big\|_F>\tau/d\Big)  \leq d^2/\tau^2 \sum_{j=1}^{d+p+1}\sum_{k=1}^{d+p+1}\mbox{var}\Big\{(nT)^{-1}\frac{\partial^2 \ell_{e}(\theta_{e0})}{\partial \theta_{ej}\partial \theta_{ek}}\Big\}\\
& \leq d^2/\tau^2\sum_{j=1}^{d+p+1}\sum_{j=1}^{d+p+1}O((nT)^{-1})=O\{d^4/(nT\tau^2)\}=o(1).
\end{align*}
Combining this with  Condition (C4-I), we obtain that
$$
\Big\|(nT)^{-1}\frac{\partial^2 \ell_{e}(\theta_{e0})}{\partial \theta_{e}\partial \theta_{e}^\top }+ \mathcal{I}^e(\theta_{e0})\Big\|_F\leq \Big\|(nT)^{-1}\frac{\partial^2 \ell_{e}(\theta_{e0})}{\partial \theta_{e}\partial \theta_{e0}^\top }+ \mathcal{I}^e_{nT}(\theta_{e0})\Big\|_F+ \laak\mathcal{I}^e(\theta_{e0})- \mathcal{I}^e_{nT}(\theta_{e0})\raak_F =o_p(1),
$$
which completes the entire proof.
\begin{la}
Under Conditions (C1)-(C3) and (C5) in Appendix A and (C4-II) and (C8-II) in the supplementary material, as $nT\to\infty$, we have the following results.
\begin{enumerate}
\item[(i)]
$$
(nTd)^{-1/2}D_2\frac{\partial \ell_{s}(\theta_{s0})}{\partial\theta_{s}}\stackrel{d}{\longrightarrow}N\lsk 0,G_2(\theta_{s0})\rsk,
$$
where $D_2$ is an arbitrary $M\times (d+1)(p+1)$ matrix  satisfying $\|D_2\|<\infty$ and $d^{-1}D_2\mathcal{J}^s(\theta_{s0})D_2^{\top}\to G_2(\theta_{s0})$, $M<\infty$, and $G_2(\theta_{s0})$ is
a positive definite matrix.
\item[(ii)]
$$
 \Big\|(nT)^{-1}\frac{\partial^2 \ell_{s}(\theta_{s0})}{\partial \theta_{s}\partial \theta_{s}^\top }+\mathcal{I}^s(\theta_{s0})\Big\|_F=o_p(1).
$$
\end{enumerate}
\end{la}

\begin{la}
Under Conditions (C1)--(C3) and (C5) in Appendix A and (C4-III) and (C8-I) in the supplementary material, as $nT\to\infty$, we have the following results.
\begin{enumerate}
\item[(i)]
$$
(nTd)^{-1/2}D_3\frac{\partial \ell_{f}(\theta_{fT})}{\partial\theta_{f}}\stackrel{d}{\longrightarrow}N\lsk 0,G_3(\theta_{f0})\rsk,
$$
where $D_3$ is an arbitrary $M\times(d+p+1)$ matrix satisfying $\|D_3\|<\infty$ and $d^{-1}D_3\mathcal{J}^f(\theta_{f0})D_3^{\top}\to G_3(\theta_{f0})$, $M<\infty$, and
$G_3(\theta_{f0})$ is a positive definite matrix.
\item[(ii)]
$$
 \Big\|(nT)^{-1}\frac{\partial^2 \ell_{f}(\theta_{fT})}{\partial \theta_{f}\partial \theta_{f}^\top }+\mathcal{I}^f(\theta_{f0})\Big\|_F=o_p(1).
$$
\end{enumerate}
\end{la}

\noindent {\textbf{Proof:}}
We first prove part (i).
After some tedious calculations, we obtain that
\begin{align*}
 \frac{\partial\ell_{f}(\theta_{fT})}{\partial\lambda_k} =&\frac{T}{(T-1)\sigma_0^2}\epsilon^{\top}(U_k-\Upsilon_k)\epsilon-tr(U_k)\\
 & +\frac{T}{(T-1)\sigma_0^2}\sum_{t=1}^{T}(\beta_0^{\top}X_t^{\top}+\omega_0^{\top})\{\Delta_t^{\top}(\lambda_0)\}^{-1}W_k^{(t)\top}(\epsilon_t-\bar{\epsilon}),\\
 \frac{\partial\ell_{f}(\theta_{fT})}{\partial\sigma^2}    = &-\frac{nT}{2(T-1)\sigma_0^2/T}+\frac{1}{2\{(T-1)/T\}^2\sigma_0^4}\epsilon^{\top}\epsilon-\frac{1}{2\{(T-1)/T\}^2\sigma_0^4}T\bar{\epsilon}^{\top}\bar{\epsilon}, \mbox{~and~}\\
 \frac{\partial\ell_{f}(\theta_{fT})}{\partial\beta_j}       = &\frac{T}{(T-1)\sigma_0^2}\sum_{t=1}^{T}(X_{tj}-\bar{X}_{j})^{\top}(\epsilon_t-\bar{\epsilon}),\
\end{align*}
 for $k=1,\cdots, d$, and $j=1,\cdots, p$, where $U_k$ is defined below Theorem 1, $\bar{\epsilon}=T^{-1}\sum_{t=1}^{T}\epsilon_t$,  $\bar{X}_{j}=T^{-1}\sum_{t=1}^{T}X_{tj}$, and
 $$
\Upsilon_k =\frac{1}{2T}\left[\begin{array}{ccc}
 \{\Delta_{1}^{\top}(\lambda_0)\}^{-1}W_k^{(1)\top}+W_k^{(1)}\Delta_{1}^{-1}(\lambda_0)&, \cdots,& \{\Delta_{1}^{\top}(\lambda_0)\}^{-1}W_k^{(1)\top}+W_k^{(T)}\Delta_{T}^{-1}(\lambda_0)\\
\vdots &  & \vdots\\
 \{\Delta_{T}^{\top}(\lambda_0)\}^{-1}W_k^{(T)\top}+W_k^{(1)}\Delta_{1}^{-1}(\lambda_0)&, \cdots,&\{\Delta_{T}^{\top}(\lambda_0)\}^{-1}W_k^{(T)\top}+W_k^{(T)}\Delta_{T}^{-1}(\lambda_0)
\end{array}\right].
$$

It can be shown that $tr(\Upsilon_k)=T^{-1}tr(U_k)$. Then, we
have $E(\frac{\partial\ell_{f}(\theta_{fT})}{\partial\theta_f})=0$.
In addition, $\frac{\partial\ell_{f}(\theta_{fT})}{\partial\theta_f}$ can be expressed as a linear-quadratic form of $\epsilon$.
Employing
similar techniques to those used in the proof of Lemma S.3, we obtain that
$$(nTd)^{-1/2}D_3\frac{\partial \ell_{f}(\theta_{fT})}{\partial\theta_{f}}\stackrel{d}{\longrightarrow}N\lsk 0,G_3(\theta_{f0})\rsk,$$
which completes the first part of the proof.

We next verify part (ii). By Condition (C4-III), it suffices to show that
 $\|(nT)^{-1}\frac{\partial^2 \ell_{f}(\theta_{fT})}{\partial \theta_{f}\partial \theta_{f}^\top }+ \mathcal{I}^f_{nT}(\theta_{fT})\|_F=o_p(1)$.
Denote
\bda
-(nT)^{-1}\frac{\partial^2\ell_{f}(\theta_{fT}) }{\partial \theta_{f}\partial \theta_{f}^\top }&=&
\begin{pmatrix}
-(nT)^{-1}\frac{\partial^2 \ell_{f}(\theta_{fT})}{\partial \lambda\partial \lambda^\top }& -(nT)^{-1}\frac{\partial^2 \ell_{f}(\theta_{fT})}{\partial \lambda \partial \sigma^2 }&  -(nT)^{-1}\frac{\partial^2 \ell_{f}(\theta_{fT})}{\partial \lambda \partial \beta^{\top} } \\
-(nT)^{-1}\frac{\partial^2 \ell_{f}(\theta_{fT})}{\partial \sigma^2\partial \lambda }& -(nT)^{-1}\frac{\partial^2 \ell_{f}(\theta_{fT})}{\partial^2 \sigma^2 }&  -(nT)^{-1}\frac{\partial^2 \ell_{f}(\theta_{fT})}{\partial \sigma^2\partial \beta^{\top} }\\
-(nT)^{-1}\frac{\partial^2 \ell_{f}(\theta_{fT})}{\partial \beta\partial \lambda^\top }& -(nT)^{-1}\frac{\partial^2 \ell_{f}(\theta_{fT})}{\partial \beta \partial \sigma^2 }&  -(nT)^{-1}\frac{\partial^2 \ell_{f}(\theta_{fT})}{\partial \beta \partial \beta^{\top} }
\end{pmatrix}.
\eda
After some simple calculations, we have that:
(i) $-(nT)^{-1}\frac{\partial^2 \ell_{f}(\theta_{fT})}{\partial^2 \sigma^2 }=\frac{1}{nT\sigma_T^6}\epsilon^\top\epsilon-\frac{1}{2\sigma_T^4}
-\frac{1}{n\sigma_T^6}\bar{\epsilon}_T^{\top}\bar{\epsilon}_T$, where
 $\sigma_{T}^2=(T-1)\sigma_0^2/T$;
(ii)
$-(nT)^{-1}\frac{\partial^2 \ell_{f}(\theta_{fT})}{\partial\lambda_k\partial\sigma^2 }=
\frac{1}{nT\sigma_T^4}\big\{\epsilon^\top (U_k-\Upsilon_k) \epsilon+\sum_{t=1}^{T}\varsigma^{(k,t)\top}(\epsilon_t-\bar{\epsilon})\big\}$, where $\varsigma^{(k,t)}=W_k^{(t)}\Delta_t(\lambda_0)^{-1}(X_t\beta_0+\omega_0)$;
(iii)
$-(nT)^{-1}\frac{\partial^2 \ell_{f}(\theta_{fT})}{\partial \beta_j \partial \sigma^2}=\frac{1}{nT\sigma_T^4}\sum_{t=1}^{T}(X_{tj}-\bar{X}_{j})^{\top}(\epsilon_t-\bar{\epsilon})$;
(iv)
$-(nT)^{-1}\frac{\partial^2 \ell_{f}(\theta_{fT})}{\partial \beta_j \partial \beta_m}=\frac{1}{nT\sigma_T^2}\sum_{t=1}^{T}(X_{tj}-\bar{X}_{j})^{\top}(X_{tm}-\bar{X}_{m})$;
(v)
$-(nT)^{-1}\frac{\partial^2 \ell_{f}(\theta_{fT})}{\partial \beta_j \partial \lambda^\top_{k} }=\frac{1}{nT\sigma_T^2}\sum_{t=1}^T(X_{tj}-\bar{X}_{j})^{\top}W_k^{(t)}\Delta_t^{-1}(\lambda_0)
(X_t\beta_0+\omega_0+\epsilon_t)$; and
(vi)
\begin{align*}
-(nT)^{-1}\frac{\partial^2 \ell_{f}(\theta_{fT})}{\partial\lambda_k\partial\lambda_l}=&\frac{1}{nT\sigma_T^2}\sum_{t=1}^{T}\Big[\varsigma^{(k,t)\top}\big\{W_l^{(t)}\Delta_t^{-1}(\lambda_0)\epsilon_t-
\frac{1}{T}\sum_{t_1=1}^{T}W_l^{(t_1)}\Delta_{t_1}^{-1}(\lambda_0)\epsilon_{t_1}\big\}\\
+&\varsigma^{(l,t)\top}\big\{W_k^{(t)}\Delta_t^{-1}(\lambda_0)\epsilon_t-\frac{1}{T}\sum_{t_1=1}^{T}W_k^{(t_1)}\Delta_{t_1}^{-1}(\lambda_0)\epsilon_{t_1}\big\}\Big]\\
+&\frac{1}{nT\sigma_T^2}\epsilon^{\top}(\mathbb{U}^{(k,l)}-\mathbb{V}^{(k,l)})\epsilon+\mathbb{C}_{k,l}, \mbox{~~where~~}
\end{align*}
$$\mathbb{U}^{(k,l)}={\rm diag}\Big[s\big\{(\Delta_1^{-1}(\lambda_0))^{\top}W_k^{(1)\top}W_l^{(1)}\Delta_1^{-1}(\lambda_0)\big\},\cdots,s\big\{(\Delta_T^{-1}(\lambda_0))^{\top}W_k^{(T)\top}W_l^{(T)}\Delta_T^{-1}(\lambda_0)\big\}\Big],$$
$$
\mathbb{V}^{(k,l)}=\frac{1}{2T}\left[\begin{array}{ccc}
\Gamma_{1,1}^{(k,l)}&,\cdots, &\Gamma_{1,T}^{(k,l)} \\
\vdots &  & \vdots\\
\Gamma_{T,1}^{(k,l)}&,\cdots, &\Gamma_{T,T}^{(k,l)} \\
\end{array}\right],
$$
$\Gamma_{t_1,t_2}^{(k,l)}=\{\Delta_{t_1}^{\top}(\lambda_0)\}^{-1}W_k^{(t_1)\top}W_l^{(t_2)}\Delta_{t_2}^{-1}(\lambda_0)+\{\Delta_{t_1}^{\top}(\lambda_0)\}^{-1}W_l^{(t_1)\top}W_k^{(t_2)}\Delta_{t_2}^{-1}(\lambda_0)$, and $\mathbb{C}_{k,l}$ is a constant.

Based on the above results, we subsequently show that the variance of each component in $-(nT)^{-1}\frac{\partial^2 \ell_{f}(\theta_{fT})}{\partial \theta_{f}\partial \theta_{f}^\top }$ is $O\{(nT)^{-1}\}$.
Employing similar techniques to those used in the proofs of Lemma S.3 and part (i) of Lemma S.5, we have $\mbox{var}(\frac{1}{nT\sigma_T^6}\epsilon^\top\epsilon)=O\{(nT)^{-1}\}$ and $\mbox{var}\big\{\frac{1}{n\sigma_T^6}\bar{\epsilon}_T^{\top}\bar{\epsilon}_T\big\}=O\{(nT^2)^{-1}\}$, which
leads to $\mbox{var}\big\{-(nT)^{-1}\frac{\partial^2 \ell_{f}(\theta_{fT})}{\partial^2 \sigma^2 }\big\}=O\{(nT)^{-1}\}$.
By Lemmas 2 and S.2,  we have that $\mbox{var}\big\{-(nT)^{-1}\frac{\partial^2 \ell_{f}(\theta_{fT})}{\partial\lambda_k\partial\sigma^2 }\big\}=O\big\{(nT)^{-2}tr\{(U_k-\Upsilon_k)^2\}+\sum_{t=1}^{T}\varsigma^{(k,t)\top}\varsigma^{(k,t)}\big\}$. Since $tr\{(U_k-\Upsilon_k)^2\}\leq (nT)(\|U_k\|^2+\|\Upsilon_k\|^2)=O(nT)$ and $\sum_{t=1}^{T}\varsigma^{(k,t)\top}\varsigma^{(k,t)}\leq c_{f}\beta_0^{\top}\sum_{t=1}^{T}{X_tX_t^\top}\beta_0=O(nT)$, we obtain $\mbox{var}\big\{-(nT)^{-1}\frac{\partial^2 \ell_{f}(\theta_{fT})}{\partial\lambda_k\partial\sigma^2 }\big\}=O\{(nT)^{-1}\}.$
Applying Lemma S.1 and Condition (C8-I), the variance of $-(nT)^{-1}\frac{\partial^2 \ell_{f}(\theta_{fT})}{\partial \beta_j \partial \sigma^2}$ is dominated by $\mbox{var}\big\{\frac{1}{nT\sigma_T^4}\sum_{t=1}^{T}(X_{tj}-\bar{X}_{j})^{\top}\epsilon_t\big\}$, which is of order $O\{(nT)^{-1}\}$.
By Condition (C8-I), we have
\begin{align*}
&(n^2T^2\sigma_T^4)^{-1}\sum_{t=1}^{T}(X_{tj}-\bar{X}_{j})^{\top}W_k^{(t)}\Delta_t^{-1}(\lambda_0)\{\Delta_t^{\top}(\lambda_0)\}^{-1}W_k^{(t)\top}(X_{tj}-\bar{X}_{j})\\
& \leq (n^2T^2\sigma_T^4)^{-1}C_w^4\sum_{t=1}^{T}(X_{tj}-\bar{X}_{j})^{\top}(X_{tj}-\bar{X}_{j})\leq =(n^2T^2\sigma_T^4)^{-1}C_w^4\sum_{t=1}^{T}X_{tj}^{\top}X_{tj}=O\{(nT)^{-1}\},
\end{align*}
which immediately leads to $\mbox{var}\big\{-(nT)^{-1}\frac{\partial^2 \ell_{f}(\theta_{fT})}{\partial \beta_j \partial \lambda^\top_{k} }\big\}=O\{(nT)^{-1}\}.$
By Lemmas 2 and S.2, we obtain that the variance  of $-(nT)^{-1}\frac{\partial^2 \ell_{f}(\theta_{fT})}{\partial\lambda_k\partial\lambda_l}$ is dominated by $O[(nT)^{-2}tr\{(\mathbb{U}^{(k,l)}-\mathbb{V}^{(k,l)})^2\}+(nT)^{-2}\sum_{t=1}^{T}\{\varsigma^{(k,t)\top}\varsigma^{(k,t)}+\varsigma^{(l,t)\top}\varsigma^{(l,t)}\}]$.
Employing similar techniques to those used in the proof of Lemma S.3, we have
$\mbox{var}\big\{(nT)^{-1}\frac{\partial^2 \ell_{f}(\theta_{fT})}{\partial\lambda_k\partial\lambda_l}\big\}=O\{(nT)^{-1}\}$.

The above results, together with Chebyshev's inequality, imply that, for any $\bar\tau>0$,
\begin{align*}
&\mbox{P}\Big(\Big\|(nT)^{-1}\frac{\partial^2 \ell_{f}(\theta_{fT})}{\partial \theta_{f}\partial \theta_{f}^\top }+ \mathcal{I}^f_{nT}(\theta_{fT})\Big\|_F>\bar{\tau}/d\Big)  \leq d^2/\bar{\tau}^2 \sum_{j=1}^{d+p+1}\sum_{k=1}^{d+p+1}\mbox{var}\Big\{(nT)^{-1}\frac{\partial^2 \ell_{f}(\theta_{fT})}{\partial \theta_{fj}\partial \theta_{fk}}\Big\}\\
& \leq d^2/\bar{\tau}^2\sum_{j=1}^{d+p+1}\sum_{j=1}^{d+p+1}O\{(nT)^{-1}\}=O\{d^4/(nT)\}=o(1).
\end{align*}
This, in conjunction with
Condition (C4-III), leads to
$$
\Big\|(nT)^{-1}\frac{\partial^2 \ell_{f}(\theta_{fT})}{\partial \theta_{f}\partial \theta_{f}^\top }+ \mathcal{I}^f(\theta_{f0})\Big\|_F\leq \Big\|(nT)^{-1}\frac{\partial^2 \ell_{f}(\theta_{fT})}{\partial \theta_{f}\partial \theta_{fT}^\top }+ \mathcal{I}^f_{nT}(\theta_{fT})\Big\|_F+ \Big\|\mathcal{I}^f(\theta_{f0})- \mathcal{I}^f_{nT}(\theta_{fT})\Big\|_F =o_p(1),
$$
which completes the entire proof.

\begin{la}\label{lem Fn}
We can obtain (i) $\Delta_t^{*-1}(\lambda)=F_{n,n-1}^{\top}\Delta_t^{-1}(\lambda)F_{n,n-1}$; (ii) $J_nW_k^{(t)}J_n=J_nW_k^{(t)}$; and (iii) $J_n\Delta_{t}^{-1}(\lambda)J_n=J_n\Delta_{t}^{-1}(\lambda)$,
where $F_{n,n-1}$, $\Delta_t^{*-1}(\lambda)$ and $J_n$ are all defined in Section 5.4.
\end{la}
\noindent {\textbf{Proof:}} By the definition of $F_{n,n-1}$ and $J_n$, we have
\begin{align*}
\begin{array}{lll}
J_nF_{n,n-1}=F_{n,n-1},& F_{n,n-1}^{\top}F_{n,n-1}=I_{n-1}, & J_n\mathbf{1}_n=0, \\
F_{n,n-1}^{\top}\mathbf{1}_n=0, & F_{n,n-1}F_{n,n-1}^{\top}=J_n, & F_{n,n-1}F_{n,n-1}^{\top}+\frac{1}{n}
\mathbf{1}_n\mathbf{1}_n^{\top}=I_{n}.
\end{array}
\end{align*}
Since $W_k^{(t)}$ is row normalized for any $k$ and $t$, we have $W_k^{(t)}\mathbf{1}_n=\mathbf{1}_n$.
Then, by the definition of $\Delta_t^{*}(\lambda)$, we obtain
\begin{align*}
&\Delta_t^{*}(\lambda)F_{n,n-1}^{\top}\Delta_t^{-1}(\lambda)F_{n,n-1}=F_{n,n-1}^{\top}\Delta_t(\lambda)J_n\Delta_t^{-1}(\lambda)F_{n,n-1}\\
=&F_{n,n-1}^{\top}F_{n,n-1}-F_{n,n-1}^{\top}(I-\lambda_1W_1^{(t)}-\cdots-\lambda_dW_d^{(t)})\frac{1}{n}
\mathbf{1}_n\mathbf{1}_n^{\top}\Delta_t^{-1}(\lambda)F_{n,n-1}
=  I_{n-1}.
\end{align*}
For $J_nW_k^{(t)}J_n$, we have
\begin{align*}
J_nW_k^{(t)}J_n=J_nW_k^{(t)}(I_n-\frac{1}{n}
\mathbf{1}_n\mathbf{1}_n^{\top})=J_nW_k^{(t)}-F_{n,n-1}F_{n,n-1}^{\top}W_k^{(t)}\frac{1}{n}
\mathbf{1}_n\mathbf{1}_n^{\top}=J_nW_k^{(t)}.
\end{align*}
For $J_n\Delta_{t}^{-1}(\lambda)J_n$, we have
\begin{align*}
J_n\Delta_{t}^{-1}(\lambda)J_n& =J_n\Delta_{t}^{-1}(\lambda)(I_n-\frac{1}{n}\mathbf{1}_n\mathbf{1}_n^{\top})=J_n\Delta_{t}^{-1}(\lambda)-J_n\Delta_{t}^{-1}\frac{1}{n}\mathbf{1}_n\mathbf{1}_n^{\top}\\
& =J_n\Delta_{t}^{-1}(\lambda)-J_n(1-\lambda_1-\ldots-\lambda_d)^{-1}\frac{1}{n}\mathbf{1}_n\mathbf{1}_n^{\top}=J_n\Delta_{t}^{-1}(\lambda).
\end{align*}
Combining the above results, we have completed the entire proof.

\begin{la}
Under Conditions (C1)--(C3) and (C5) in Appendix A and (C4-IV) and (C8-I) in the supplementary material,  as $nT\to\infty$, we have the following results.
\begin{enumerate}
\item[(i)]
$$
(nTd)^{-1/2}D_4\frac{\partial \ell_{g}(\theta_{gT})}{\partial\theta_{g}}\stackrel{d}{\longrightarrow}N\lsk 0,G_4(\theta_{g0})\rsk,
$$
where $D_4$ is an arbitrary $M\times(d+p+1)$ matrix satisfying $\|D_4\|<\infty$ and $d^{-1}D_4\mathcal{J}^g(\theta_{g0})D_4^{\top}\to G_4(\theta_{g0})$, $M<\infty$, and
$G_4(\theta_{f0})$ is a positive definite matrix.
\item[(ii)]
$$
 \Big\|(nT)^{-1}\frac{\partial^2 \ell_{g}(\theta_{gT})}{\partial \theta_{g}\partial \theta_{g}^\top }+\mathcal{I}^g(\theta_{g0})\Big\|_F=o_p(1).
$$
\end{enumerate}
\end{la}

\noindent {\textbf{Proof:}}
We first prove part (i).
After some tedious calculations, we obtain that
\begin{align*}
 \frac{\partial\ell_{g}(\theta_{gT})}{\partial\lambda_k} =&\frac{T}{(T-1)\sigma_0^2}\epsilon^{\top}(U_k^{*}-\Upsilon_k^{*})\epsilon-tr(U_k^{*})\\
 & +\frac{T}{(T-1)\sigma_0^2}\sum_{t=1}^{T}(\beta_0^{\top}X_t^{\top}+\omega_0^{\top})\{\Delta_t^{\top}(\lambda_0)\}^{-1}W_k^{(t)\top}J_n(\epsilon_t-\bar{\epsilon}),\\
 \frac{\partial\ell_{g}(\theta_{gT})}{\partial\sigma^2}    = &\frac{T}{2(T-1)\sigma_0^2}\Big[-T(n-1)+\frac{T}{(T-1)\sigma_0^2}\epsilon^{\top}\big\{(I_T-\frac{1}{T}\mathbf{1}_T\mathbf{1}_T^{\top})\otimes J_n\big\}\epsilon\Big],
  \mbox{~and~}\\
 \frac{\partial\ell_{g}(\theta_{gT})}{\partial\beta_j}       = &\frac{T}{(T-1)\sigma_0^2}\sum_{t=1}^{T}(X_{tj}-\bar{X}_{j})^{\top}J_n(\epsilon_t-\bar{\epsilon}),\
\end{align*}
 for $k=1,\cdots, d$ and $j=1,\cdots, p$, where $\mathbf{1}_T=(1,\cdots,1)^{\top}\in \mathbb{R}^{T}$, ``$\otimes$" denotes the Kronecker product,
 $
 U_k^{*}={\rm diag}\big\{s(J_nW_k^{(1)}\Delta_1^{-1}(\lambda_0)),\cdots,s(J_nW_k^{(T)}\Delta_T^{-1}(\lambda_0))\big\},
 $ and
 \footnotesize
 $$
\Upsilon_k^* =\frac{1}{2T}\left[\begin{array}{ccc}
 \{\Delta_{1}^{\top}(\lambda_0)\}^{-1}W_k^{(1)\top}J_n+J_nW_k^{(1)}\Delta_{1}^{-1}(\lambda_0)&, \cdots,& \{\Delta_{1}^{\top}(\lambda_0)\}^{-1}W_k^{(1)\top}J_n+J_nW_k^{(T)}\Delta_{T}^{-1}(\lambda_0)\\
\vdots &  & \vdots\\
 \{\Delta_{T}^{\top}(\lambda_0)\}^{-1}W_k^{(T)\top}J_n+J_nW_k^{(1)}\Delta_{1}^{-1}(\lambda_0)&, \cdots,&\{\Delta_{T}^{\top}(\lambda_0)\}^{-1}W_k^{(T)\top}J_n+J_nW_k^{(T)}\Delta_{T}^{-1}(\lambda_0)
\end{array}\right].
$$
\normalsize

It can be shown that $tr(\Upsilon_k^*)=T^{-1}tr(U_k^*)$ and $tr\big\{(I_T-\frac{1}{T}\mathbf{1}_T\mathbf{1}_T^{\top})\otimes J_n\big\}=(n-1)(T-1)$.
Then, we have $E(\frac{\partial\ell_{g}(\theta_{gT})}{\partial\theta_g})=0$. In addition,  $\frac{\partial\ell_{g}(\theta_{gT})}{\partial\theta_g}$ can be expressed as a linear-quadratic form of $\epsilon$.
Employing
similar techniques to those used in the proof of Lemma S.3, we obtain that
$$(nTd)^{-1/2}D_4\frac{\partial \ell_{g}(\theta_{gT})}{\partial\theta_{g}}\stackrel{d}{\longrightarrow}N\lsk 0,G_4(\theta_{g0})\rsk,$$
which completes the first part of the proof.

We next verify part (ii) of this lemma. By Condition (C4-IV), it suffices to show that
 $\|(nT)^{-1}\frac{\partial^2 \ell_{g}(\theta_{gT})}{\partial \theta_{g}\partial \theta_{g}^\top }+ \mathcal{I}^g_{nT}(\theta_{gT})\|_F=o_p(1)$.
Denote
\bda
-(nT)^{-1}\frac{\partial^2\ell_{g}(\theta_{gT}) }{\partial \theta_{g}\partial \theta_{g}^\top }&=&
\begin{pmatrix}
-(nT)^{-1}\frac{\partial^2 \ell_{g}(\theta_{gT})}{\partial \lambda\partial \lambda^\top }& -(nT)^{-1}\frac{\partial^2 \ell_{g}(\theta_{gT})}{\partial \lambda \partial \sigma^2 }&  -(nT)^{-1}\frac{\partial^2 \ell_{g}(\theta_{gT})}{\partial \lambda \partial \beta^{\top} } \\
-(nT)^{-1}\frac{\partial^2 \ell_{g}(\theta_{gT})}{\partial \sigma^2\partial \lambda }& -(nT)^{-1}\frac{\partial^2 \ell_{g}(\theta_{gT})}{\partial^2 \sigma^2 }&  -(nT)^{-1}\frac{\partial^2 \ell_{g}(\theta_{gT})}{\partial \sigma^2\partial \beta^{\top} }\\
-(nT)^{-1}\frac{\partial^2 \ell_{g}(\theta_{gT})}{\partial \beta\partial \lambda^\top }& -(nT)^{-1}\frac{\partial^2 \ell_{g}(\theta_{gT})}{\partial \beta \partial \sigma^2 }&  -(nT)^{-1}\frac{\partial^2 \ell_{g}(\theta_{gT})}{\partial \beta \partial \beta^{\top} }
\end{pmatrix}.
\eda
 After some tedious calculations, we have that:
(i)
$-(nT)^{-1}\frac{\partial^2 \ell_{g}(\theta_{gT})}{\partial^2 \sigma^2 }=-\frac{(n-1)}{2n\sigma_T^4}+\frac{1}{nT\sigma_T^6}\epsilon^{\top}\big\{(I_T-\frac{1}{T}\mathbf{1}_T\mathbf{1}_T^{\top})\otimes J_n\big\}\epsilon$, where
$\sigma_{T}^2=(T-1)\sigma_0^2/T$ was defined in the proof of Lemma S.5;
(ii)
$-(nT)^{-1}\frac{\partial^2 \ell_{g}(\theta_{gT})}{\partial\lambda_k\partial\sigma^2 }=\frac{1}{nT\sigma_T^4}\big\{\epsilon^\top (U_k^{*}-\Upsilon_k^{*}) \epsilon+\sum_{t=1}^{T}\varsigma^{(k,t)\top}_{*}(\epsilon_t-\bar{\epsilon})\big\}$, where $\varsigma^{(k,t)}_{*}=J_nW_k^{(t)}\Delta_t(\lambda_0)^{-1}(X_t\beta_0+\omega_0)$;
(iii)
$-(nT)^{-1}\frac{\partial^2 \ell_{g}(\theta_{gT})}{\partial \beta_j \partial \sigma^2}=\frac{1}{nT\sigma_T^4}\sum_{t=1}^{T}(X_{tj}-\bar{X}_{j})^{\top}J_n(\epsilon_t-\bar{\epsilon})$;
(iv)
$-(nT)^{-1}\frac{\partial^2 \ell_{g}(\theta_{gT})}{\partial \beta_j \partial \beta_m}=\frac{1}{nT\sigma_T^2}\sum_{t=1}^{T}(X_{tj}-\bar{X}_{j})^{\top}J_n(X_{tm}-\bar{X}_{m})$;
(v)
$-(nT)^{-1}\frac{\partial^2 \ell_{g}(\theta_{gT})}{\partial \beta_j \partial \lambda^\top_{k} }=\frac{1}{nT\sigma_T^2}\sum_{t=1}^T(X_{tj}-\bar{X}_{j})^{\top}J_nW_k^{(t)}\Delta_t^{-1}(\lambda_0)
(X_t\beta_0+\omega_0+\epsilon_t)$; and
(vi)
\begin{align*}
-(nT)^{-1}\frac{\partial^2 \ell_{g}(\theta_{gT})}{\partial\lambda_k\partial\lambda_l}=&\frac{1}{nT\sigma_T^2}\sum_{t=1}^{T}\Big[\varsigma^{(k,t)\top}\big\{J_nW_l^{(t)}\Delta_t^{-1}(\lambda_0)\epsilon_t-\frac{1}{T}\sum_{t_1=1}^{T}J_nW_l^{(t_1)}\Delta_{t_1}^{-1}(\lambda_0)\epsilon_{t_1}\big\}\\
+&\varsigma^{(l,t)\top}\big\{J_nW_k^{(t)}\Delta_t^{-1}(\lambda_0)\epsilon_t-\frac{1}{T}\sum_{t_1=1}^{T}J_nW_k^{(t_1)}\Delta_{t_1}^{-1}(\lambda_0)\epsilon_{t_1}\big\}\Big]\\
+&\frac{1}{nT\sigma_T^2}\epsilon^{\top}(\mathbb{U}^{(k,l)}_{*}-\mathbb{V}^{(k,l)}_{*})\epsilon+\mathbb{C}_{k,l}^{*},\mbox{~~where~~}
\end{align*}
$$\mathbb{U}^{(k,l)}_{*}={\rm diag}\Big[s\big\{\{\Delta_1^{-1}(\lambda_0)\}^{\top}W_k^{(1)\top}J_nW_l^{(1)}\Delta_1^{-1}(\lambda_0)\big\},\cdots,s\big\{\{\Delta_T^{-1}(\lambda_0)\}^{\top}W_k^{(T)\top}J_nW_l^{(T)}\Delta_T^{-1}(\lambda_0)\big\}\Big],$$
$$
\mathbb{V}^{(k,l)}_{*}=\frac{1}{2T}\left[\begin{array}{ccc}
\Gamma_{1,1}^{*(k,l)}&,\cdots, &\Gamma_{1,T}^{*(k,l)} \\
\vdots &  & \vdots\\
\Gamma_{T,1}^{*(k,l)}&,\cdots, &\Gamma_{T,T}^{*(k,l)} \\
\end{array}\right],
$$
$\Gamma_{t_1,t_2}^{*(k,l)}=\{\Delta_{t_1}^{\top}(\lambda_0)\}^{-1}W_k^{(t_1)\top}J_nW_l^{(t_2)}\Delta_{t_2}^{-1}(\lambda_0)+\{\Delta_{t_1}^{\top}(\lambda_0)\}^{-1}W_l^{(t_1)\top}J_nW_k^{(t_2)}\Delta_{t_2}^{-1}(\lambda_0)$, and $\mathbb{C}_{k,l}^{*}$ is a constant.

 Employing similar techniques to those used in the proof of Lemma S.5, we  can verify that
the variance of each element in $-(nT)^{-1}\frac{\partial^2 \ell_{g}(\theta_{gT})}{\partial \theta_{g}\partial \theta_{g}^\top }$ is $O\{(nT)^{-1}\}$.
This, together with Chebyshev's inequality, implies that, for any $\bar\tau^*>0$,
\begin{align*}
&\mbox{P}\Big(\Big\|(nT)^{-1}\frac{\partial^2 \ell_{g}(\theta_{gT})}{\partial \theta_{g}\partial \theta_{g}^\top }+ \mathcal{I}^g_{nT}(\theta_{gT})\Big\|_F>\bar{\tau}^*/d\Big)  \leq d^2/\bar{\tau}^{*2} \sum_{j=1}^{d+p+1}\sum_{k=1}^{d+p+1}\mbox{var}\Big\{(nT)^{-1}\frac{\partial^2 \ell_{g}(\theta_{gT})}{\partial \theta_{gj}\partial \theta_{gk}}\Big\}\\
& \leq d^2/\bar{\tau}^{*2}\sum_{j=1}^{d+p+1}\sum_{j=1}^{d+p+1}O\{(nT)^{-1}\}=O\{d^4/(nT)\}=o(1).
\end{align*}
Using this result in conjunction with
Condition (C4-IV), we obtain that
$$
\Big\|(nT)^{-1}\frac{\partial^2 \ell_{g}(\theta_{gT})}{\partial \theta_{g}\partial \theta_{g}^\top }+ \mathcal{I}^g(\theta_{g0})\Big\|_F\leq \Big\|(nT)^{-1}\frac{\partial^2 \ell_{g}(\theta_{gT})}{\partial \theta_{g}\partial \theta_{gT}^\top }+ \mathcal{I}^g_{nT}(\theta_{gT})\Big\|_F+ \Big\|\mathcal{I}^g(\theta_{g0})- \mathcal{I}^g_{nT}(\theta_{gT})\Big\|_F =o_p(1),
$$
which completes the entire proof.

\csection{Proof of Theorem 11}

To prove this theorem, we take the following three steps:
(i) showing that $\hat\theta_f-\theta_{fT}$ is $\{nT/d\}^{1/2}$-consistent;
 (ii) verifying that  $\hat\theta_f$ is asymptotically normal; (iii) showing that
 $\sqrt{T}(\hat\omega_{i}-\omega_{0i})$
is asymptotically normal with mean 0 and variance $\sigma^2_{0}$ for $i=1,\cdots,n$, and  asymptotically independent from $\sqrt{T}(\hat\omega_{j}-\omega_{0j})$ for $j\ne i$.

{\sc Step I.}
To show consistency, it suffices to
follow the technique of Fan and Li (2001) to
demonstrate that, for an arbitrarily small positive constant
$\hbar>0$, there exists a finite constant $\bar C_{\hbar}>0$ such that
\beq \label{eq:A1}
\rmP\Big\{ \sup_{u\in \mR^{d+p+1}:\|u\|=\bar C_{\hbar}} \ell_{f}\{ \theta_{fT}+ (nT/d)^{-1/2}u\}<\ell_{f}(\theta_{fT}) \Big\}\geq 1-\hbar
\eeq
as $nT$ becomes sufficiently large.
To this end,  we employ the Taylor series expansion and obtain that
\bea
&&\sup_{u\in \mR^{d+p+1}:\|u\|=\bar C_{\hbar}}  \ell_{f}\big\{\theta_{fT}+ (nT/d)^{-1/2}u\big\}-\ell_{f}(\theta_{fT})\nn\\
&=&\sup_{u\in \mR^{d+p+1}:\|u\|=\bar C_{\hbar}} \lmk \frac{d^{1/2}}{(nT)^{1/2}}u^\top \frac{\partial\ell_{f}(\theta_{fT})}{\partial \theta_{f}}-\frac{d}{2nT}u^\top \Big\{-\frac{\partial^2\ell_{f}(\theta_{fT})}{\partial\theta_{f}\partial \theta_f^\top}\Big\} u+R^f_n(u) \rmk, \nonumber
\eea
where
$R^f_n(u)$ is a negligible term that satisfies $R^f_n(u)=o_p(d)$.
By Lemma S.5 and Condition (C4-III), we have ${(nT/d)^{-1/2}u^\top \frac{\partial\ell_{f}(\theta_{fT})}{\partial \theta_f}}=d\bar C_{\hbar}O_p(1)$ and
$$
-\frac{d}{2nT}u^\top \lbk-\frac{\partial^2\ell_{f}(\theta_{fT})}{\partial\theta_{f}\partial \theta_{f}^\top}\rbk u=-\frac{d}{2}u^{\top}\mathcal{I}^f(\theta_{f0})u+o_p(d)\leq -\frac{1}{2}d{c_{f,1}}\bar C_{\hbar}^2.
$$
Note that $d\bar C_{\hbar}O_{p}(1)-\frac{1}{2}d{c_{f,1}}\bar C_{\hbar}^2$ is a quadratic function of $\bar C_{\hbar}$. Hence, as long as $\bar C_{\hbar}$ is sufficient large,
$d\bar C_{\hbar}O_{p}(1)-\frac{1}{2}d{c_{f,1}}\bar C_{\hbar}^2<0$.
Accordingly, we have
\beq\label{eq:supell}
\sup_{u\in \mR^{d+p+1}:\|u\|=\bar C_{\hbar}} \Big[ \ell_{f}\big\{ \theta_{fT}+ (nT/d)^{-1/2}u\big\}-\ell_{f}(\theta_{fT})\Big]<0,
\eeq
with probability tending to 1, which demonstrates (\ref{eq:A1}).
Based on the result of (\ref{eq:supell}), there exists a local maximizer $\hat\theta_{f}$ such that $\|\hat\theta_{f}-\theta_{fT}\|\leq {(nT/d)}^{-1/2}\bar C_{\hbar}$ for  sufficiently large $T$.
This, in conjunction with
(\ref{eq:A1}), implies
\begin{align*}
&P\lsk \| \hat\theta_f-{\theta_{fT}}\|\leq (nT/d)^{-1/2}\bar C_{\hbar}\rsk\\
 \geq &  P\Big\{\sup_{u\in \mR^{d+p+1}:\|u\|=\bar C_{\hbar}} \ell_{f}\lsk \theta_{fT}+ (nT/d)^{-1/2}u\rsk<\ell_{f}(\theta_{fT})\Big\}\geq 1-\hbar.
\end{align*}
As a result, $(nT/d)^{1/2}\|\hat\theta_f-\theta_{fT}\|=O_p(1)$, which completes the proof of Step I.

{\sc Step II.}
Using the result of STEP I and applying the Taylor series expansion,  we have that $0={\partial \ell_{f}({\hat{\theta}_f})}/{\partial {\theta_f}}={\partial \ell_{f}({\theta_{fT}})}/{\partial {\theta_f}}+\{{\partial^2 \ell_{f}({\theta_{fT}})}/{\partial {\theta_f\partial \theta_f^\top}}\}{(\hat\theta_f-\theta_{fT})+o_p(d)}$. Hence,
$$
(nTd)^{-1/2} {\partial \ell_{f}(\theta_{fT})}/{\partial \theta_f}=-(nT/d)^{1/2} \frac{1}{nT}\big\{{\partial^2 \ell_{f}(\theta_{fT})}/{\partial \theta_f\partial \theta_f^\top}\big\}(\hat{\theta}_f-\theta_{fT})+o_p(1).
$$
By Lemma S.5 (ii), we have that
\begin{align*}
& \Big\|\left(\frac{nT}{d}\right)^{1/2}\Big(\frac{1}{nT}\frac{\partial^2 \ell_{f}(\theta_{fT})}{\partial \theta_f\partial \theta_f^\top}+\mathcal{I}^f(\theta_{f0})\Big)(\hat{\theta}_{f}-\theta_{fT})\Big\| \\
& \leq \left(\frac{nT}{d}\right)^{1/2}\big\|\frac{1}{nT}\frac{\partial^2 \ell_{f}(\theta_{fT})}{\partial \theta_f\partial \theta_f^\top}+\mathcal{I}^f(\theta_{f0})\big\|\big\|\hat{\theta}_{f}-\theta_{fT}\big\|=o_p(1).
\end{align*}
Thus, we obtain
$$
-(nT/d)^{1/2} \frac{1}{nT}\frac{\partial^2 \ell_{f}(\theta_{fT})}{\partial \theta_f\partial \theta_f^\top}(\hat{\theta}_f-\theta_{fT})=(nT/d)^{1/2}\mathcal{I}^f(\theta_{f0})(\hat{\theta}_f-\theta_{fT})+o_p(1).
$$
This, together with Lemma S.5 (i), implies
$$
\sqrt{nT/d}D_3\mathcal{I}^f(\theta_{f0})(\hat{\theta}_f-\theta_{fT})\stackrel{d}{\longrightarrow}N\lsk 0,G_3(\theta_{f0})\rsk,
$$
which completes the proof of Step II.

{\sc Step III.}
Note that $\hat{\omega}=T^{-1}\sum_{t=1}^{T}\big\{\Delta_t(\hat{\lambda})Y_t-X_t\hat{\beta}\big\}$, which can be rewritten as
$$
\hat{\omega}=T^{-1}\sum_{t=1}^{T}\Big[\big\{I_n+\sum_{k=1}^{d}(\lambda_{0k}-\hat{\lambda}_{k})W_k^{(t)}\Delta_t^{-1}(\lambda_0)\big\}({X_t}\beta_{0}+{\omega_0}+\epsilon_t)-X_t\hat{\beta}\Big].
$$
Thus, we have that
\begin{align*}
\hat{\omega}-{\omega_0}=& T^{-1}\sum_{t=1}^{T}\Big[\big\{I_n+\sum_{k=1}^{d}(\lambda_{0k}-\hat{\lambda}_{k})W_k^{(t)}\Delta_t^{-1}(\lambda_0)\big\}\epsilon_t\Big]\\
& +T^{-1}\sum_{t=1}^{T}\Big[X_t(\beta_0-\hat{\beta})+\big\{\sum_{k=1}^{d}(\lambda_{0k}-\hat{\lambda}_{k})W_k^{(t)}\big\}\Delta_t^{-1}(\lambda_0)(X_t\beta_0+{\omega_0})]\\
{\triangleq} & I_1+I_2.
\end{align*}
Note that the second term in the brackets of $I_1$ satisfies
\begin{align*}
\|\sum_{k=1}^{d}(\lambda_{0k}-\hat{\lambda}_{k})W_k^{(t)}\Delta_t^{-1}(\lambda_0)\|\leq & \sum_{k=1}^{d}|\lambda_{0k}-\hat{\lambda}_{k}|\|W_k^{(t)}\Delta_t^{-1}(\lambda_0)\|\\
\leq& \sqrt{d}\|\lambda_0-\hat{\lambda}\|C_w=O_p(d/\sqrt{nT})\to 0.
\end{align*}
Hence, the dominant term of $I_1$ is $T^{-1}\sum_{t=1}^{T}{\epsilon_{t}}$.
To study the property of $I_2$, define $\iota=T^{-1}\sum_{t=1}^{T}\big\{X_{1,i}^{(t)},\cdots,X_{p,i}^{(t)},W_{1,i\cdot}^{(t)}\Delta_{t}^{-1}(\lambda_0)
(X_t\beta_0+\omega_0),\cdots,W_{d,i\cdot}^{(t)}\Delta_{t}^{-1}(\lambda_0)(X_t\beta_0+\omega_0)\big\}^\top\in\mathbb{R}^{p+d}$,
where $W_{k,i\cdot}^{(t)}$ is the $i$-th row of $W_{k}^{(t)}$ for $k=1,\cdots,d$.
Then, for any $i =1,\cdots,n$,
\begin{align*}
I_{2i}& =\iota^{\top}\big\{(\beta_0-\hat{\beta})^{\top},(\lambda_0-\lambda)^{\top}\big\}^{\top}\leq \|\iota\|\|\theta_{fT}-\hat{\theta}_{f}\|\\
        &\leq \|\theta_{fT}-\hat{\theta}_{f}\|\sup_{t,i}\Big\|\big\{X_{1,i}^{(t)},\cdots,X_{p, i}^{(t)},W_{1,i\cdot}^{(t)}\Delta_{t}^{-1}(\lambda_0)(X_t\beta_0+\omega_0),\cdots,W_{d,i\cdot}^{(t)}\Delta_{t}^{-1}(\lambda_0)(X_t\beta_0+\omega_0)\big\}\Big\|.
\end{align*}
By Conditions (C3) and (C8-I), we have $
W_{k,i\cdot}^{(t)}\Delta_t^{-1}(\lambda_0)(X_t\beta_0+\omega_0)\leq C_w\Delta_t^{-1}(\lambda_0)(X_t\beta_0+\omega_0)\leq C_w\|\{\Delta_{t}^{\top}(\lambda_0)\}^{-1}\|_{R}\sup_{t,i}|(X_t\beta_0)+\omega_0)_i|< c_{\delta,x}
$ for some positive constant $c_{\delta,x}$.
This, together with the result
$\|\theta_{fT}-\hat{\theta}_{f}\|=O_p(\sqrt{d/nT})$ proved in Step I, implies that
$I_{2i}=O_{p}(d/\sqrt{nT})=o_p(1/\sqrt T)$.
Accordingly, $I_2$ is negligible compared with $T^{-1}\sum_{t=1}^{T}{\epsilon_{t}}$, and
the dominant term of $\hat\omega-\omega_{0}$ is $T^{-1}\sum_{t=1}^{T}{\epsilon_{t}}$.
This leads to
$\sqrt{T}(\hat\omega_{i}-\omega_{0i})$
is asymptotically normal with mean 0 and variance $\sigma^2_{0}$ for $i=1,\cdots,n$, and asymptotically independent from $\sqrt{T}(\hat\omega_{j}-\omega_{0j})$ for $j\ne i$.

\csection{Additional Simulation Results}

In this section, we consider five settings for additional simulation studies. Setting I is used to study
a larger weight matrices with
$d=12$. Setting II is constructed to compare EBIC with the Deviance information criterion (DIC) for weight matrix selection.
Settings III and IV introduce distributions with non-normal errors (i.e., exponential and mixture normal) to demonstrate the robustness of our proposed methods.
Setting V is designed to demonstrate the MIR model with exogenous covariates.

\noindent\textbf{Setting I: Weight Matrices with $d=12$.} In addition to $d=2$ and 6 presented in the paper,
we consider  weight matrices with
$d=12$. The Monte Carlo settings are the same as those in Section 5 of the manuscript, and
the simulation results are summarized in Table S.1. We find that the results yield similar patterns to those in Table 1 of the manuscript, which indicates
that our method is applicable for larger weight matrices.

\noindent\textbf{Setting II: Comparison with DIC.}
We compare EBIC with the Deviance information criterion (DIC)  of Spiegelhalter et al. (2002)
for the selection of  relevant matrices. The DIC is a Bayesian method for model comparison and popularly used in the literature (see, e.g., Spiegelhalter et al., 2002 and Li et al., 2020).
The simulation settings are the same as those in Section 5 of the manuscript.
To save space, we only present the results with  full model size
$d=8$ and true model size $|\mS_T|=3$. This is because $d=6$ and $d=12$ show
similar patterns.
Table S.2 indicates that the false positive rate of DIC is
higher than that of EBIC, and DIC is slightly overfitting compared to EBIC based on its
larger average size of the selected model and slightly higher true positive rate.
In  contrast, EBIC can select the model consistently, which is consistent with our theoretical findings.

\noindent\textbf{Setting III: Exponential Errors.}
The simulation settings are the same as those in Section 5 of the manuscript, except that the random error terms are iid from an exponential distribution
($\exp(1)-1$). The simulation results for assessing the finite sample performance of the QMLE estimates, the EBIC criterion and the influence matrix test
are presented in Tables S.3-S.5, respectively. We find that the results yield similar patterns to those in Tables 1--3 of the manuscript, which demonstrate the robustness of the
parameter estimate, weight matrix selection  and the test statistic against non-normal errors.

\noindent\textbf{Setting IV: Mixture Normal Errors.}
The simulation settings are the same as those in Section 5 of the manuscript, except that the random error terms are iid from a mixture normal distribution
($0.9N(0,5/9)+0.1 N(0,5)$). The simulation results for assessing the finite sample performance of the QMLE estimates, the EBIC criterion and the influence matrix test
are given in Tables S.6-S.8, respectively. The results show similar patterns to those in Tables 1--3 of the manuscript,
which further demonstrate the robustness of our proposed methods.

\noindent\textbf{Setting V: Exogenous Covariates.}
The  weight matrices $W_k^{(t)}$ and random error terms $\epsilon_t$, for $k=1, \cdots, d$ and $t=1, \cdots, T$,
are generated from the same settings as those in Section 5 of the manuscript.  In addition, the exogenous covariates $X_{tj}$s for $j=1,\cdots, p$ are independently and identically generated from a standard normal distribution with the regression coefficients
$\beta_1=\cdots=\beta_p=1$. Finally, the response vectors $Y_t$ are generated by
$Y_t=(I_n-\lambda_1W_1^{(t)}-\cdots-\lambda_d W_d^{(t)})^{-1}\epsilon_t+(I_n-\lambda_1W_1^{(t)}-\cdots-\lambda_d W_d^{(t)})^{-1}\sum_{j=1}^p X_{tj}\beta_j$ for $t=1,\cdots,T$.
 Table S.9 presents the simulation results for QMLE estimates with $d=6$ and $p=3$. The results  yield  similar patterns to those in Table 1 of the manuscript,
which demonstrates that the proposed estimation method performs well even with additional exogenous covariates.

\scsection{REFERENCES}

\begin{description}
\newcommand{\enquote}[1]{``#1''}
\expandafter\ifx\csname natexlab\endcsname\relax\def\natexlab#1{#1}\fi

\bibitem[{Bao and Ullah(2010)}]{Bao:2010}
Bao, Y. and Ullah, A. (2010).
\enquote{Expectation of quadratic forms in normal and nonnormal variables with applications,}
 \textit{Journal of Statistical Planning and Inference}, 140, 1193--1205.

 \bibitem[{Fan and Li(2001)}]{Fan:Li:2001}
Fan, J. and Li, R. (2001).
\enquote{Variable selection via nonconcave penalized likelihood and its oracle properties,}
 \textit{Journal of the American Statistical Association}, 96, 1348--1360.


\bibitem[Hanson and Wright(1971)]{Hanson:1971:A}
Hanson, D.~L. and Wright, E.~T. (1971). \enquote{A bound on tail probabilities for quadratic forms in independent random variables},
\textit{Annals of Mathematical Statistics}, 42,1079--1083.

\bibitem[{Kelejian and Prucha(2001)}]{Kelejian:Prucha:2001}
Kelejian, H. and Prucha, I. (2001). \enquote{On the asymptotic distribution of the Moran I test statistic with applications,}\textit{Journal of Econometrics}, 104, 219--257.

 \bibitem[{Li et al.(2020)}]{Li:2020}
Li, Y., Yu, J. and Zeng, T. (2020). \enquote{Deviance information criterion for latent variable models and misspecified models,} \textit{Journal of Econometrics}, 216(2), 450--493.


\bibitem[{Spiegelhalter et al.(2002)}]{Spiegelhalter:2002}
Spiegelhalter, D., Best, N., Carlin, B. and  van der Linde, A. (2002). \enquote{Bayesian measures of model complexity and fit,} \textit{Journal of the Royal Statistical Society, Series B}, 64, 583--639.

\bibitem[Wright(1973)]{Wright:1973:A}
Wright, E.~T. (1973). \enquote{A bound on tail probabilities for quadratic forms in independent random variables whose distributions are not necessarily symmetric,}
\textit{Annals of Probability}, 1, 1068--1070.

\end{description}

\begin{landscape}
\begin{table}[!h]
\begin{center} \caption{\label{tab:t1} The bias and standard error of the parameter estimates when the true parameters are
$\lambda_k=0.2$ for $k=1,\cdots,12$, and the random errors are normally distributed.
BIAS: the average bias; SE: the average of the estimated standard errors via Theorem 1; SE$^*$: the standard error of parameter estimates calculated from 500 realizations.}
\vspace{0.18 cm}
\begin{tabular}{ccc|rrrrrrrrrrrr}
\hline
\hline
$n$ & $T$ & &   $\lambda_1$ & $\lambda_2$ & $\lambda_3$ & $\lambda_4$ & $\lambda_5$ & $\lambda_6$ & $\lambda_7$  & $\lambda_8$ & $\lambda_9$  & $\lambda_{10}$ & $\lambda_{11}$ & $\lambda_{12}$\\
\hline
 25 &  25   &BIAS        &   -0.000 &   -0.006 &    0.001 &   -0.002 &   -0.005 &   -0.001 &   -0.001 &    0.004 &   -0.003 &   -0.001 &   -0.003 &   -0.004 \\
            &&SE          &    0.052 &    0.052 &    0.052 &    0.052 &    0.052 &    0.052 &    0.052 &    0.052 &    0.052 &    0.052 &    0.052 &    0.052 \\
            &&SE$^*$      &    0.055 &    0.050 &    0.053 &    0.053 &    0.048 &    0.055 &    0.053 &    0.051 &    0.054 &    0.053 &    0.056 &    0.053 \\
\hline
 25 &  50   &BIAS        &   -0.002 &   -0.004 &   -0.001 &   -0.000 &   -0.003 &   -0.003 &    0.002 &   -0.001 &   -0.001 &   -0.001 &    0.002 &   -0.002 \\
            &&SE          &    0.037 &    0.037 &    0.037 &    0.037 &    0.037 &    0.037 &    0.037 &    0.037 &    0.037 &    0.037 &    0.037 &    0.037 \\
            &&SE$^*$      &    0.039 &    0.036 &    0.037 &    0.035 &    0.038 &    0.037 &    0.037 &    0.035 &    0.037 &    0.036 &    0.037 &    0.037 \\
\hline
 25 &  100  &BIAS        &    0.000 &    0.001 &   -0.002 &   -0.002 &   -0.000 &   -0.001 &   -0.003 &    0.001 &    0.001 &   -0.001 &    0.001 &   -0.002 \\
            &&SE          &    0.026 &    0.026 &    0.027 &    0.026 &    0.026 &    0.026 &    0.026 &    0.026 &    0.026 &    0.027 &    0.026 &    0.027 \\
            &&SE$^*$      &    0.027 &    0.026 &    0.026 &    0.026 &    0.027 &    0.027 &    0.025 &    0.028 &    0.027 &    0.026 &    0.026 &    0.027 \\
\hline
 50 &  25   &BIAS        &   -0.001 &   -0.004 &    0.001 &   -0.004 &   -0.002 &    0.001 &   -0.004 &   -0.000 &   -0.004 &    0.002 &   -0.003 &   -0.003 \\
            &&SE          &    0.032 &    0.032 &    0.032 &    0.032 &    0.032 &    0.032 &    0.032 &    0.032 &    0.032 &    0.032 &    0.032 &    0.032 \\
            &&SE$^*$      &    0.032 &    0.033 &    0.032 &    0.032 &    0.033 &    0.035 &    0.031 &    0.033 &    0.034 &    0.032 &    0.033 &    0.031 \\
\hline
 50 &  50   &BIAS        &    0.000 &   -0.002 &   -0.001 &   -0.001 &   -0.001 &   -0.002 &   -0.000 &   -0.001 &   -0.000 &   -0.001 &   -0.000 &   -0.001 \\
            &&SE          &    0.023 &    0.023 &    0.023 &    0.023 &    0.023 &    0.023 &    0.023 &    0.023 &    0.023 &    0.023 &    0.023 &    0.023 \\
            &&SE$^*$      &    0.024 &    0.023 &    0.024 &    0.023 &    0.023 &    0.021 &    0.022 &    0.022 &    0.022 &    0.022 &    0.023 &    0.022 \\
\hline
 50 &  100  &BIAS        &   -0.001 &    0.001 &   -0.002 &    0.000 &   -0.001 &   -0.000 &   -0.001 &   -0.000 &    0.000 &    0.000 &   -0.002 &    0.000 \\
             &&SE          &    0.016 &    0.016 &    0.016 &    0.016 &    0.016 &    0.016 &    0.016 &    0.016 &    0.016 &    0.016 &    0.016 &    0.016 \\
            &&SE$^*$      &    0.016 &    0.017 &    0.016 &    0.016 &    0.016 &    0.016 &    0.017 &    0.016 &    0.016 &    0.016 &    0.017 &    0.016 \\
\hline
 100 &  25  &BIAS        &    0.000 &   -0.000 &   -0.001 &   -0.001 &   -0.003 &   -0.002 &    0.001 &   -0.001 &   -0.001 &   -0.001 &   -0.001 &   -0.000 \\
            &&SE          &    0.021 &    0.021 &    0.021 &    0.021 &    0.021 &    0.021 &    0.021 &    0.021 &    0.021 &    0.021 &    0.021 &    0.021 \\
            &&SE$^*$      &    0.022 &    0.021 &    0.022 &    0.022 &    0.022 &    0.021 &    0.020 &    0.022 &    0.022 &    0.021 &    0.021 &    0.021 \\
\hline
 100 &  50  &BIAS        &   -0.001 &   -0.000 &   -0.000 &   -0.001 &   -0.001 &   -0.001 &   -0.002 &   -0.000 &   -0.001 &    0.001 &   -0.001 &   -0.000 \\
            &&SE          &    0.015 &    0.015 &    0.015 &    0.015 &    0.015 &    0.015 &    0.015 &    0.015 &    0.015 &    0.015 &    0.015 &    0.015 \\
            &&SE$^*$      &    0.015 &    0.015 &    0.015 &    0.015 &    0.016 &    0.015 &    0.014 &    0.015 &    0.015 &    0.014 &    0.016 &    0.014 \\
\hline
 100 &  100 &BIAS        &    0.000 &   -0.000 &   -0.000 &   -0.001 &    0.000 &    0.000 &   -0.000 &    0.000 &   -0.001 &    0.000 &   -0.000 &    0.000 \\
            &&SE          &    0.011 &    0.011 &    0.011 &    0.011 &    0.011 &    0.011 &    0.011 &    0.011 &    0.011 &    0.011 &    0.011 &    0.011 \\
            &&SE$^*$      &    0.011 &    0.011 &    0.010 &    0.011 &    0.011 &    0.010 &    0.011 &    0.010 &    0.011 &    0.011 &    0.010 &    0.010 \\
\hline\hline
\end{tabular}
\end{center}
\end{table}
\end{landscape}

 \begin{table}[!h]
\begin{center} \caption{Model selection results via EBIC and DIC when
$d=8$ and the random errors are normally distributed. AS: the average size of the selected model; CT: the average percentage of the correct fit;
TPR: the average true positive rate; FPR: the average false positive rate.} \vspace{0.28 cm}
\begin{tabular}{cc|cccc|cccc}
\hline
\hline
&&\multicolumn{4}{|c}{EBIC} & \multicolumn{4}{|c}{DIC} \\
$n$ & $T$ &  AS & CT & TPR & FPR &  AS & CT & TPR & FPR   \\
\hline
25  &     25&      3.3 &    74.5 &    92.6  &    9.7  & 5.5   & 20.2 & 100.0 & 20.5\\
    &     50&      3.2 &    79.2 &    96.8  &    8.2  & 4.7   & 23.4 & 100.0 & 18.2\\
    &    100&      3.1 &    82.1 &    100.0 &    5.5  & 4.2   & 25.1 & 100.0 & 17.3\\
    \hline
50  &     25&      3.2 &    79.3 &    95.1  &    7.2  & 4.7   & 21.7 & 100.0 & 19.4\\
    &     50&      3.1 &    82.5 &    98.4  &    5.3  & 4.2   & 24.8 & 100.0 & 17.6\\
    &    100&      3.1 &    85.7 &    100.0 &    4.7  & 4.1   & 24.5 & 100.0 & 17.9\\
    \hline
100 &     25&      3.1 &    83.1 &    100.0 &    6.2  & 4.6   & 22.7 & 100.0 & 18.6\\
    &     50&      3.1 &    85.2 &    100.0 &    4.5  & 4.3   & 25.4 & 100.0 & 16.5\\
    &    100&      3.0 &    88.4 &    100.0 &    3.9  & 4.0   & 24.3 & 100.0 & 17.2\\
\hline
\end{tabular}
\end{center}
\end{table}

\begin{table}[!h]
\begin{center} \caption{The bias and standard error of the parameter estimates when the true parameters are
$\lambda_k=0.2$ for $k=1,\cdots,d$, and the random errors follow a standardized exponential distribution.
BIAS: the average bias; SE: the average of the estimated standard errors via Theorem 1; SE$^*$: the standard error of parameter estimates calculated from 500 realizations.}
\vspace{0.18 cm}

\begin{tabular}{ccc|rr|rrrrrr}
\hline
\hline
&&&\multicolumn{2}{|c}{$d=2$ } & \multicolumn{6}{|c}{$d=6$}\\
$n$ & $T$ &&  $\lambda_1$ & $\lambda_2$ & $\lambda_1$ & $\lambda_2$ & $\lambda_3$ & $\lambda_4$ & $\lambda_5$ & $\lambda_6$\\
\hline
 25 &  25   &BIAS        &   -0.002 &   -0.001 &   -0.017 &   -0.016 &   -0.011 &   -0.015 &   -0.017 &   -0.010 \\
            &&SE          &    0.050 &    0.050 &    0.048 &    0.048 &    0.048 &    0.048 &    0.048 &    0.048 \\
            &&SE$^*$      &    0.048 &    0.049 &    0.053 &    0.053 &    0.056 &    0.057 &    0.054 &    0.052 \\
\hline
 25 &  50   &BIAS        &   -0.002 &   -0.002 &   -0.004 &    0.005 &   -0.003 &    0.002 &   -0.004 &   -0.006 \\
            &&SE          &    0.036 &    0.036 &    0.034 &    0.034 &    0.034 &    0.034 &    0.034 &    0.034 \\
            &&SE$^*$      &    0.036 &    0.036 &    0.036 &    0.035 &    0.033 &    0.035 &    0.037 &    0.036 \\
\hline
 25 &  100  &BIAS        &   -0.000 &   -0.003 &   -0.002 &   0.004 &   -0.003 &   0.002 &   0.004 &   0.002 \\
            &&SE          &    0.025 &    0.025 &    0.024 &    0.024 &    0.024 &    0.024 &    0.024 &    0.024 \\
            &&SE$^*$      &    0.024 &    0.024 &    0.026 &    0.027 &    0.025 &    0.026 &    0.028 &    0.027 \\
\hline
50 &  25   &BIAS        &   -0.003 &   0.001 &   -0.003 &   0.004 &   0.001 &   0.004 &   -0.004 &   -0.001 \\
           &&SE          &    0.035 &    0.035 &    0.032 &    0.032 &    0.032 &    0.032 &    0.032 &    0.032 \\
           &&SE$^*$      &    0.034 &    0.032 &    0.035 &    0.033 &    0.035 &    0.034 &    0.036 &    0.034 \\
\hline
 50 &  50   &BIAS        &   -0.001 &   -0.002 &   -0.003 &   -0.003 &   -0.001 &   -0.002 &   -0.004 &   -0.003 \\
            &&SE          &    0.025 &    0.025 &    0.023 &    0.023 &    0.023 &    0.023 &    0.023 &    0.023 \\
            &&SE$^*$      &    0.024 &    0.024 &    0.027 &    0.026 &    0.025 &    0.025 &    0.025 &    0.026 \\
\hline
 50 &  100  &BIAS        &    0.001 &    0.000 &   0.003 &   -0.005 &   -0.003 &   0.001 &   0.003 &   -0.004 \\
           &&SE          &    0.018 &    0.018 &    0.016 &    0.016 &    0.016 &    0.016 &    0.016 &    0.016 \\
            &&SE$^*$      &    0.017 &    0.017 &    0.020 &    0.017 &    0.018 &    0.017 &    0.018 &    0.018 \\
\hline
 100 &  25  &BIAS        &   -0.001 &    0.000 &   -0.001 &   -0.002 &   -0.000 &   -0.000 &   -0.002 &   -0.002 \\
            &&SE          &    0.025 &    0.025 &    0.022 &    0.022 &    0.022 &    0.022 &    0.022 &    0.022 \\
            &&SE$^*$      &    0.024 &    0.025 &    0.023 &    0.024 &    0.022 &    0.024 &    0.022 &    0.024 \\
\hline
 100 &  50  &BIAS        &   -0.001 &   0.001 &    0.000 &   -0.001 &   -0.000 &   -0.001 &    0.000 &   -0.001 \\
            &&SE          &    0.018 &    0.018 &    0.016 &    0.016 &    0.016 &    0.016 &    0.016 &    0.016 \\
            &&SE$^*$      &    0.018 &    0.017 &    0.016 &    0.017 &    0.016 &    0.017 &    0.016 &    0.016 \\
\hline
 100 &  100 &BIAS        &   -0.000 &   -0.000 &   0.000 &   0.001 &   -0.000 &   -0.002 &   -0.002 &   0.001 \\
            &&SE          &    0.012 &    0.012 &    0.011 &    0.011 &    0.011 &    0.011 &    0.011 &    0.011 \\
            &&SE$^*$      &    0.012 &    0.013 &    0.013 &    0.014 &    0.013 &    0.013 &    0.014 &    0.013 \\
        \hline\hline
\end{tabular}
\end{center}
\end{table}

\newpage

 \begin{table}[!h]
\begin{center} \caption{\label{tab:t1} Model selection via {EBIC} when
$d=8$ and the random errors are distributed as standardized exponential. AS: the average size of the selected model; CT: the average percentage of the correct fit;
TPR: the average true positive rate; FPR: the average false positive rate.} \vspace{0.28 cm}
\begin{tabular}{cc|cccc}
\hline
\hline
$n$ & $T$ &  AS & CT & TPR & FPR    \\
\hline
25  &     25&      3.2 &    75.8 &    90.6  &    9.9 \\
    &     50&      3.2 &    80.4 &    96.3  &    7.6 \\
    &    100&      3.1 &    83.7 &    99.8  &    5.4 \\
    \hline
50  &     25&      3.2 &    77.6 &    92.7  &    8.2 \\
    &     50&      3.1 &    82.9 &    98.2  &    6.8 \\
    &    100&      3.1 &    85.4 &    100.0 &    4.7 \\
    \hline
100 &     25&      3.2 &    80.3 &    96.8  &    7.2 \\
    &     50&      3.1 &    85.0 &    100.0 &    4.8 \\
    &    100&      3.0 &    88.5 &    100.0 &    3.9 \\
\hline
\end{tabular}
\end{center}
\end{table}

\begin{table}[!h]
\begin{center} \caption{\label{tab:t1} The empirical sizes and powers of the influence matrix test.
The case of ${\kappa}=0$ corresponds to the null model and
${\kappa}>0$ represents alternative models. The random errors are distributed as standardized exponential, and the full model sizes are $d=2$ and 6. } \vspace{0.28 cm}
\begin{tabular}{cc|ccc|ccc}
\hline\hline
&&\multicolumn{3}{|c}{$d$=2} & \multicolumn{3}{|c}{$d$=6}  \\
$n$ & $T$ &  {$\kappa$}=0 & {$\kappa$}=0.1 & {$\kappa$}=0.2 & {$\kappa$}=0 & {$\kappa$}=0.1 & {$\kappa$}=0.2  \\
\hline
25  &     25&      0.023 &     0.267 &     0.654  &      0.025 &     0.231 &     0.563\\
    &     50&      0.028 &     0.481 &     0.780  &      0.030 &     0.377 &     0.724\\
    &    100&      0.040 &     0.602 &     0.902  &      0.037 &     0.548 &     0.819\\
    \hline
50  &     25&      0.027 &     0.386 &     0.745  &      0.029 &     0.345 &     0.655\\
    &     50&      0.032 &     0.561 &     0.872  &      0.038 &     0.480 &     0.775\\
    &    100&      0.044 &     0.642 &     0.946  &      0.044 &     0.623 &     0.922\\
    \hline
100 &     25&      0.032 &     0.522 &     0.901  &      0.030 &     0.423 &     0.876\\
    &     50&      0.040 &     0.701 &     0.987  &      0.036 &     0.567 &     0.928\\
    &    100&      0.056 &     0.879 &     1.000  &      0.047 &     0.822 &     1.000\\
\hline
\end{tabular}
\end{center}
\end{table}

\newpage

\

\

\

 \begin{table}[!h]
\begin{center} \caption{The bias and standard error of the parameter estimates when the true parameters are
$\lambda_k=0.2$ for $k=1,\cdots, d$, and the random errors follow a mixture normal distribution.
BIAS: the average bias; SE: the average of the estimated standard errors via Theorem 1; SE$^*$: the standard error of parameter estimates calculated from 500 realizations.}
\vspace{0.18 cm}

\begin{tabular}{ccc|rr|rrrrrr}
\hline
\hline
&&&\multicolumn{2}{|c}{$d=2$ } & \multicolumn{6}{|c}{$d=6$}\\
$n$ & $T$ &&  $\lambda_1$ & $\lambda_2$ & $\lambda_1$ & $\lambda_2$ & $\lambda_3$ & $\lambda_4$ & $\lambda_5$ & $\lambda_6$\\
\hline
 25 &  25   &BIAS        &   -0.002 &   -0.001 &   -0.016 &   -0.015 &   -0.010 &   -0.009 &   -0.014 &   -0.011 \\
            &&SE          &    0.049 &    0.051 &    0.049 &    0.049 &    0.048 &    0.047 &    0.047 &    0.050 \\
            &&SE$^*$      &    0.047 &    0.048 &    0.052 &    0.052 &    0.055 &    0.056 &    0.055 &    0.053 \\
\hline
 25 &  50   &BIAS        &   -0.002&   -0.003 &   -0.003 &    0.004 &   -0.004 &    0.003 &   -0.005 &   -0.007 \\
            &&SE          &    0.037 &    0.036 &    0.035 &    0.034 &    0.033 &    0.035 &    0.033 &    0.034 \\
            &&SE$^*$      &    0.036 &    0.037 &    0.037 &    0.036 &    0.034 &    0.036 &    0.036 &    0.035 \\
\hline
 25 &  100  &BIAS        &   -0.001 &   -0.002 &   -0.002 &   0.003 &   0.004 &   0.001 &   0.003 &   0.001 \\
            &&SE          &    0.024 &    0.024 &    0.024 &    0.024 &    0.024 &    0.024 &    0.025 &    0.024 \\
            &&SE$^*$      &    0.025 &    0.024 &    0.025 &    0.026 &    0.026 &    0.027 &    0.027 &    0.026 \\
\hline
50 &  25   &BIAS         &   -0.002 &   0.002 &   -0.003 &   0.005 &   0.002 &   0.003 &   -0.003 &   -0.002 \\
           &&SE          &    0.036 &    0.034 &    0.033 &    0.031 &    0.033 &    0.034 &    0.035 &    0.036 \\
           &&SE$^*$      &    0.035 &    0.036 &    0.034 &    0.032 &    0.034 &    0.033 &    0.034 &    0.035 \\
\hline
 50 &  50   &BIAS        &   -0.002 &   -0.001 &   -0.003 &   -0.002 &   -0.002 &   -0.001 &   -0.003 &   -0.002 \\
            &&SE          &    0.024 &    0.024 &    0.023 &    0.024 &    0.023 &    0.025 &    0.023 &    0.024 \\
            &&SE$^*$      &    0.024 &    0.024 &    0.026 &    0.025 &    0.024 &    0.026 &    0.024 &    0.026 \\
\hline
 50 &  100  &BIAS        &    0.002 &    0.000 &   0.003 &   -0.004 &   -0.003 &   0.002 &   0.003 &   -0.003 \\
           &&SE          &    0.019 &    0.017 &    0.016 &    0.017 &    0.018 &    0.016 &    0.015 &    0.017 \\
            &&SE$^*$      &    0.018 &    0.017 &    0.018 &    0.018 &    0.017 &    0.019 &    0.017 &    0.019 \\
\hline
 100 &  25  &BIAS        &   -0.001 &    0.002 &   -0.001 &   -0.002 &   -0.000 &   -0.001 &   -0.002 &   -0.002 \\
            &&SE          &    0.024 &    0.024 &    0.023 &    0.024 &    0.022 &    0.021 &    0.024 &    0.022 \\
            &&SE$^*$      &    0.024 &    0.025 &    0.022 &    0.023 &    0.024 &    0.023 &    0.023 &    0.023 \\
\hline
 100 &  50  &BIAS        &   -0.001 &   0.001 &    0.001 &   -0.002 &   -0.000 &   -0.002 &    0.001 &   -0.001 \\
            &&SE          &    0.016 &    0.017 &    0.017 &    0.016 &    0.015 &    0.016 &    0.017 &    0.017 \\
            &&SE$^*$      &    0.016 &    0.016 &    0.017 &    0.017 &    0.016 &    0.015 &    0.018 &    0.016 \\
\hline
 100 &  100 &BIAS        &   -0.000 &   -0.001 &   0.001 &   0.001 &   -0.000 &   -0.001 &   0.001 &   0.001 \\
            &&SE          &    0.011 &    0.014 &    0.012 &    0.013 &    0.012 &    0.012 &    0.012 &    0.012 \\
            &&SE$^*$      &    0.011 &    0.013 &    0.013 &    0.013 &    0.013 &    0.014 &    0.013 &    0.012 \\
        \hline\hline
\end{tabular}
\end{center}
\end{table}

 \begin{table}[!h]
\begin{center} \caption{\label{tab:t1} Model selection via {EBIC} when
$d=8$ and the random errors are distributed as mixture normal. AS: the average size of the selected model; CT: the average percentage of the correct fit;
TPR: the average true positive rate; FPR: the average false positive rate.} \vspace{0.28 cm}
\begin{tabular}{cc|cccc}
\hline
\hline
$n$ & $T$ &  AS & CT & TPR & FPR    \\
\hline
25  &     25&      3.2 &    79.2 &    94.6  &    8.1 \\
    &     50&      3.1 &    83.4 &    98.7  &    6.3 \\
    &    100&      3.0 &    87.6 &    100.0 &    4.4 \\
    \hline
50  &     25&      3.2 &    83.5 &    97.1  &    6.6 \\
    &     50&      3.1 &    86.1 &    100.0 &    5.1 \\
    &    100&      3.0 &    89.3 &    100.0 &    3.7 \\
    \hline
100 &     25&      3.1 &    86.5 &    100.0 &    4.7 \\
    &     50&      3.0 &    88.7 &    100.0 &    4.0 \\
    &    100&      3.0 &    91.1 &    100.0 &    3.1 \\
\hline
\end{tabular}
\end{center}
\end{table}

\begin{table}[!h]
\begin{center} \caption{\label{tab:t1} The empirical sizes and powers of the influence matrix test.
The case of {$\kappa=0$} corresponds to the null model and
{$\kappa>0$} represents alternative models. The random errors are distributed as mixture normal, and the full model sizes are $d=2$ and 6. } \vspace{0.28 cm}
\begin{tabular}{cc|ccc|ccc}
\hline\hline
&&\multicolumn{3}{|c}{$d$=2} & \multicolumn{3}{|c}{$d$=6}  \\
$n$ & $T$ &  {$\kappa$}=0 & {$\kappa$}=0.1 & {$\kappa$}=0.2 & {$\kappa$}=0 & {$\kappa$}=0.1 & {$\kappa$}=0.2  \\
\hline
25  &     25&      0.025 &     0.266 &     0.640  &      0.024 &     0.242 &     0.573\\
    &     50&      0.032 &     0.481 &     0.805  &      0.029 &     0.427 &     0.714\\
    &    100&      0.041 &     0.592 &     0.899  &      0.040 &     0.569 &     0.803\\
    \hline
50  &     25&      0.028 &     0.371 &     0.748  &      0.029 &     0.377 &     0.619\\
    &     50&      0.037 &     0.542 &     0.856  &      0.032 &     0.504 &     0.768\\
    &    100&      0.044 &     0.661 &     0.986  &      0.045 &     0.644 &     0.953\\
    \hline
100 &     25&      0.030 &     0.447 &     0.918  &      0.033 &     0.408 &     0.879\\
    &     50&      0.034 &     0.650 &     0.944  &      0.041 &     0.564 &     0.924\\
    &    100&      0.048 &     0.842 &     1.000  &      0.039 &     0.801 &     1.000\\
\hline
\end{tabular}
\end{center}
\end{table}

\begin{landscape}
\begin{table}[!h]
\begin{center} \caption{\label{tab:t1} The bias and standard error of the parameter estimates for MIR with covariates,
when the true parameters are
$\lambda_k=0.2$ for $k=1,\cdots,6$ and $\beta_j=1$ for $j=1, \cdots, 3$, and the random errors are normally distributed.
BIAS: the average bias; SE: the average of the estimated standard errors via Theorem 1; SE$^*$: the standard error of parameter estimates calculated from 500 realizations. }
\vspace{0.18 cm}
\begin{tabular}{ccc|rrrrrrrrr}
\hline
\hline
$n$ & $T$ & &   $\lambda_1$ & $\lambda_2$ & $\lambda_3$ & $\lambda_4$ & $\lambda_5$ & $\lambda_6$ & $\beta_1$  & $\beta_2$ & $\beta_3$  \\
\hline
 25 &  25   &BIAS        &   0.002 &  -0.000 &  -0.000 &   0.000 &   0.000 &   0.001 &   0.001 &  -0.001 &  -0.002 \\
            &&SE          &   0.028 &   0.028 &   0.028 &   0.028 &   0.028 &   0.028 &   0.040 &   0.040 &   0.040 \\
            &&SE$^*$      &   0.029 &   0.030 &   0.029 &   0.028 &   0.030 &   0.028 &   0.040 &   0.043 &   0.039 \\
\hline
 25 &  50   &BIAS        &   0.003 &   0.001 &  -0.000 &  -0.001 &  -0.001 &  -0.001 &  -0.001 &   0.000 &  -0.001 \\
            &&SE          &   0.020 &   0.020 &   0.020 &   0.020 &   0.020 &   0.020 &   0.028 &   0.028 &   0.028 \\
            &&SE$^*$      &   0.020 &   0.020 &   0.020 &   0.019 &   0.020 &   0.021 &   0.029 &   0.029 &   0.028 \\
\hline
 25 &  100  &BIAS        &  -0.001 &   0.001 &  -0.000 &  -0.000 &   0.001 &  -0.000 &  -0.001 &  -0.001 &  -0.000 \\
            &&SE          &   0.014 &   0.014 &   0.014 &   0.014 &   0.014 &   0.014 &   0.020 &   0.020 &   0.020 \\
            &&SE$^*$      &   0.014 &   0.014 &   0.014 &   0.015 &   0.015 &   0.013 &   0.020 &   0.020 &   0.019 \\
\hline
 50 &  25   &BIAS        &  -0.001 &  -0.000 &  -0.000 &   0.000 &   0.000 &   0.002 &  -0.001 &  -0.000 &   0.000 \\
            &&SE          &   0.019 &   0.019 &   0.019 &   0.019 &   0.019 &   0.019 &   0.028 &   0.028 &   0.028 \\
            &&SE$^*$      &   0.019 &   0.019 &   0.019 &   0.019 &   0.018 &   0.020 &   0.027 &   0.029 &   0.029 \\
\hline
 50 &  50   &BIAS        &   0.001 &   0.001 &  -0.000 &  -0.001 &   0.001 &   0.000 &  -0.001 &   0.001 &   0.001 \\
            &&SE          &   0.014 &   0.014 &   0.014 &   0.014 &   0.014 &   0.014 &   0.020 &   0.020 &   0.020 \\
            &&SE$^*$      &   0.013 &   0.013 &   0.013 &   0.013 &   0.013 &   0.013 &   0.020 &   0.020 &   0.020 \\
\hline
 50 &  100  &BIAS        &   0.000 &   0.00 &   0.000 &   0.000 &  -0.000 &  -0.001 &   0.000 &  -0.001 &  -0.001 \\
            &&SE          &   0.010 &   0.010 &   0.010 &   0.010 &   0.010 &   0.010 &   0.014 &   0.014 &   0.014 \\
            &&SE$^*$      &   0.009 &   0.009 &   0.010 &   0.010 &   0.010 &   0.010 &   0.014 &   0.015 &   0.014 \\
\hline
 100 &  25  &BIAS        &   0.000 &   0.001 &   0.001 &  -0.001 &   0.001 &   0.001 &   0.000 &  -0.001 &  -0.0003 \\
            &&SE          &   0.014 &   0.014 &   0.014 &   0.014 &   0.014 &   0.014 &   0.020 &   0.020 &   0.020 \\
            &&SE$^*$      &   0.014 &   0.014 &   0.014 &   0.013 &   0.013 &   0.013 &   0.020 &   0.020 &   0.019 \\
\hline
 100 &  50  &BIAS        &  -0.000 &   0.001 &   0.001 &  -0.001 &   0.001 &  -0.001 &  -0.001 &  -0.001 &   0.001 \\
            &&SE          &   0.010 &   0.010 &   0.010 &   0.010 &   0.010 &   0.010 &   0.014 &   0.014 &   0.014 \\
            &&SE$^*$      &   0.010 &   0.010 &   0.010 &   0.010 &   0.009 &   0.009 &   0.014 &   0.015 &   0.014 \\
\hline
 100 &  100 &BIAS        &   0.000 &   0.000 &  -0.000 &  -0.000 &  -0.001 &   0.000 &   0.000 &   0.000 &   0.000 \\
            &&SE          &   0.007 &   0.007 &   0.007 &   0.007 &   0.007 &   0.007 &   0.010 &   0.010 &   0.010 \\
            &&SE$^*$      &   0.007 &   0.007 &   0.007 &   0.007 &   0.007 &   0.007 &   0.011 &   0.010 &   0.010 \\
\hline\hline
\end{tabular}
\end{center}
\end{table}
\end{landscape}